\documentclass[journal]{IEEEtran}

\usepackage{pdfpages}
\usepackage{graphicx}
\usepackage{framed}
\usepackage{colorweb}

\usepackage{amsmath, amsfonts, amssymb}

\usepackage{url}
\usepackage{cite}
\usepackage{multirow}

\usepackage{caption}
\usepackage{subcaption}

\usepackage{array}

\usepackage{algorithm}
\usepackage{algorithmicx}
\usepackage{algpseudocode}

\usepackage{bbding}
\usepackage{pifont}
\usepackage{wasysym}
\usepackage{amssymb}

\usepackage{booktabs}
\newcommand{\bhline}[1]{\noalign{\hrule height #1}}

\newcommand{\argmin}{\mathop{\rm argmin}\limits}
\newcommand*{\defeq}{\stackrel{\text{def}}{=}}

\newcommand{\la}{\lambda}

\def\x{{\mathbf x}}
\def\y{{\mathbf y}}
\def\u{{\mathbf u}}
\def\M{{\mathbf M}}
\def\I{{\mathbf I}}

\definecolor{module-N1}{cmyk}{0.00, 0.42, 0.14, 0.0}
\definecolor{module-N2}{cmyk}{0.10, 0.20, 0.00, 0.0}
\definecolor{module-N3}{cmyk}{0.30, 0.00, 0.80, 0.0}
\definecolor{module-D1}{cmyk}{0.00, 0.20, 0.70, 0.0}
\definecolor{module-D2}{cmyk}{0.60, 0.00, 0.00, 0.0}
\definecolor{module-D3}{cmyk}{0.00, 0.05, 0.75, 0.0}


\title{Graph Signal Restoration Using \\ Nested Deep Algorithm Unrolling}

\author{Masatoshi Nagahama,~\IEEEmembership{Student~Member,~IEEE,}
Koki~Yamada,~\IEEEmembership{Student~Member,~IEEE,}
Yuichi~Tanaka,~\IEEEmembership{Senior~Member,~IEEE,}
Stanley~H.~Chan,~\IEEEmembership{Senior~Member,~IEEE,}
and Yonina~C.~Eldar,~\IEEEmembership{Fellow,~IEEE}\vspace{-0.2in}

\thanks{Preliminary results of this work was presented in \cite{nagahama2021GraphSignal}.}
\thanks{M. Nagahama and Y. Tanaka are with the Department of Electrical Engineering and Computer Science, Tokyo University of Agriculture and Technology, Koganei, Tokyo 184--8588, Japan. Y. Tanaka is also with PRESTO, Japan Science and Technology Agency, Kawaguchi, Saitama 332--0012, Japan (email: nagahama@msp-lab.org; ytnk@cc.tuat.ac.jp).}
\thanks{K. Yamada was with the Department of Electrical Engineering and Computer Science, Tokyo University of Agriculture and Technology, Koganei, Tokyo 184--8588, Japan. He is now with the Department of Electrical Engineering, Tokyo University of Science, Katsushika, Tokyo 125--8585, Japan (email: k-yamada@rs.tus.ac.jp).}
\thanks{S. H. Chan is with School of Electrical and Computer Engineering, Purdue University, West West Lafayette, IN 47907, USA (email: stanchan@purdue.edu).}
\thanks{Y. C. Eldar is with Faculty of Mathematics and Computer Science, The Weizmann Institute of Science, Rehovot 7610001, Israel (email: yonina.eldar@weizmann.ac.il).}
\thanks{Y. Tanaka is funded in part by JST PRESTO under Grant JPMJPR1935 and JSPS KAKENHI under Grant 20H02145.}
\thanks{S. H. Chan is supported, in part, by the US National Science Foundation under the grants 1763896, 2030570, 2134209, and 2133032.}
}

\begin{document}

\maketitle
\begin{abstract}
Graph signal processing is a ubiquitous task in many applications such as sensor, social, transportation and brain networks, point cloud processing, and graph neural networks.
Often, graph signals are corrupted in the sensing process, thus requiring restoration.
In this paper, we propose two graph signal restoration methods based on deep algorithm unrolling (DAU).
First, we present a graph signal denoiser by unrolling iterations of the alternating direction method of multiplier (ADMM).
We then suggest a general restoration method for linear degradation by unrolling iterations of  Plug-and-Play ADMM (PnP-ADMM).
In the second approach, the unrolled ADMM-based denoiser is incorporated as a submodule, leading to a nested DAU structure.
The parameters in the proposed denoising/restoration methods are trainable in an end-to-end manner.
Our approach is interpretable and keeps the number of parameters small since we only tune graph-independent regularization parameters.
We overcome two main challenges in existing graph signal restoration methods:
1) limited performance of convex optimization algorithms due to fixed parameters which are often determined manually.
2) large number of parameters of graph neural networks that result in difficulty of training.
Several experiments for graph signal denoising and interpolation are performed on synthetic and real-world data.
The proposed methods show performance improvements over several existing techniques in terms of root mean squared error in both tasks.
\end{abstract}
\begin{IEEEkeywords}
  Graph signal processing, 
  signal restoration,
  deep algorithm unrolling,
  Plug-and-Play ADMM
\end{IEEEkeywords}

\section{Introduction}
Signal restoration is a ubiquitous task in many applications.
Depending on the types of signals, the interconnectivity among samples can often be exploited, for example, signals residing on sensor networks, social networks, transportation networks, and brain networks, power grids, 3D meshes, and point clouds, all have various connectivities which can often be represented as graphs.

A \textit{graph signal} is defined as a signal whose domain is the nodes of the graph.
The relations between the samples, i.e., nodes, are given by the edges.
In contrast to standard signals on a regular grid such as audio and image signals, graph signal processing (GSP) explicitly exploits the underlying structure of the signal \cite{sandryhaila2013,shuman2013,ortega2018}.
GSP has been used in a wide range of applications for irregularly-structured data such as compression \cite{hu2015}, sampling and restoration \cite{chen2015b,anis2016,tanaka2020SamplingSignals, hara2020GeneralizedGraph, tanaka2020GeneralizedSampling}, and analysis of graph signals \cite{sakiyama2016SpectralGraph,shuman2020LocalizedSpectral}.

Graph signal restoration is an important task aiming to address the problems of  noise and missing values.
For example in sensor networks, some sensors may not work properly resulting in missing values, and samples on the nodes are often noisy \cite{narang2013Signalprocessing}.
Many approaches for graph signal restoration have been proposed based on regularized optimization \cite{chen2015a}, graph filters and filter banks \cite{chen2014a, onuki2016, tanaka2014a}, and deep learning on graphs \cite{do2020}.
These existing works can be classified into two main approaches:
1) model-based restoration and
2) neural network-based restoration.

\begin{figure*}[!t]
  \begin{minipage}[t]{1.0\linewidth} \centering
    \centerline{\includegraphics[width=14cm]{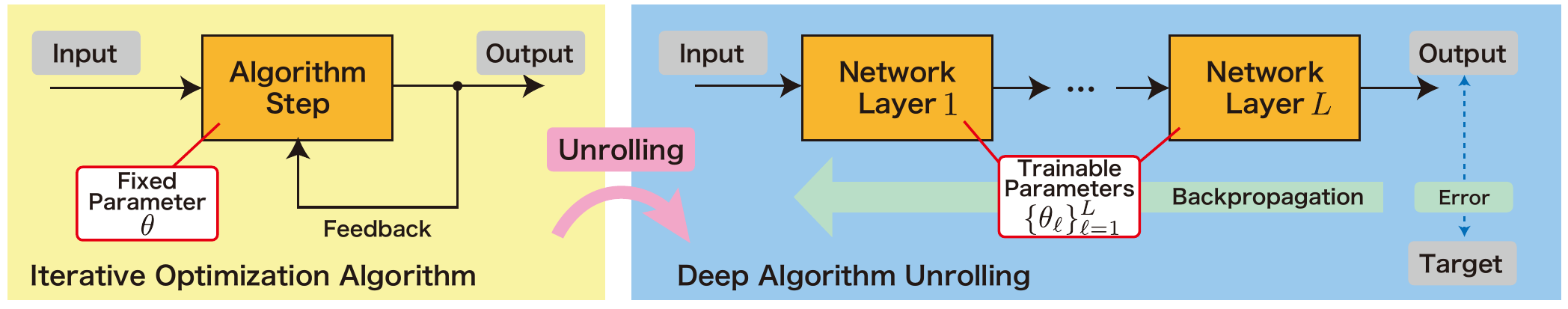}}
  \end{minipage}
  \caption{Conventional iterative optimization algorithm (left) and deep algorithm unrolling (right).}
  \label{fig:overview}
\end{figure*}

\textbf{Model-based restoration:}
Model-based approaches often rely on convex optimization whose objective function contains a data fidelity term and a regularization term \cite{chen2015a}.
Signal priors are often required in such tasks because the problem is ill-posed.
For example, a smoothness prior like graph total variation (GTV) has demonstrated effectiveness in graph signal denoising, whereas graph spectral filters have been shown to satisfy certain quadratic optimization solutions \cite{chan2001a,ono2015,yazaki2019,gadde2013}.
A limitation of model-based restoration methods is that they are often iterative as illustrated in Fig.~\ref{fig:overview} (left).
Performance and speed of the algorithm depend on the hyper-parameters $\theta$ (e.g., step size and regularization strength) whose values are determined manually and are fixed throughout the iterations.

\textbf{Neural network-based restoration:}
Graph convolutional networks (GCNs) are considered as a counterpart of the convolutional neural networks for image processing \cite{kipf2017SemiSupervisedClassification}.
GCNs can automatically learn network parameters to minimize a loss function.
However, GCNs have two drawbacks:
1) lack of interpretability
and 2) the requirement of a large dataset for training.
Furthermore, as reported in \cite{defferrard2016a, kipf2017SemiSupervisedClassification,chen2020Graphunrolling}, deeper networks cannot always achieve good performance in the graph settings, in contrast to the remarkable success of convolutional networks for signals on a regular grid \cite{ulyanov2020}.
Therefore, many GCNs are limited to a small number of layers \cite{zhang2018EndtoEndDeep, wang2019DynamicGraph}.

As a hybrid approach of the model- and neural network-based restoration methods, we utilize \textit{deep algorithm unrolling} (DAU) by integrating learnable parameters into the iterative algorithm \cite{zhang2018Dynamicallyunfolding,li2020EfficientInterpretable,bertocchi2020,monga2021AlgorithmUnrollinga, chen2020Graphunrolling}.
As illustrated in Fig.~\ref{fig:overview} (right), DAU unrolls the iterations of the iterative algorithm and deploys the trainable parameters at each unrolled iteration \cite{gregor2010LearningFast,monga2021AlgorithmUnrollinga}.
Instead of manually choosing the parameters as in the conventional iterative approach, parameters in each unrolled iteration are determined from the training data so as to minimize a loss function.
The practical advantages of DAU against the classical iterative solver are faster convergence and performance improvement since the parameters are learned to fit the target signals.
Advantages compared with fully parameterized neural networks are the interpretability and a small number of parameters.
Hence, the networks can be trained with a small number of training data.

An extension of DAU for graph signal denoising was recently developed in \cite{chen2020Graphunrolling}, which proposed unrolled GCNs based on two optimization problems of sparse coding and trend filtering.
Although the formulation itself allows the network to be arbitrarily deep, the number of layers is set to be very small (typically one middle layer) in its practical implementation.
This is because deeper networks do not result in better performance in this case.
Additionally, GCNs often assume a fixed graph both in the training and testing phases.
However, the underlying graphs are often slightly perturbed in practice.
Hence, restoration algorithms should be robust to (small) perturbations of graphs.
A detailed comparison between \cite{chen2020Graphunrolling} and our approach is further discussed in Section \ref{subsec:difference}.

In this work, we first propose a simple yet efficient graph signal \textit{denoising} method that utilizes DAU of the alternating direction method of multiplier (ADMM) to solve a minimization problem with two regularizers based on graph total variation and elastic net.
In contrast to \cite{chen2020Graphunrolling}, we only train the graph-independent regularization parameters in the model-based iterative algorithms.
The resulting denoising algorithm contains a significantly smaller number of parameters than neural network-based methods while showing better denoising results.

Next, we propose a nested version of DAU based on unrolling the iterations of Plug-and-Play ADMM (PnP-ADMM) \cite{venkatakrishnan2013a,ahmad2020PlugandPlayMethodsa,sreehari2016, chan2017, chan2019a}.
This version is designed for general graph signal restoration problems with linear degradation.
In this approach, the ADMM-based denoiser is plugged into the unrolled PnP-ADMM algorithm leading to a nested DAU structure.
All of the parameters in the algorithm are trained in an end-to-end fashion \cite{gregor2010LearningFast, monga2021AlgorithmUnrollinga, li2020EfficientInterpretable}.

In contrast to GCN-based methods, parameters to be tuned in the proposed techniques are graph-independent
leading to the following advantages:
\begin{enumerate}
    \item Interpretability: All internal modules are designed based on (convex) optimization algorithms.
    \item Ease to train: Our techniques do not require large training data due to the small number of parameters.
    \item Transferability: Since our methods only tune graph-independent parameters, we can immediately use the same parameter set for graphs with different sizes.
\end{enumerate}
We also avoid large matrix inversion by using popular acceleration techniques in GSP:
1) precomputing graph Fourier bases and 2) polynomial approximation.
Through comprehensive experiments on denoising and interpolation for synthetic and real-world data, our proposed methods are shown to achieve better performance than existing restoration methods including graph low-pass filters, model-based iterative optimization, and \cite{chen2020Graphunrolling}, in terms of root mean squared error (RMSE).

The remainder of this paper is organized as follows.
Signal restoration algorithms using ADMM and PnP-ADMM are introduced in Section \ref{sec:pnp_admm} along with notation used throughout the paper.
The proposed two restoration methods are introduced in Section \ref{sec:proposed_methods}.
Experimental results comparing denoising and interpolation performances with existing methods are shown in Sections \ref{sec:denoising} and \ref{sec:interpolation}.
Section \ref{sec:conclusion} concludes this paper.

\section{Signal restoration with ADMM}

\label{sec:pnp_admm}
In this section, we first present notations and the problem formulation.
Then, we review ADMM and PnP-ADMM which are the fundamental building blocks of our algorithms.

\subsection{Notation}
Throughout the paper, vectors and matrices are written in bold style and sets are written as calligraphic letters.
An undirected graph $\mathcal{G}=(\mathcal{V},\mathcal{E},\mathbf{W})$ consists of a collection of undirected vertices $\mathcal{V}=\{v_{i}\}_{i=1}^{N}$ and edges $\mathcal{E}=\{(e_{i,j},w_{i,j})\}$.
The number of vertices and edges is $|\mathcal{V}|=N$ and $|\mathcal{E}|$, respectively;
$w_{i,j} \in \mathbb{R}_{\geq 0}$ denotes the edge weight between $v_i$ and $v_j$.
We define a weighted adjacency matrix of $\mathcal{G}$ as an $N \times N$ matrix with $[\mathbf{W}]_{ij} = w_{i,j}$; $[\mathbf{W}]_{ij} = 0$ represents unconnected vertices.
In this paper, we consider a graph that does not have self-loops, i.e., $[\mathbf{W}]_{ii}=0$ for all $i$.
The degree matrix of $\mathcal{G}$ is defined as a diagonal matrix $[\mathbf{D}]_{ii} = \sum_{j} w_{i,j}$.
The combinatorial graph Laplacian matrix of $\mathcal{G}$ is given by $\mathbf{L=D-W}$.
Since $\mathbf{L}$ is a real symmetric matrix, $\mathbf{L}$ always has an eigendecomposition.
Let the eigendecomposition of the graph Laplacian matrix be $\mathbf{L}=\mathbf{U}\mathbf{\Lambda}\mathbf{U}^{\top}$, where $\mathbf{U}$ is an eigenvector matrix and $\mathbf{\Lambda}=\text{diag}(\la_1,\ldots,\la_N)$.
The weighted graph incidence matrix is denoted as $\mathbf{M} \in \mathbb{R}^{|\mathcal{E}| \times N}$.
We index integers to the set of edges as $\mathcal{E}=\{e_{s}\}_{s=1}^{|\mathcal{E}|}$.
Then, the $s$th row and $t$th column of $\mathbf{M}$ corresponding to $e_{s}$ and $v_{t}$ is
\begin{equation}
  [\mathbf{M}]_{s,t} = \left\{
    \begin{array}{ll}
       \sqrt{w_{i,j}} & e_{s} = (v_i,v_j)~\text{and}~t=i, \\
      -\sqrt{w_{i,j}} & e_{s} = (v_i,v_j)~\text{and}~t=j, \\
      0 & \text{otherwise}.
    \end{array}
    \right. \nonumber
\end{equation}

A graph signal $x: \mathcal{V} \rightarrow \mathbb{R}$ is a function that assigns a value to each vertex. It can be written as a vector $\mathbf{x} \in \mathbb{R}^N$ in which the $i$th element $x[i]$ represents the signal value at the $i$th vertex.

\subsection{General Restoration Problem}
Consider an observed graph signal $\mathbf{y} \in \mathbb{R}^{N}$ which is related to an input graph signal $\mathbf{x} \in \mathbb{R}^{N}$ as
\begin{align}
  \mathbf{y}=\mathbf{Hx+n},
  \label{eq:problem}
\end{align}
where $\mathbf{H} \in \mathbb{R}^{N \times N}$ is a degradation matrix
and $\mathbf{n} \sim \mathcal{N}(0, \sigma^2\mathbf{I})$ is an i.i.d.~additive white Gaussian noise (AWGN).

Throughout the paper, we assume that $\mathbf{x}$ is a graph signal, i.e., its domain is given by $\mathcal{G}$.
This graph structure will be exploited to provide a prior for the recovery problem.
The degradation model \eqref{eq:problem} generally appears in restoration problems such as denoising, interpolation, deblurring, and super-resolution, to name a few.
The main objective of many restoration problems is estimating an unknown $\mathbf{x}$ from a given degraded signal $\mathbf{y}$.
We assume that $\mathbf{H}$ is known a priori.
In this paper, we perform two representative experiments with the following $\mathbf{H}$: 1) $\mathbf{H} = \mathbf{I}$ (denoising), and 2) a binary $\mathbf{H}$ matrix (interpolation).

\subsection{Plug-and-Play ADMM}
\subsubsection{ADMM}
Many inverse problems are posed as the following unconstrained minimization problem:
\begin{align}
  \underset{\mathbf{x} \in \mathbb{R}^{N}} {\text{min}} ~
  \frac{1}{2} \|\mathbf{y} - \mathbf{Hx}\|_2^{2} + \la g(\mathbf{Ax}),
  \label{eq:general_problem}
\end{align}
where $g$ is some regularization function, $\lambda \in \mathbb{R}_{\ge 0}$ is the regularization parameter, and $\mathbf{A} \in \mathbb{R}^{M \times N}$ is an arbitrary matrix.
A widely used algorithm to solve \eqref{eq:general_problem} is the alternating direction of multipliers (ADMM) which has been used to solve generic unconstrained optimization problems with nondifferentiable convex functions (see \cite{boyd2010} for details).
Through variable splitting, the general problem \eqref{eq:general_problem} is rewritten as the following constrained minimization problem:
\begin{align}
    (\mathbf{\tilde{x}}, \mathbf{\tilde{s}}) =&
    \underset{\x \in \mathbb{R}^{N}, \mathbf{s} \in \mathbb{R}^{M}} {\text{argmin}}~ \frac{1}{2} \|\mathbf{y} - \mathbf{Hx}\|_2^{2} + \la g(\mathbf{s}), \nonumber \\
    &\text{subject to} ~ \mathbf{s=Ax}.
  \label{eq:general_problem2}
\end{align}
Applying ADMM to \eqref{eq:general_problem2} leads to the following sequence of subproblems:
\begin{subequations}

    \begin{align} %
\mathbf{x}^{(p+1)} &=
	  \left(\mathbf{H}^{\top}\mathbf{H} + {\rho}\mathbf{A}^{\top}\mathbf{A}\right)^{-1}
	  \left(\mathbf{H}^{\top}\mathbf{y} + \rho \mathbf{A}^{\top} \bar{\mathbf{x}}^{(p)} \right),
	  \label{eq:general_ADMM1}
    \\
	  \mathbf{s}^{(p+1)} &= \argmin_{ \mathbf{s} \in \mathbb{R}^{M}}
    \lambda g(\mathbf{s}) + \frac{\rho}{2} \|\mathbf{s} - \bar{\mathbf{s}}^{(p)}\|_2^{2}, \label{eq:general_ADMM2}
    \\
    \mathbf{t}^{(p+1)} &= \mathbf{t}^{(p)} + \mathbf{A}\mathbf{x}^{(p+1)} - \mathbf{s}^{(p+1)},
    	  \label{eq:general_ADMM3}
	\end{align}
\end{subequations}
where $\mathbf{t}^{(p)} \in \mathbb{R}^{M}$ is the Lagrangian multiplier, $\mathbf{s}^{(p)} \in \mathbb{R}^{M}$ is an auxiliary variable, $g$ is the regularization function in \eqref{eq:general_problem},
$\bar{\mathbf{x}}^{(p)} \defeq \mathbf{s}^{(p)} - \mathbf{t}^{(p)}$
and $\bar{\mathbf{s}}^{(p)} \defeq \mathbf{Ax}^{(p+1)} + \mathbf{t}^{(p)}$.

\subsubsection{Plug-and-Play ADMM}
PnP-ADMM is a variation of the classical ADMM \cite{venkatakrishnan2013a} for the problem of \eqref{eq:general_problem2} with $\mathbf{A} = \mathbf{I}$.
Oftentimes, \eqref{eq:general_ADMM1} and \eqref{eq:general_ADMM2} are called the \textit{inverse step} and \textit{denoising step} (i.e., denoiser), respectively \cite{ahmad2020PlugandPlayMethodsa}.
A notable feature is that any off-the-shelf denoiser, including deep neural networks, can be used instead of naively solving \eqref{eq:general_ADMM2} without explicitly specifying regularization terms $g$ before implementation.
Such examples are found in \cite{zhang2017LearningDeep, zhang2019DeepPlugAndPlay, wei2020TuningfreePlugandPlay}.

Empirically, PnP-ADMM has demonstrated improved performance over the standard ADMM with explicit regularization in some image restoration tasks \cite{sreehari2016, kamilov2017PlugandPlayPriors, shastri2020AutotuningPlugandPlay}.
Graph signal restoration with PnP-ADMM is also studied in \cite{yazaki2019} showing improved restoration performance over the existing model-based techniques.

In this paper, we follow an approach of PnP-ADMM proposed in \cite{chan2017}.
Suppose that two initial variables $\mathbf{s}^{(0)},\mathbf{t}^{(0)} \in \mathbb{R}^{N}$ are set.
The algorithm of PnP-ADMM corresponding to \eqref{eq:general_ADMM1}--\eqref{eq:general_ADMM3} (again, assuming $\mathbf{A} = \mathbf{I}$) is represented as
\begin{subequations}
	\begin{align} %
	  \mathbf{x}^{(p+1)} &=
	  \left(\mathbf{H}^{\top}\mathbf{H} + {\rho}\mathbf{I}\right)^{-1}
	  \left(\mathbf{H}^{\top}\mathbf{y} + \rho \left(\mathbf{s}^{(p)} - \mathbf{t}^{(p)}\right) \right),
	  \label{eq:normal_pnp_admm1}
	\\
	  \mathbf{s}^{(p+1)} &= \mathcal{D}_{g}\left(\mathbf{x}^{(p+1)} + \mathbf{t}^{(p)}\right),
	  \label{eq:normal_pnp_admm2}
	\\
	  \mathbf{t}^{(p+1)} &= \mathbf{t}^{(p)} + \mathbf{x}^{(p+1)} - \mathbf{s}^{(p+1)},
	  \label{eq:normal_pnp_admm3}
	\end{align}
\end{subequations}
where $\mathcal{D}_{g}$ is an off-the-shelf denoiser.
Note that we still need to determine the parameter $\rho$ and the off-the-shelf graph signal denoiser $\mathcal{D}_{g}$ (and its internal parameters) prior to running the algorithm.

The key idea of our proposed method is to unroll the ADMM and PnP-ADMM for graph signal processing.

\section{Graph Signal Restoration Algorithms} \label{sec:proposed_methods}
\begin{figure*}[!tb]
  \begin{minipage}[t]{1.0\linewidth} \centering
    \centerline{\includegraphics[width=0.7\linewidth]{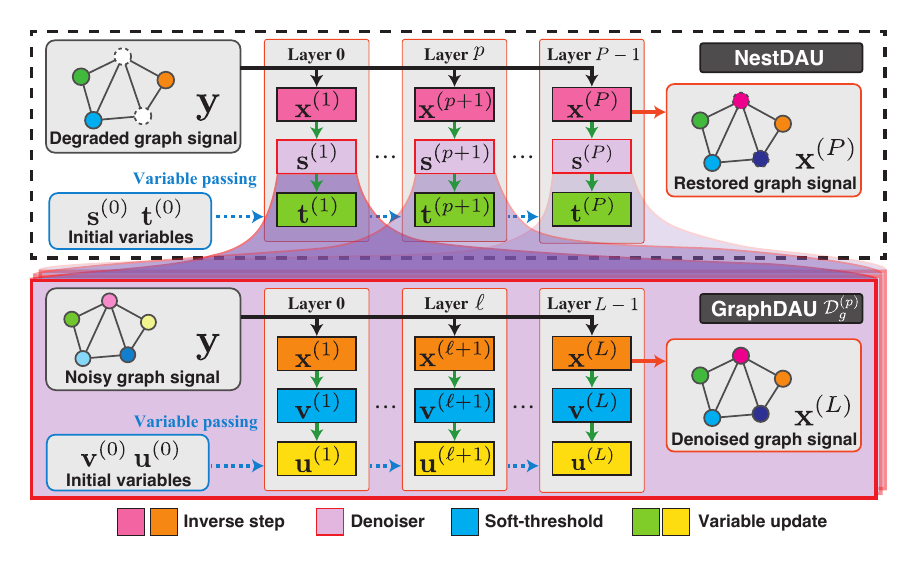}}
  \end{minipage}
  \caption{Overview of GraphDAU and NestDAU for graph signal restoration. The top represents NestDAU: General restoration based on PnP-ADMM. The bottom represents GraphDAU: Denoiser based on ADMM. GraphDAU can be used as the off-the-shelf denoiser in NestDAU.}
  \label{fig:archtecture}
\end{figure*}

In this section, we propose the following two graph signal restoration methods, both based on DAU.
\begin{enumerate}
    \item GraphDAU: Graph signal denoiser by unrolling ADMM to address the problem $\mathbf{H = I}$. We consider a mixture of $\ell_1$ and $\ell_2$ regularization terms like the elastic net \cite{zou2005Regularizationvariable}.
    GraphDAU works as a better independent denoiser than the model-based and deep-learning-based approaches.
    \item NestDAU: General graph signal restoration algorithm by unrolling PnP-ADMM to  handle a generic $\mathbf{H}$.
    We plug the GraphDAU into each layer of an unrolled PnP-ADMM as a denoiser.
\end{enumerate}
Our methods are illustrated in Fig.~\ref{fig:archtecture}.

\subsection{GraphDAU} \label{ssec:denoiser}
GraphDAU considers the case where $\mathbf{H = I}$ due to signal denoising and $\mathbf{A = M}$ in \eqref{eq:general_problem}.
It combines the regularization terms of graph total variation (GTV) and graph Laplacian regularization \cite{pang2017GraphLaplacian}, leading to
\begin{align}
  \mathcal{D}_{g}(\mathbf{y}) &=
  \underset{\x \in \mathbb{R}^{N}}{\text{argmin}} ~ \frac{1}{2} \|\x - \y\|_{2}^{2}
  + \la_{1} \|\M \x\|_{1} + \frac{\la_{2}}{2} \|\M \x\|_{2}^{2}, \label{eq:min_prob}
\end{align}
where $\|\M\x\|_{1}$ and $\|\M \x\|_{2}^{2}=\x^{\top} \mathbf{L} \x$ (since $\mathbf{L}=\M^{\top}\M$) are the regularization terms for first-order and second-order differences, respectively, and $\la_1$ and $\la_2$ are nonnegative regularization parameters.
The second and third terms in \eqref{eq:min_prob} can be written explicitly as
\begin{align}
  \|\mathbf{M}\mathbf{x}\|_{1} &= \sum_{j \in \mathcal{N}_{i}}\sqrt{w_{ij}}~|x_{i}-x_{j}|, \label{eq:l1_regularizer}\\
  \|\mathbf{M}\mathbf{x}\|_{2}^{2} &= \sum_{j \in \mathcal{N}_{i}} w_{ij}~(x_{i}-x_{j})^{2}, \label{eq:l2_regularizer}
\end{align}
where $\mathcal{N}_{i}$ is a set of vertices connecting with $v_i$.
The norms \eqref{eq:l1_regularizer} and \eqref{eq:l2_regularizer} are effective regularization functions for piecewise constant and smooth graph signals \cite{ono2015}.

In this paper, GraphDAU is only applied for denoising and not used for the general restoration problems in \eqref{eq:general_problem}.
This is because we use various acceleration techniques introduced in Section \ref{denoiser_acceleration} under the assumption $\mathbf{H} = \mathbf{I}$.

We utilize ADMM as a baseline iterative solver of \eqref{eq:min_prob}.
The variable splitting is applied to \eqref{eq:min_prob} with $\mathbf{v} = \M\x$, leading to the following constrained minimization problem:
\begin{align}
  \mathcal{D}_{g}(\mathbf{y}) = &
  \underset{\x \in \mathbb{R}^{N}, \mathbf{v} \in \mathbb{R}^{|\mathcal{E}|}} {\text{argmin}}~ \frac{1}{2}\|\x-\y\|_{2}^2 + \la_{1}\|\mathbf{v}\|_{1} + \frac{\la_{2}}{2}\|\mathbf{v}\|_{2}^{2}, \nonumber
  \\
  &\text{subject to} ~ \mathbf{v}=\M\x.
  \label{eq:optim_tv3}
\end{align}
The solution of \eqref{eq:optim_tv3} can be found by solving a sequence of the following subproblems\cite{parikh2014ProximalAlgorithms}:
\begin{subequations}
\begin{align}
  \x^{(\ell+1)}
  &= \left(\I + \frac{1}{\gamma} \M^{\top}\M \right)^{-1} \left(\y + \frac{1}{\gamma} \M^{\top} (\mathbf{v}^{(\ell)} - \u^{(\ell)}) \right), \label{eq:admm_graph_re1}
\\
  \mathbf{v}^{(\ell+1)}
  &= \frac{1}{1 + \la_2\gamma} S_{\la_1\gamma}(\M\x^{(\ell+1)} + \u^{(\ell)}), \label{eq:admm_graph_re2}
\\
  \u^{(\ell+1)}
  &= \u^{(\ell)} + \M\x^{(\ell+1)} - \mathbf{v}^{(\ell+1)}, \label{eq:admm_graph_re3}
\end{align}
\end{subequations}
where $\gamma$ is the step size of the algorithm
and $S_{\la_{1} \gamma}$ is the soft-thresholding operator
\begin{align}
  [S_{\la_{1} \gamma}(\mathbf{x})]_{i} = \text{sgn}(x_{i})\max\{|x_{i}|-\la_{1} \gamma, 0\},
\end{align}
where $\text{sgn}(\cdot)$ denotes the signum function.

Next, we unroll the iteration of \eqref{eq:admm_graph_re1}--\eqref{eq:admm_graph_re3} to design a trainable $\mathcal{D}_{g}$.
In other words, instead of using fixed parameters in \eqref{eq:admm_graph_re1}--\eqref{eq:admm_graph_re2}, we deploy trainable parameters in each iteration.
The terms including $\M$ and $\M^{\top}\M$ in \eqref{eq:admm_graph_re1}--\eqref{eq:admm_graph_re3} are graph filters, i.e., \textit{graph convolution}, and are fixed:
We only tune three parameters, $\gamma$, $\lambda_1$, and $\lambda_2$, in each unrolled iteration.
This is because we aim to construct an interpretable and easy-to-train graph signal restoration algorithm.
The training configurations are described later in Section \ref{sssec:training_config}\footnote{The detailed gradient computations for training the parameters are given in the Appendix.}.
In the following sections, we propose two forms of GraphDAU and introduce its acceleration techniques.
\subsubsection{GraphDAU-TV}
In this method, we only consider the $\ell_{1}$ term of \eqref{eq:optim_tv3} by setting $\la_{2}=0$.
Then, we choose $\gamma$ and $\gamma \la_1$ to be learnable, i.e., $\gamma \to \{\gamma_{\ell}\}_{\ell=0}^{L-1}$ and $\gamma \la_{1} \to \{\beta_{\ell}\}_{\ell=0}^{L-1}$.
This regularization is based on the assumption that the signal is piecewise constant.

\subsubsection{GraphDAU-EN}
This GraphDAU is based on a combination of the $\ell_{1}$ and $\ell_{2}$ regularizations in \eqref{eq:optim_tv3} like the elastic net (EN), defined by the weighted incidence matrix $\mathbf{M}$.
We introduce a set of trainable parameters $1/(1 + \la_{2}\gamma) \to \{\alpha_{\ell}\}_{\ell=0}^{L-1}$ in addition to $\{\gamma_{\ell}\}_{\ell=0}^{L-1}$ and $\{\beta_{\ell}\}_{\ell=0}^{L-1}$.
This method automatically controls piecewise and smoothness terms at each layer.

\subsubsection{Algorithm Acceleration} \label{denoiser_acceleration}
The graph filter $(\I+\frac{1}{\gamma}\M^{\top} \M)^{-1}$ in \eqref{eq:admm_graph_re1} requires matrix inversion and its computational complexity is typically $\mathcal{O}(N^3)$ (for a dense matrix).
If each layer requires calculating the inversion, the complexity becomes $\mathcal{O}(N^3 L)$.
We consider accelerating GraphDAU by the following two popular techniques:
1) eigendecomposition of $\mathbf{L}$, and
2) Chebyshev polynomial approximation of a graph filter.

\noindent
\textbf{Precomputing Eigendecomposition:}
In this approach, we precompute the eigendecomposition (EVD) of $\mathbf{L}$.
The inverse matrix in \eqref{eq:admm_graph_re1} can be decomposed as
\begin{align}
  \left(\mathbf{I}+\frac{1}{\gamma_{\ell}} \mathbf{L}\right)^{-1}
  &=\mathbf{U}\left(\mathbf{I}+\frac{1}{\gamma_{\ell}}\mathbf{\Lambda}\right)^{-1}\mathbf{U}^{\top}.
  \label{eq:eigen_decom1}
\end{align}
Since $\mathbf{I}+(1/\gamma_{\ell})\mathbf{\Lambda}$ is a diagonal matrix,
the inversion has $\mathcal{O}(N)$ complexity.
If $\mathcal{G}$ does not change frequently throughout the iterations (which often is the case), the eigenvalues $\mathbf{\Lambda}$ and eigenvectors $\mathbf{U}$ are fixed.
Therefore, the eigendecomposition of the graph Laplacian is performed only once.
This GraphDAU with acceleration is represented with the suffix -E and is summarized in Algorithm~\ref{alg:deep_denoiser}.

\noindent
\textbf{Chebyshev Polynomial Approximation:}
This technique approximates \eqref{eq:eigen_decom1} with a polynomial,
for example, using the Chebyshev polynomial approximation (CPA) (see \cite{mallat2009,onuki2017} for details).

First, we rewrite the inverse step at the $\ell$th layer corresponding to \eqref{eq:eigen_decom1} as
\begin{align}
  \mathbf{x}^{(\ell + 1)}
  &=\mathcal{H}^{(\ell)}(\mathbf{L})\widetilde{\mathbf{y}}^{(\ell)},
  \label{eq:cpf}
\end{align}
where $\widetilde{\mathbf{y}}^{(\ell)}=\mathbf{y}+\frac{1}{\gamma_{\ell}} \mathbf{M}^{\top}\left(\mathbf{v}^{(\ell)}-\mathbf{u}^{(\ell)}\right)$
and $\mathcal{H}^{(\ell)}(\mathbf{L}):= \mathbf{U}\mathcal{H}^{(\ell)}(\mathbf{\Lambda})\mathbf{U}^\top$ is the filter function.
This filter kernel has the following graph frequency response:
\begin{align}
  \mathcal{H}^{(\ell)}(\mathbf{\Lambda})
  =\text{diag}\left(h^{(\ell)}(\lambda_{1}),\ldots,h^{(\ell)}(\lambda_{N})\right),
\end{align}
where $h^{(\ell)}(x)={\gamma_{\ell}}/(\gamma_{\ell}+x)$ is the filter kernel which acts as a graph low-pass filter.
By performing $K$-truncated Chebyshev approximations to $h^{(\ell)}(x)$, the approximated version $\widetilde{\mathcal{H}}^{(\ell)}(\mathbf{L})$ is represented as:
\begin{align}
  \mathbf{x}^{(\ell + 1)}
  &=\widetilde{\mathcal{H}}^{(\ell)}(\mathbf{L})\widetilde{\mathbf{y}}^{(\ell)}.
  \label{eq:cpf_app}
\end{align}
GraphDAU with Chebyshev polynomial approximation is specified by a suffix -C in Algorithm~\ref{alg:deep_denoiser}.

\begin{algorithm}[!tbp]
	\caption{GraphDAU for graph signal denoising $\mathcal{D}_{g}^{(p)}$\\
		\quad \textit{NOTE}: Background colors correspond to those in Fig.~\ref{fig:archtecture}.
	} \label{alg:deep_denoiser}
	\begin{algorithmic}[1]
		\renewcommand{\algorithmicrequire}{\textbf{Input:}}
		\renewcommand{\algorithmicensure}{\textbf{Output:}}
		\Require noisy graph signal $\mathbf{y}$; graph incidence matrix $\mathbf{M}$; initial variables $\mathbf{v}^{(0)},\mathbf{u}^{(0)}$; network layers $L$; polynomial order $K$ (for GraphDAU-TV-C and GraphDAU-EN-C).
		\Ensure denoised graph signal $\mathbf{x}^{(L)}$.
		\State compute the graph Laplacian $\mathbf{L}=\mathbf{M}^{\top}\mathbf{M}$
		\State \textbf{If} {GraphDAU-TV-E or GraphDAU-EN-E} \textbf{then}
		\State \hspace{0.5cm} compute the eigendecomposition $\mathbf{L}=\mathbf{U}\mathbf{\Lambda}\mathbf{U}^{\top}$
		\For{$\ell=0,\cdots,L-1$}
		\colorlet{shadecolor}{module-D1}
		\State
		\begin{minipage}{0.9\linewidth}\begin{shaded}
			$\widetilde{\mathbf{y}}^{(\ell)} \leftarrow
			\mathbf{y}+\frac{1}{\gamma_{\ell}} \mathbf{M}^{\top}\left(\mathbf{v}^{(\ell)}-\mathbf{u}^{(\ell)}\right)$
		\end{shaded}
		\end{minipage}
		\colorlet{shadecolor}{module-D1}
		\State \begin{minipage}{0.9\linewidth}\begin{shaded}
			$\mathbf{x}^{(\ell + 1)} \leftarrow
			\begin{cases}
				\mathbf{U} \left(\mathbf{I}+\frac{1}{\gamma_{\ell}} \mathbf{\Lambda}\right)^{-1}\mathbf{U}^{\top} \widetilde{\mathbf{y}}^{(\ell)}\\
				\hspace{0.5cm} \text{for} \ \text{GraphDAU-TV-E and} \\
				\hspace{0.9cm} \text{ GraphDAU-EN-E}
				\\
				\widetilde{\mathcal{H}}^{(\ell)}(\mathbf{L})\widetilde{\mathbf{y}}^{(\ell)}\\
				\hspace{0.5cm} \text{for} \ \text{GraphDAU-TV-C and} \\
				\hspace{1.04cm} \text{GraphDAU-EN-C}
			\end{cases}$ \end{shaded}
		\end{minipage}
\colorlet{shadecolor}{module-D2}
\State \begin{minipage}{0.9\linewidth}\begin{shaded}
	$\mathbf{v}^{(\ell+1)} \leftarrow
	\begin{cases} S_{\beta_{\ell}}\left(\mathbf{M}\mathbf{x}^{(\ell+1)}+\mathbf{u}^{(\ell)}\right)\\
		\hspace{0.5cm} \text{for} \ \text{GraphDAU-TV}\\
		\alpha_{\ell} S_{\beta_{\ell}}\left(\mathbf{M}\mathbf{x}^{(\ell+1)}+\mathbf{u}^{(\ell)}\right)\\
		\hspace{0.5cm} \text{for} \ \text{GraphDAU-EN}
	\end{cases}$
\end{shaded}
		\end{minipage}
		\colorlet{shadecolor}{module-D3}
		\State \begin{minipage}{0.9\linewidth}\begin{shaded}
			$\mathbf{u}^{(\ell+1)} \leftarrow \mathbf{u}^{(\ell)} +\mathbf{M}\mathbf{x}^{(\ell+1)} -\mathbf{v}^{(\ell+1)}$
		\end{shaded} \end{minipage}
		\EndFor
		\State \Return $\mathbf{x}^{(L)}$
	\end{algorithmic}
\end{algorithm}

\subsection{NestDAU: Unrolled PnP-ADMM with GraphDAU as the Denoiser} \label{ssec:DAU_pnp}
Next, we develop a restoration algorithm for general $\mathbf{H}$ in \eqref{eq:general_problem}.
The baseline algorithm we consider is PnP-ADMM introduced in \eqref{eq:general_ADMM1}--\eqref{eq:general_ADMM3} because it is able to adapt to general $\mathbf{H}$.
In addition, any denoiser can be used in its internal algorithm to boost performance.

Suppose that the iteration number $P$ is given.
We then unroll \eqref{eq:normal_pnp_admm1}--\eqref{eq:normal_pnp_admm3} of the PnP-ADMM iterations to construct $P$ layer networks.
That is, we set $\rho$ in \eqref{eq:normal_pnp_admm1} to be learnable, i.e., $\rho \rightarrow \{\rho_{p}\}_{p=0}^{P-1}$ in which $p$ indicates the layer number.
The restoration steps are equivalent to those in PnP-ADMM with $P$ iterations, but each iteration is conducted with different regularization parameters.

The important part of the restoration algorithm is the design of the off-the-shelf denoiser $\mathcal{D}_{g}$ in \eqref{eq:normal_pnp_admm2} since \eqref{eq:normal_pnp_admm1} and \eqref{eq:normal_pnp_admm3} are independent of the underlying graph.
In this paper, we aim to keep the algorithm fully interpretable and the number of parameters small for efficient training, and thereby, we utilize GraphDAU in each layer as $\mathcal{D}_{g}^{(p)}$.
As a result, the restoration algorithm has a nested unrolled structure as shown in Fig.~\ref{fig:archtecture}.
Based on this structure, we refer to the proposed method as \textit{NestDAU}.
Note that all the parameters in NestDAU, including those in GraphDAU, can be trained in an end-to-end fashion from a training set.
The training details are presented in Section \ref{sssec:training_config}.

Algorithm~\ref{alg1} shows the details of NestDAU.
Note that we perform two representative signal restoration experiments (i.e., denoising and interpolation) in this paper, but NestDAU can be applicable to other cases as well, e.g., deblurring \cite{yamamoto2016DeblurringPoint} and point cloud super-resolution \cite{dinesh2020SuperResolution3D}.

\begin{algorithm}[!tbp] %
    \caption{NestDAU for graph signal restoration\\
    \quad \textit{NOTE}: Background colors correspond to those in Fig.~\ref{fig:archtecture}.
    } \label{alg1}
  \begin{algorithmic}[1]
    \renewcommand{\algorithmicrequire}{\bf{Input:}}
    \renewcommand{\algorithmicensure}{\bf{Output:}}
  \Require degraded graph signal $\mathbf{y}$;
  initial variables $\mathbf{s}^{(0)},\mathbf{t}^{(0)}$; network layers $P$;
  trainable denoiser $\mathcal{D}_{g}^{(p)}$.
  \Ensure restored graph signal $\mathbf{x}^{(P)}$.
  \For{$p=0,\cdots,P-1$}
    \colorlet{shadecolor}{module-N1}
    \State \begin{minipage}{0.9\linewidth}\begin{shaded}
      $\mathbf{x}^{(p+1)} \leftarrow \left(\mathbf{H}^{\top}\mathbf{H}+ {\rho_p} \mathbf{I}\right)^{-1}
    \left(\mathbf{H}^{\top}\mathbf{y} + \rho_p \left(\mathbf{s}^{(p)}-\mathbf{t}^{(p)}\right)\right)$
    \end{shaded}
    \end{minipage}
    \colorlet{shadecolor}{module-N2}
    \State \begin{minipage}{0.9\linewidth}\begin{shaded}
      $\mathbf{s}^{(p+1)} \leftarrow \mathcal{D}_{g}^{(p)}\left(\mathbf{x}^{(p+1)} + \mathbf{t}^{(p)}\right)$
    \end{shaded}
    \end{minipage}
    \colorlet{shadecolor}{module-N3}
    \State \begin{minipage}{0.9\linewidth}\begin{shaded}
      $\mathbf{t}^{(p+1)} \leftarrow \mathbf{t}^{(p)}+ \mathbf{x}^{(p+1)}-\mathbf{s}^{(p+1)}$
    \end{shaded}
    \end{minipage}
  \EndFor
  \State \Return $\mathbf{x}^{(P)}$
  \end{algorithmic}
\end{algorithm}

\begin{table*}[!tb] \centering
\caption{Comparison among the proposed methods.} \label{tab:summary}
\begin{tabular}{c||c|c|c|c|l} \bhline{1.0pt}
Methods      & Task & Regularization & Acceleration & \# of params & Computational complexity \\ \hline\hline
GraphDAU-TV-E & \multirow{4}{*}{Denoising} & \multirow{2}{*}{Graph total variation} & Precomputing EVD  & \multirow{2}{*}{$2L$} & $\mathcal{O}(N^{3}+N^{2}L)$ \\
GraphDAU-TV-C & & & CPA &  & $\mathcal{O}(K|\mathcal{E}|L)$ \\ \cline{3-6}
GraphDAU-EN-E & & \multirow{2}{*}{Elastic net} & Precomputing EVD & \multirow{2}{*}{$3L$} & $\mathcal{O}(N^{3}+N^{2}L)$ \\
GraphDAU-EN-C & &  & CPA & & $\mathcal{O}(K|\mathcal{E}|L)$ \\ \hline
NestDAU-TV-E & \multirow{4}{*}{General restoration}& \multirow{2}{*}{Graph total variation} & Precomputing EVD & \multirow{2}{*}{$(2L+1)P$} & $\mathcal{O}(N^{3}+N^{2}LP)$ \\
NestDAU-TV-C & & & CPA & & $\mathcal{O}(K|\mathcal{E}|LP)$ \\ \cline{3-6}
NestDAU-EN-E & & \multirow{2}{*}{Elastic net} & Precomputing EVD & \multirow{2}{*}{$(3L+1)P$} & $\mathcal{O}(N^{3}+N^{2}LP)$ \\
NestDAU-EN-C & & & CPA & & $\mathcal{O}(K|\mathcal{E}|LP)$ \\ \bhline{1.0pt}
\end{tabular}
\end{table*}

\subsection{Summary of Computation Issues}
In Table~\ref{tab:summary}, we compare the proposed methods in terms of the regularization function, the acceleration technique, the number of parameters, and the computational complexity.
NestDAUs are classified based on its GraphDAU specifications and have the same suffix as the corresponding GraphDAU.

The number of parameters linearly increases in proportion to the number of layers $L$ but is independent of $N$.
The complexity mainly depends on the use of EVD.
The methods with EVD have complexities depending on the number of nodes $N$,
while those with CPA only rely on the number of edges $|\mathcal{E}|$ and the polynomial order $K$;
$K|\mathcal{E}|$ is generally much smaller than $N^{2}$ when $N$ becomes large.

As mentioned, the proposed methods require training data (i.e., a set of ground-truth and degraded data) to tune parameters.
They come from the hyperparameter(s) of the original (PnP-)ADMM algorithms.
Note that, even for a regular ADMM, we need to determine the optimal hyperparameter(s) for practical applications: This often needs training data.

In general, many trainable parameters in deep learning require a large dataset to avoid overfitting.
This implies that GNNs require many training data. 
In contrast, NestDAU and GraphDAU have significantly fewer parameters than representative deep learning methods. 
This leads to that the proposed method can train with the small number of training data, which is beneficial for practical applications.
This is experimentally verified in Sections \ref{sec:denoising} and \ref{sec:interpolation}.

\subsection{Comparison to \cite{chen2020Graphunrolling}}
\label{subsec:difference}
Two approaches for graph signal denoising based on DAU, called graph unrolling sparse coding (GUSC) and graph unrolling trend filtering (GUTF), were proposed in \cite{chen2020Graphunrolling}.
Since they have the same objective as that for GraphDAU,
we compare the details of DAU-based graph signal restoration methods in Table~\ref{tab:methods_comparison}.

First, GUSC/GUTF only consider the problem of graph signal denoising.
This is the same objective as that of GraphDAU, while NestDAU focuses on a generic restoration problem in \eqref{eq:problem}.
This is possible by employing the PnP-ADMM as a prototype of the iterative algorithm.
Second, the regularization of GUSC/GUTF only contains the $\ell_1$ term, while GraphDAU also includes an $\ell_2$ term $\|\M \x\|_{2}^{2}=\x^{\top} \mathbf{L} \x$, which is beneficial for globally smooth signals.
GraphDAU-EN can automatically control the regularization weights between the $\ell_1$ and $\ell_2$ terms, leading to flexibility in capturing signal characteristics.
Third, GUSC and GUTF train parameters in an unsupervised setting while our proposed methods train the network in a supervised way.
In the following experiments, we train GUSC/GUTF in a supervised setting for a fair comparison.
Extending GraphDAU and NestDAU to the unsupervised setting is left for future work.

In \eqref{eq:admm_graph_re1}, we keep the structure of the original graph filter $h(\mathbf{L}) = (\I + \frac{1}{\gamma} \mathbf{L})^{-1}$ of the ADMM algorithm and only train a graph-independent parameter $\gamma$.
As such, GraphDAU performs stably with many layers (typically $L= 10$ in the experiments).
In contrast, GUSC/GUTF use GCNs for its internal algorithm.
Therefore, they result in few middle layers (as reported in \cite{chen2020Graphunrolling}, they have only one middle layer in the experiment).
They reduce many learnable parameters compared to usual GCNs thanks to their edge-weight-sharing convolution, however, they still contain many parameters.
A detailed comparison of the number of parameters is presented along with the restoration performance in Section~\ref{sec:denoising}.

\begin{table*}[!tpb] \centering
  \caption{Comparison between related works and ours.} \label{tab:methods_comparison}
  \begin{tabular}{l | c | c | c} \bhline{1.0pt}
                             & GUSC/GUTF \cite{chen2020Graphunrolling} & GraphDAU & NestDAU \\ \hline \hline
  Optimization algorithm for unrolling & Half-quadratic splitting \cite{wang2008} & ADMM \cite{boyd2010} & PnP-ADMM \cite{venkatakrishnan2013a}\\ \hline
  Problem setting & unsupervised \& supervised & \multicolumn{2}{c}{supervised} \\ \hline
  Considered restoration problem & denoising & denoising & general restoration \\
  \bhline{1.0pt}
  \end{tabular}
\end{table*}

\section{Experimental Results: Denoising} \label{sec:denoising}
In the following two sections, we compare graph signal restoration performances of NestDAU and GraphDAU with existing methods using synthesized and real-world data.
In both sections, parameters of the proposed and neural network-based methods are trained by setting the mean squared error (MSE) $\frac{1}{N}\|\widehat{\mathbf{x}}-\mathbf{x^{*}}\|_{2}^2$ as a loss function,
where $\widehat{\mathbf{x}} \in \mathbb{R}^{N}$ is the restored signal
and $\mathbf{x^{*}} \in \mathbb{R}^{N}$ is the ground-truth signal available during the training phase.

In this section, we consider denoising corresponding to $\mathbf{H}=\mathbf{I}$ in \eqref{eq:problem}.

We conduct three experiments:
\begin{enumerate}
    \item Denoising on fixed graphs;
    \item Denoising on graphs with perturbation;
    \item Transferring tuned parameters to different $N$.
\end{enumerate}
In the following subsections, we describe the details of the denoising experiment.
We also show an in-depth analysis of the proposed methods in terms of the number of layers (i.e., $L$ or $P$) and the polynomial order $K$.

\subsection{Methods and Training Configurations}
\subsubsection{Alternative Methods}
We compare the denoising performance with several existing methods using smoothing filters and optimization approaches:
\begin{itemize}
    \item Graph spectral diffusion with heat kernel (HD) \cite{zhang2008};
    \item Spectral graph bilateral filter (SGBF) \cite{gadde2013, wang2014};
    \item ADMM-based smoothing with a fixed parameter; (\eqref{eq:admm_graph_re1}--\eqref{eq:admm_graph_re3}) with 10 iterations;
    \item PnP-ADMM-based smoothing with fixed parameters with 8 iterations \cite{yazaki2019}: Its formulation is given in Section~\ref{ssec:DAU_pnp} and off-the-shelf denoisers are HD or SGBF.
\end{itemize}
Filtering operations of the algorithms are partly implemented by \texttt{pygsp}~\cite{defferrard2017}.
For a fair comparison, their fixed parameters are tuned by performing a grid search on the validation data to minimize RMSE.

We also include the following deep learning-based methods for comparison:
\begin{itemize}
  \item Multi-layer perceptron (MLP);
  \item Graph convolutional network (GCN) and that with residual connections (GCN-R) \cite{kipf2017SemiSupervisedClassification};
  \item Graph attention networks (GAT) \cite{velickovic2018GraphAttention};
  \item Graph unrolling-based trend filtering (GUTF) \cite{chen2020Graphunrolling};
  \item Graph unrolling-based sparse coding (GUSC) \cite{chen2020Graphunrolling}.
\end{itemize}
These existing methods are set to 64 dimensions as a hidden layer of neural nets as in the setting in \cite{chen2020Graphunrolling}.
These methods and ours are implemented with \texttt{Pytorch} \cite{paszke2019}.
MLP, GCN, GCN-R, and GAT are trained for 30 epochs that lead to convergence of the loss function.
GUTF and GUSC are trained with the same hyper-parameters as \cite{chen2020Graphunrolling}, but they are trained in the supervised setting in this paper.

\subsubsection{Training Configuration} \label{sssec:training_config}
On the basis of preliminary experiments, hyper-parameters used for training of the proposed methods are summarized in Table~\ref{tab:training_configuration}.
Training scheduler StepLR in \texttt{Pytorch} is used to gradually decay the learning rate by multiplying 0.6 each epoch.
Since our proposed methods have a small number of parameters, training usually converges in no more than three epochs.
A detailed performance analysis is discussed in Section~\ref{ssec:analysis}.
\begin{table}[t]
    \centering
    \caption{Training configuration.}
    \label{tab:training_configuration}
    \begin{tabular}{c|c} \bhline{1.0pt}
        Batch size        & $1$ \\
        Epochs            & $\leq 3$ \\
        Weight decay      & $1.0 \times 10^4$ \\
        Optimizer         & Adam \cite{kingma2015AdamMethod}\\
        Learning rate     & $0.02$ \\
        Scheduler         & StepLR \\
        \bhline{1.0pt}
    \end{tabular}
\end{table}

\subsection{Datasets and Setup}\label{sec:denoising_setup}
Here, we describe the details of the experiments and datasets.
The dataset specifications are summarized in Table~\ref{tab:summary_dataset}.
\begin{table*}[!t] \centering
\caption{Summary of data used in Section \ref{sec:denoising}} \label{tab:summary_dataset}
\begin{tabular}{l|llll}
\bhline{1.0pt}
Experiment & Graph & Signal characteristic & Type & Data splitting (train/valid/test) \\  \hline \hline
\multirow{3}{*}{Denoising on fixed graphs} & Community graph & Piecewise constant & Synthetic data & 500/50/50 \\
 & Sensor graph & Piecewise smooth & Synthetic data & 500/50/50 \\
 & U.S. geometry & Daily temperature (time-series) & Real data & 304/30/31 \\ \hline
\multirow{4}{*}{Denoising on perturbed graphs} & \multirow{3}{*}{Sensor graph} & Piecewise constant & \multirow{3}{*}{Synthetic data} & 500/50/50 \\
 &  & Piecewise smooth &  & 500/50/50 \\
 &  & Globally smooth &  & 500/50/50 \\ \cline{2-5}
 & 3D Point clouds & RGB color attributes & Real data & 129/43/44 \\ \hline
Parameter transfer & 3D Point clouds & RGB color attributes for different $N$ & Real data & N/A \\
\bhline{1.0pt}
\end{tabular}
\end{table*}
\begin{table*}[!tb]
  \centering
  \caption{Denoising results on fixed graphs (average RMSEs for test data)}
  \label{tab:result_denoising}
  \begin{tabular}{c|r|ccc|cc|cc|cccc} \bhline{1.0pt}
   &  &  &  &  & \multicolumn{2}{c|}{Community graph} & \multicolumn{2}{c|}{Random sensor graph} & \multicolumn{4}{c}{U.S. temperature} \\
   &  &  &  &  & \multicolumn{2}{c|}{(Piecewise constant)} & \multicolumn{2}{c|}{(Piecewise smooth)} & \multicolumn{4}{c}{(Globally smooth)}\\\hline
Methods & \# params & $L$ & $K$ & $P$ & $\sigma=0.5$ & $1.0$ & $0.5$ & $1.0$ & $3.0$ & $5.0$ & $7.0$ & $9.0$ \\ \hline \hline
Noisy       & -     & -  & -  & - & 0.495 & 1.002 & 0.499 & 0.996 & 2.986 & 5.032 & 7.029 & 9.012 \\
HD          & -     & -  & -  & - & 0.230 & 0.325 & 0.405 & 0.598 & 1.712 & 2.149 & 2.475 & 2.751 \\
SGBF        & -     & -  & -  & - & 0.195 & 0.256 & 0.394 & 0.588 & 1.731 & 2.167 & 2.470 & 2.762 \\
ADMM (GTV)  & -     & 10 & -  & - & 0.114 & 0.233 & 0.378 & 0.603 & 1.784 & 2.249 & 2.483 & 2.805 \\
PnP-HD      & -     & -  & -  & 8 & 0.218 & 0.324 & 0.400 & 0.592 & 1.706 & 2.168 & 2.467 & 2.745 \\
PnP-SGBF    & -     & -  & -  & 8 & 0.195 & 0.294 & 0.399 & 0.586 & 1.730 & 2.164 & 2.457 & 2.749 \\ \hline
MLP         & 4,353  & -  & -  & - & 0.436 & 0.795 & 0.454 & 0.739 & 2.892 & 4.588 & 6.634 & 7.816 \\
GCN         & 4,353  & -  & -  & - & 0.258 & 0.283 & 0.649 & 0.705 & 2.208 & 2.380 & 2.653 & 2.948 \\
GCN-R     & 4,353  & -  & -  & - & 0.272 & 0.305 & 0.718 & 0.759 & 2.285 & 2.411 & 2.642 & 2.931 \\
GAT         & 2,050  & -  & -  & - & 0.236 & 0.247 & 0.560 & 0.752 & 2.213 & 2.776 & 3.042 & 3.698 \\
GUTF        & 19,397 & -  & -  & - & 0.100 & 0.158 & 0.419 & 0.523 & 2.127 & 2.293 & 2.551 & 2.891 \\
GUSC        & 11,205 & -  & -  & - & 0.139 & 0.206 & 0.383 & \textbf{0.512} & 2.069 & 2.248 & 2.587 & 2.935 \\ \hline
GraphDAU-TV-E & 20    & 10 & -  & - & 0.060 & 0.117 & 0.364 & 0.583 & 1.688 & 2.154 & 2.436 & 2.775 \\
GraphDAU-TV-C & 20    & 10 & 10 & - & 0.073 & 0.142 & 0.364 & 0.583 & 1.693 & 2.170 & 2.536 & 2.743 \\
GraphDAU-EN-E & 30    & 10 & -  & - & 0.081 & 0.138 & 0.340 & 0.547 & \textbf{1.652} & 2.123 & 2.411 & 2.723 \\
GraphDAU-EN-C & 30    & 10 & 10  & - & 0.095 & 0.165 & 0.339 & 0.554 & 1.685 & 2.151 & 2.479 & 2.745 \\
NestDAU-TV-E  & 168   & 10 & -  & 8 & \textbf{0.054} & 0.106 & \textbf{0.324} & 0.559 & 1.661 & 2.153 & 2.429 & 2.736 \\
NestDAU-TV-C  & 168   & 10 & 10 & 8 & 0.056 & \textbf{0.103} & 0.374 & 0.631 & 1.665 & 2.124 & 2.458 & 2.730 \\
NestDAU-EN-E  & 248   & 10 & -  & 8 & 0.072 & 0.105 & \textbf{0.324} & 0.528 & 1.656 & \textbf{2.087} & \textbf{2.409} & 2.713 \\
NestDAU-EN-C  & 248   & 10 & 10 & 8 & 0.061 & 0.110 & 0.330 & 0.530 & 1.654 & 2.100 & 2.434 & \textbf{2.674} \\
\bhline{1.0pt}
  \end{tabular}
\end{table*}

\subsubsection{Denoising on Fixed Graphs}
The first experiment is graph signal denoising for the following fixed graphs:
\begin{itemize}
    \item Synthetic signals on a community graph having three clusters ($N=250$);
    \item Synthetic signals on a random sensor graph ($N=150$);
    \item Temperature data in the United States ($N=614$).
\end{itemize}
We assume that the graph is consistent in all of the training, validation, and testing phases.

\noindent
\textbf{Characteristics of Graphs and Graph Signals:}
The community graph is generated by \texttt{pygsp}~\cite{defferrard2017} and is shown in Fig.~\ref{sfig:com_gt}.
We synthetically create piecewise constant graph signals based on the cluster labels of the community graph.
Note that the cluster labels are different while the graph itself is fixed.
Each cluster in the graph is assigned an integer value between $1$ to $6$ randomly as its cluster label.
Then, AWGN ($\sigma= \{0.5, 1.0\}$) is added to the ground-truth signals.

The random sensor graph is also obtained by \texttt{pygsp}~\cite{defferrard2017} and is shown in Fig.~\ref{sfig:sen_gt}.
On the random sensor graph, piecewise-smooth signals are synthesized in the following manner.
First, vertices on a graph are partitioned into eight non-overlapping subgraphs $\{\mathcal{G}_k\}_{k = 1}^8$.
Then, smooth signals on $\mathcal{G}_k$ are synthesized based on the first three eigenvectors of the graph Laplacian of $\mathcal{G}_k$.
Let $\mathbf{L}_{k}$ and $\mathbf{U}_{k}$ be the graph Laplacian of $\mathcal{G}_k$ and its eigenvector matrix, respectively.
Then, a smooth signal on $\mathcal{G}_k$ is given by
\begin{equation}
    \mathbf{x}_{k} = \mathbf{U}_{k, 3} \mathbf{d},
    \label{eq:pws1}
\end{equation}
where $\mathbf{U}_{k, 3}$ is the first three eigenvectors in $\mathbf{U}_{k}$ and $\mathbf{d} \in \mathbb{R}^3$ are expansion coefficients whose element is randomly selected from $[0,5]$.
Finally, a piecewise-smooth signal on $\mathcal{G}$ is obtained by combining eight $\mathbf{x}_{k}$'s as follows:
\begin{equation}
    \mathbf{x} = \sum_k \mathbf{1}_{\mathcal{C}_k} \mathbf{x}_k
    \label{eq:pws2}
\end{equation}
where $\mathbf{1}_{\mathcal{C}_k} \in \{0, 1\}^{N\times |\mathcal{C}_k|}$ is the indicator matrix in which $[\mathbf{1}_{\mathcal{C}_k}]_{i,j} = 1$ when the node $i$ in $\mathcal{G}$ corresponds to the node $j$ in $\mathcal{G}_k$ and $0$ otherwise.
AWGN ($\sigma= \{0.5, 1.0\}$) is added to the ground-truth signals.

In order to demonstrate the effectiveness of our method for real-world data, we use daily average temperature data in the United States in 2017,
provided by QCLCD\footnote{\url{https://www.ncdc.noaa.gov/orders/qclcd/}}~\cite{nationaloceanicandatmosphericadministration}.
The data contain local temperatures recorded at weather stations, yet they include missing observations.
To obtain the completed data (as the ground truth) for a year, we conduct the following preprocessing:
1) 614 stations (out of 7501 ones) having relatively few missing values are selected.
2) Missing values in these stations are filled using the average temperatures observed at the same station in the previous and subsequent days.
For experiment, we split the dataset into three parts: 304 training (January to October), 30 validation (November), and 31 testing (December) data.
In this experiment, we study four noise strengths of AWGN, i.e., $\sigma=\{3.0,5.0,7.0,9.0\}$.
The weighted graph is constructed by an $8$-nearest neighbor (NN) graph based on the stations' geographical coordinates.

\subsubsection{Denoising on Graphs with Perturbation}
The second experiment is conducted for signals on graphs with perturbation to verify the robustness of the proposed method to small perturbations of the underlying graph.
Indeed, the tuned parameters for one graph are not expected to work properly for a completely different graph because the topologies and graph Fourier basis on different graphs are different.
However, signals on similar graphs, in terms of their edge weights, could have similar characteristics and therefore, it is expected that the learned parameters for one graph could work satisfactorily for the signal on another graph if these two graphs are similar enough.

Note that many graph neural network-based methods assume the graph is fixed, while our approach based on DAU only needs to tune graph-independent parameters.
Thus, we can use different graphs in each epoch for training, validation, and testing.
In this experiment, we only showcase the performance with a comparison to the model-based methods because the model-based approaches are applicable even if the graphs are different.

We used the following graph signals:
\begin{itemize}
    \item Synthetic signals on random sensor graphs ($N=150$) having piecewise-constant, piecewise-smooth, and globally-smooth characteristics;
    \item RGB color attributes on 3D point clouds ($N=1,000$).
\end{itemize}

\noindent
\textbf{Characteristics of Graphs and Graph Signals:}
For the experiment on random sensor graphs, each graph is synthetically generated by using a different seed of \texttt{graphs.Sensor} from \texttt{pygsp}~\cite{defferrard2017}.
This results in that all graphs have different topologies and edge weights, but their characteristics are similar.

We then synthetically generate the following graph signals:
\begin{enumerate}
 \renewcommand{\theenumi}{\alph{enumi}}
    \item Piecewise constant signals: We first partition each graph into five clusters with non-overlapping nodes and randomly assign an integer for each cluster between $1$ to $6$. The cluster labels are used as a graph signal.
    \item Piecewise smooth signal: Similar to the piecewise constant case, we first partition each graph into five clusters with non-overlapping nodes.
    The signal is generated by \eqref{eq:pws1} and \eqref{eq:pws2}.
    \item Globally smooth signal: The signal is obtained with a linear combination of the first five graph Fourier basis with random expansion coefficients $\mathbf{d} \in \mathbb{R}^5$ like \eqref{eq:pws1}.
\end{enumerate}
In this experiment, we study two noise strengths of AWGN, i.e., $\sigma=\{0.5, 1.0\}$.

As real data on graphs with perturbation, we use the color attributes of 3D point clouds from JPEG Pleno Database~\cite{loop}, where the human motions are captured as point clouds.
We randomly sample $1,000$ points from the original data.
Then, weighted graphs are constructed using a $4$-NN method whose weights are determined based on the Euclidean distance.
Graphs in this dataset are, therefore, not fixed because the Euclidean distances between points are different due to random sampling.
AWGN with $\sigma =\{20,30,40\}$ is added to each sample to yield a noisy signal.
Note that the implementation of the proposed methods are conducted channel-wise so that the parameters are adjusted to each channel.

\subsubsection{Parameter Transfer for Different $N$}
The number of nodes of a graph directly influences computation complexities for all (training, validation, and testing) phases.
To apply the proposed methods to a signal with large $N$, naive training results in  large computational burden.
Motivated by this, in the third experiment, we consider transferring the learned parameters to graph signals having different $N$.
That is, parameters trained with signals on a small graph $\mathcal{G}^{\prime}$ with $N^{\prime}~ (\ll N)$ nodes are reused for evaluation with signals on $\mathcal{G}$ with $N$ nodes.
This approach can be easily realized with the proposed method since its parameters are independent of $N$.

We first train the GraphDAU-TV-C (i.e., Chebyshev polynomial version of GraphDAU based on the GTV regularization) on the 3D point cloud datasets with $N^\prime = 1,000$ points.
After that, the pre-trained parameters are applied to the datasets with larger $N = \{2,000,\  5,000,\  10,000\}$.

\begin{figure*}[!t]  %
  \centering
  \begin{minipage}[t]{.24\linewidth} \centering
    \includegraphics[width=\linewidth]{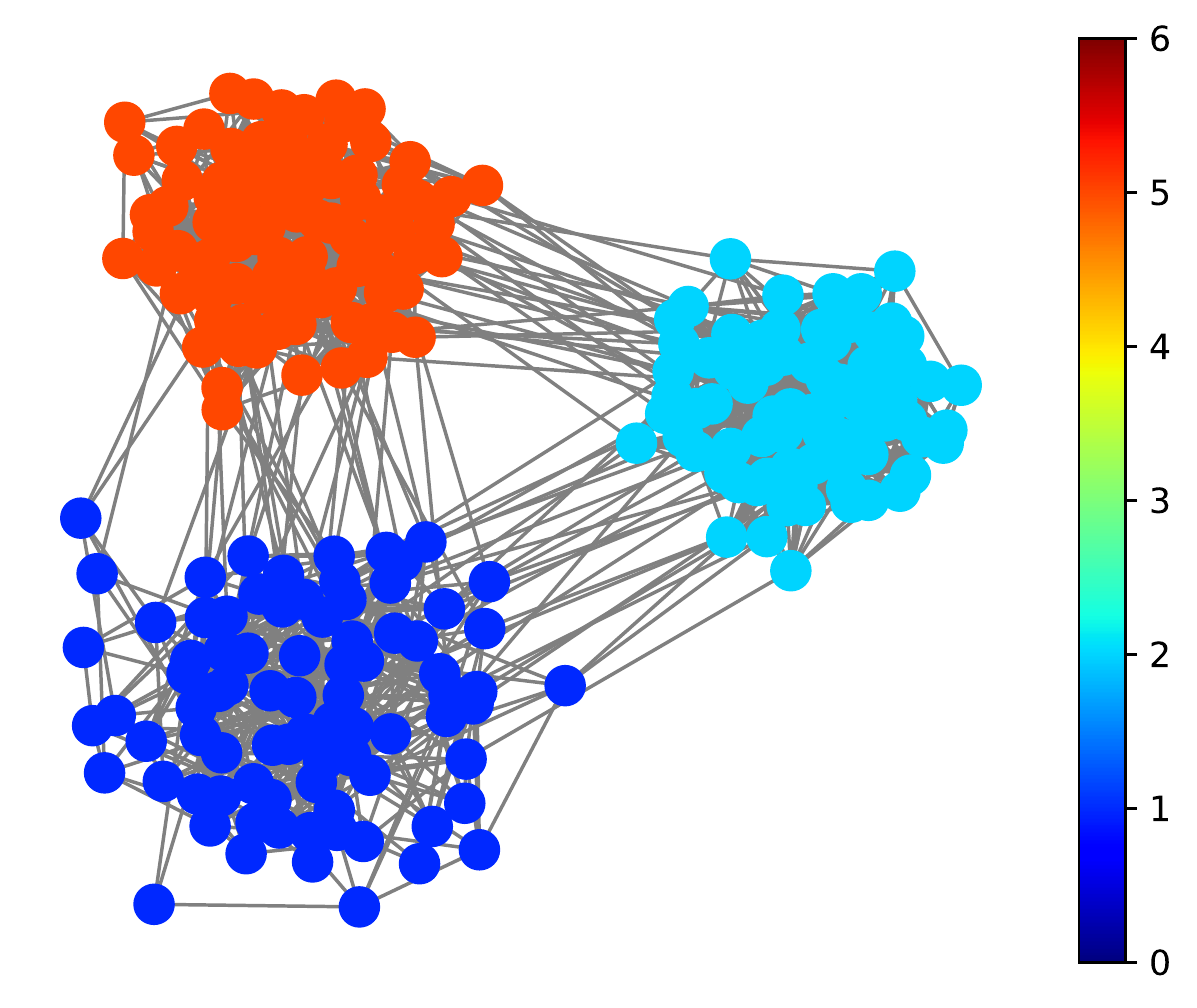}
    \subcaption{Ground truth} \label{sfig:com_gt}
  \end{minipage}
  \begin{minipage}[t]{.24\linewidth} \centering
    \includegraphics[width=\linewidth]{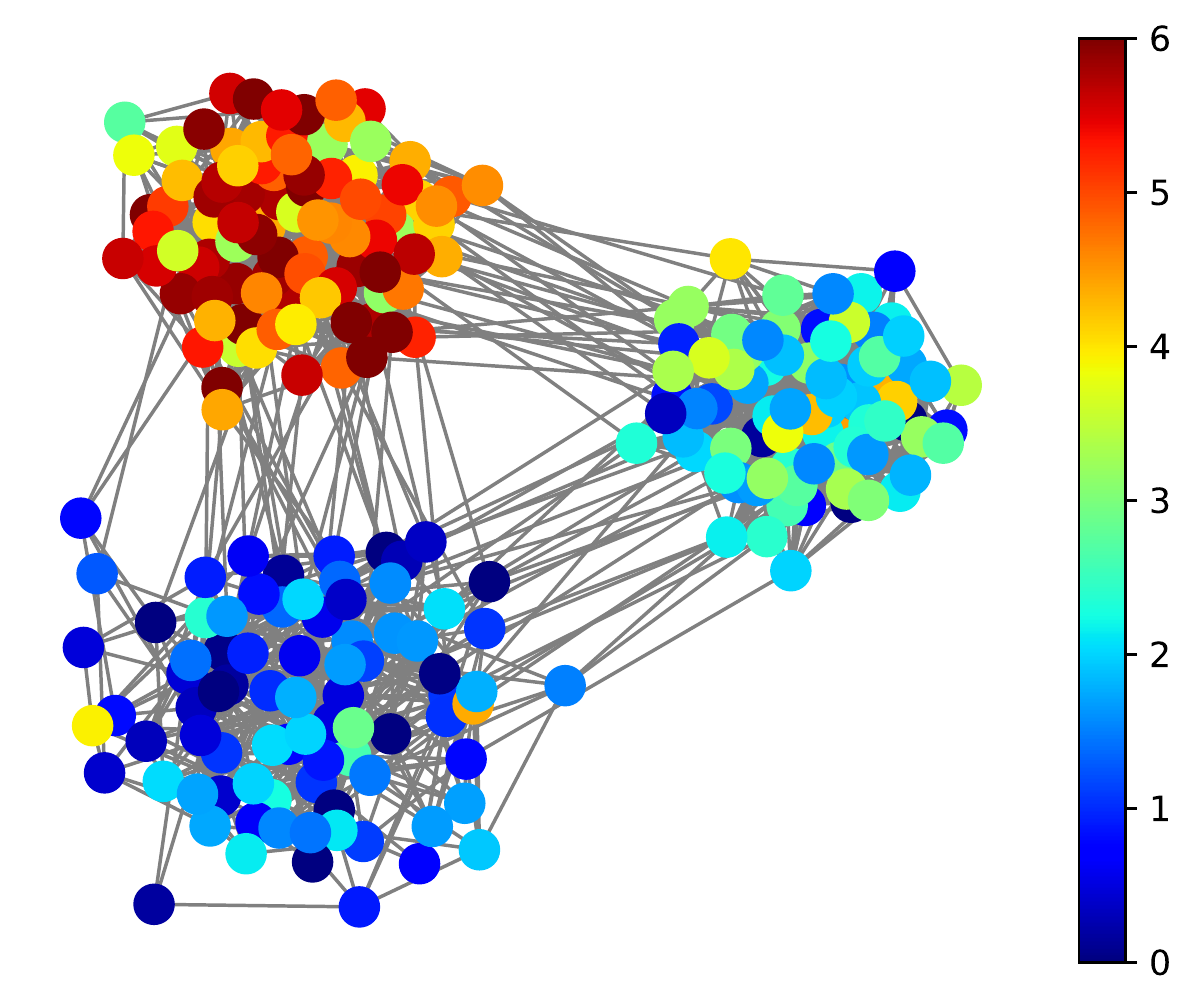}
    \subcaption{Noisy} \label{sfig:com_noisy}
  \end{minipage}
  \begin{minipage}[t]{.24\linewidth} \centering
    \includegraphics[width=\linewidth]{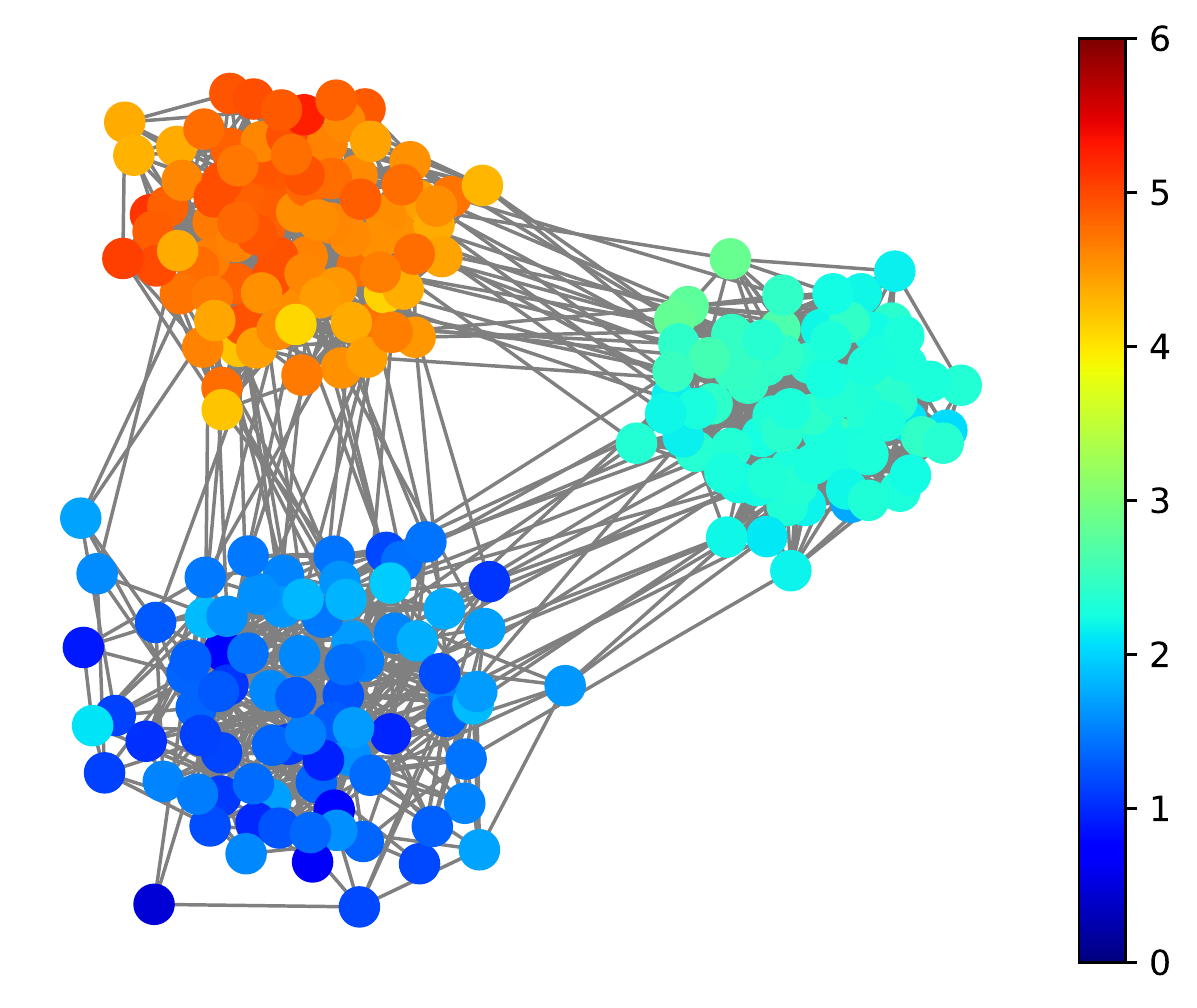}
    \subcaption{PnP-HD}
  \end{minipage}
  \begin{minipage}[t]{.24\linewidth} \centering
    \includegraphics[width=\linewidth]{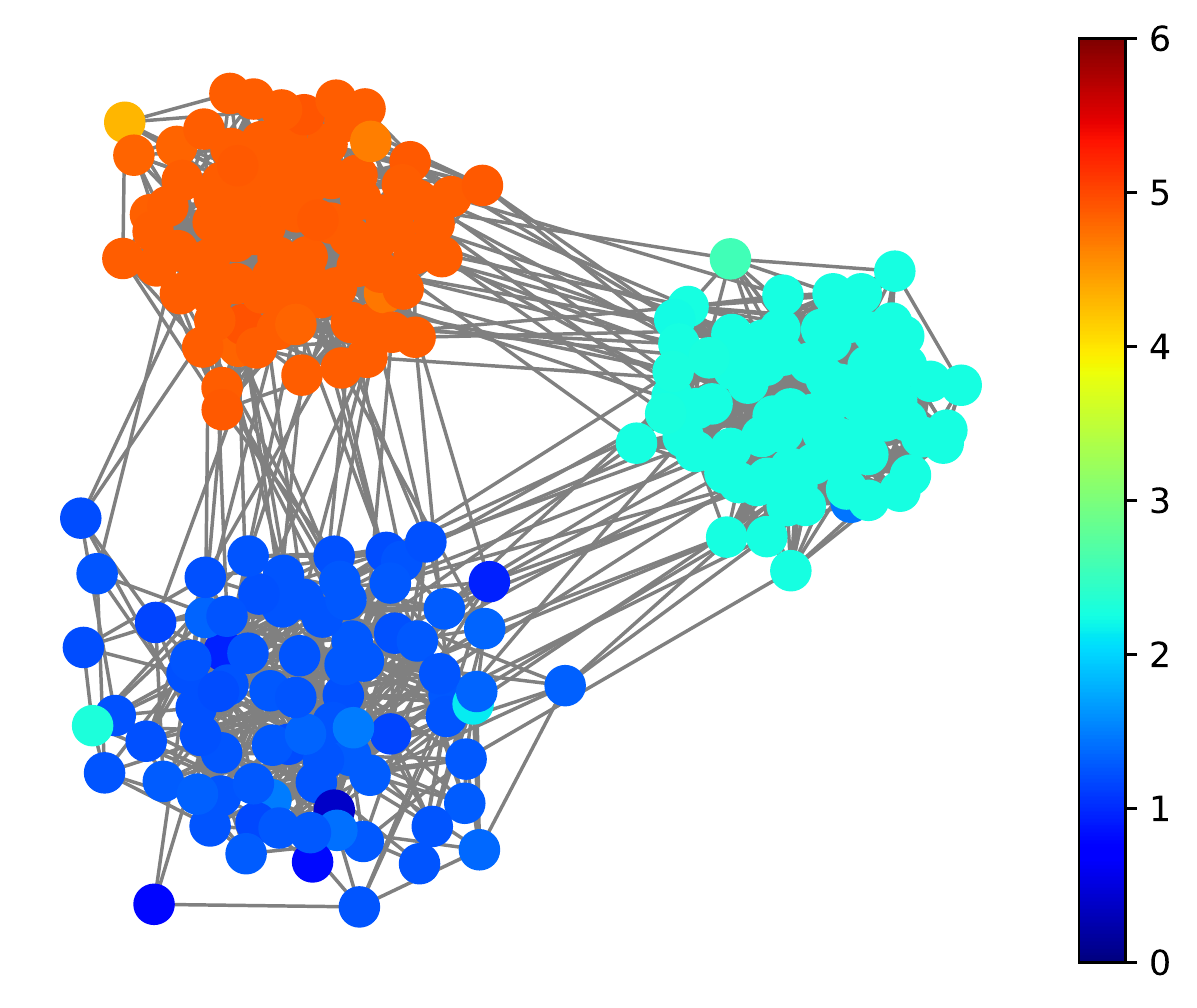}
    \subcaption{ADMM (GTV)}
  \end{minipage}
  \\
  \begin{minipage}[t]{.24\linewidth} \centering
    \includegraphics[width=\linewidth]{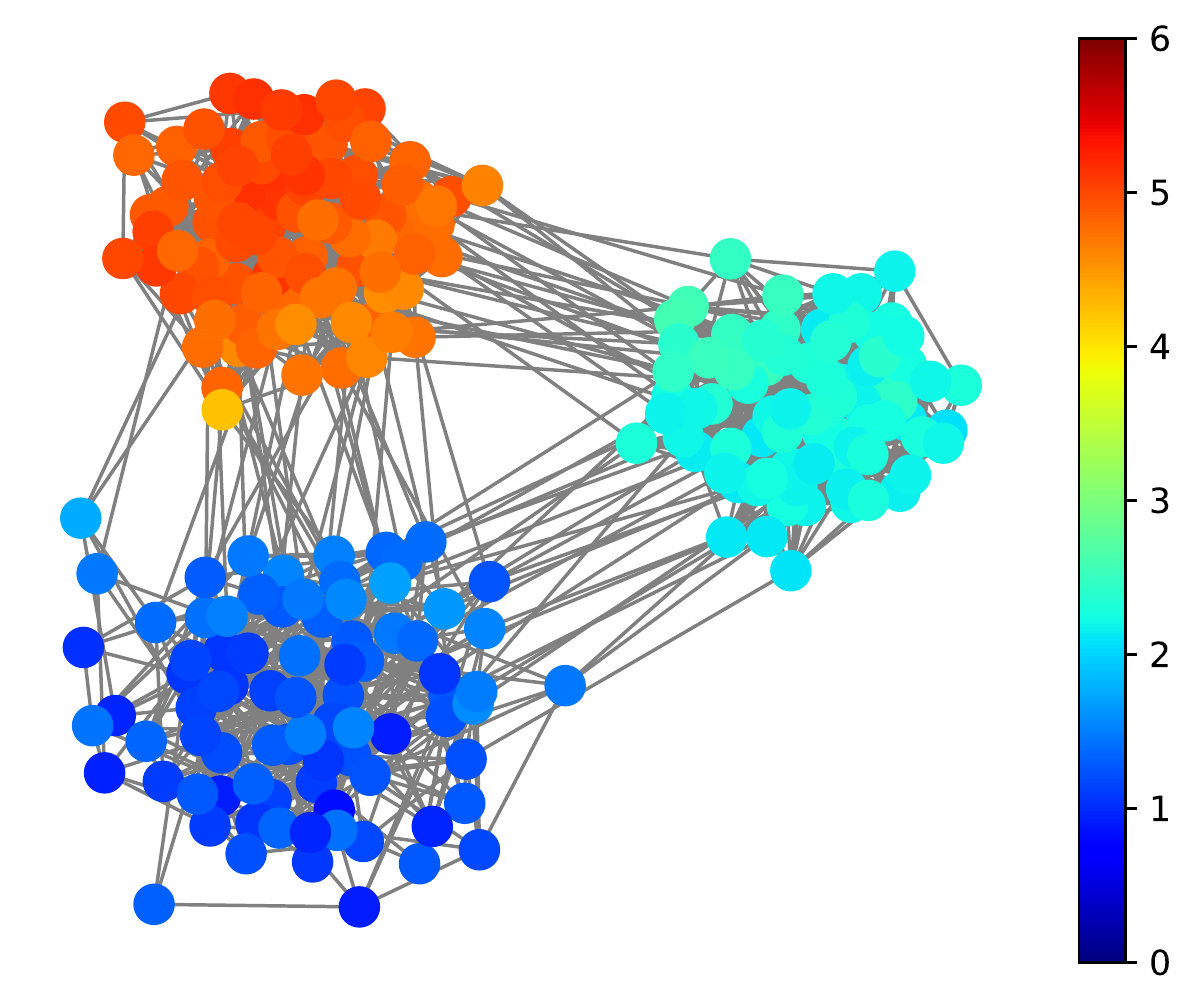}
    \subcaption{GUSC}
  \end{minipage}
  \begin{minipage}[t]{.24\linewidth} \centering
    \includegraphics[width=\linewidth]{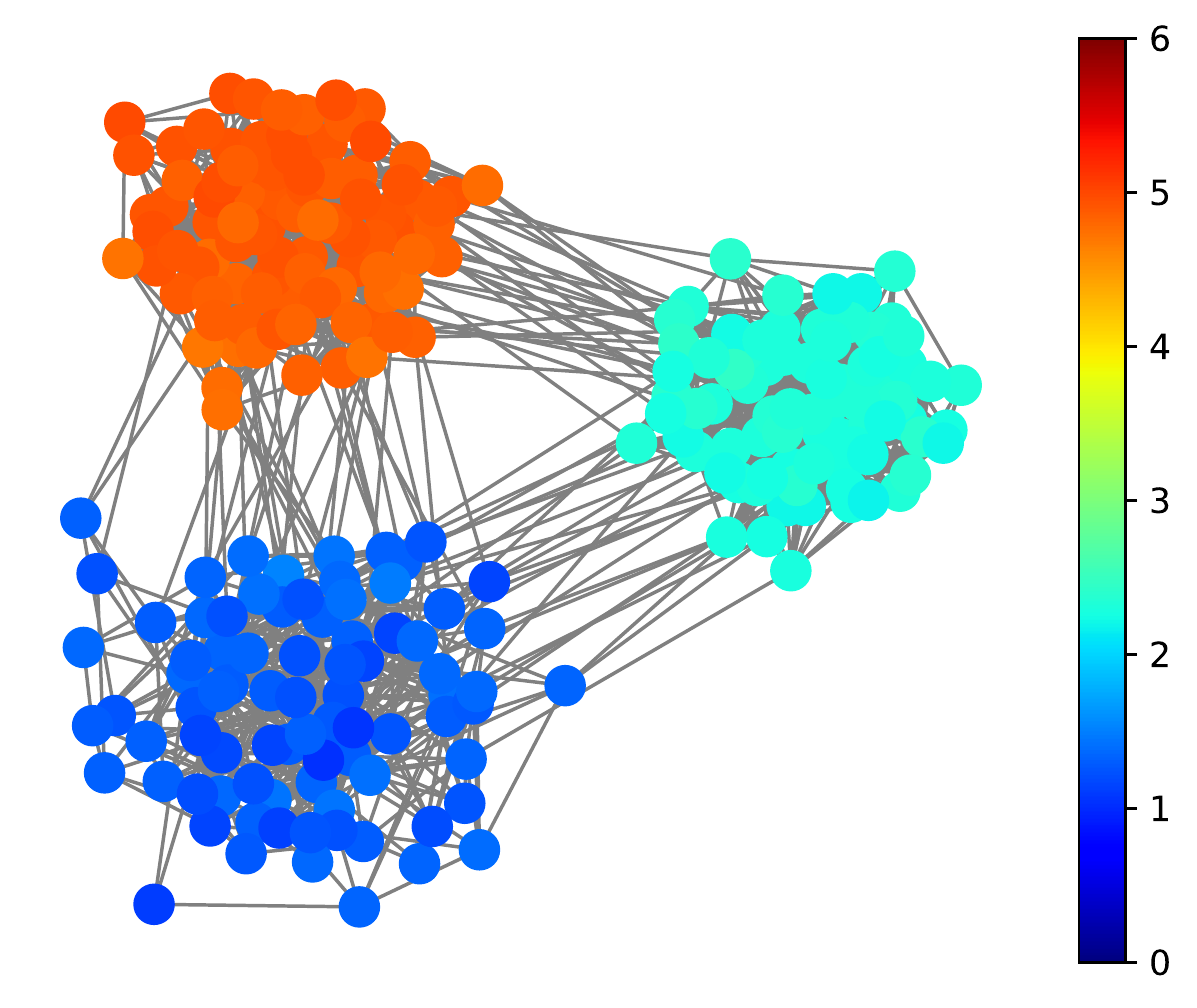}
    \subcaption{GUTF}
  \end{minipage}
  \begin{minipage}[t]{.24\linewidth} \centering
    \includegraphics[width=\linewidth]{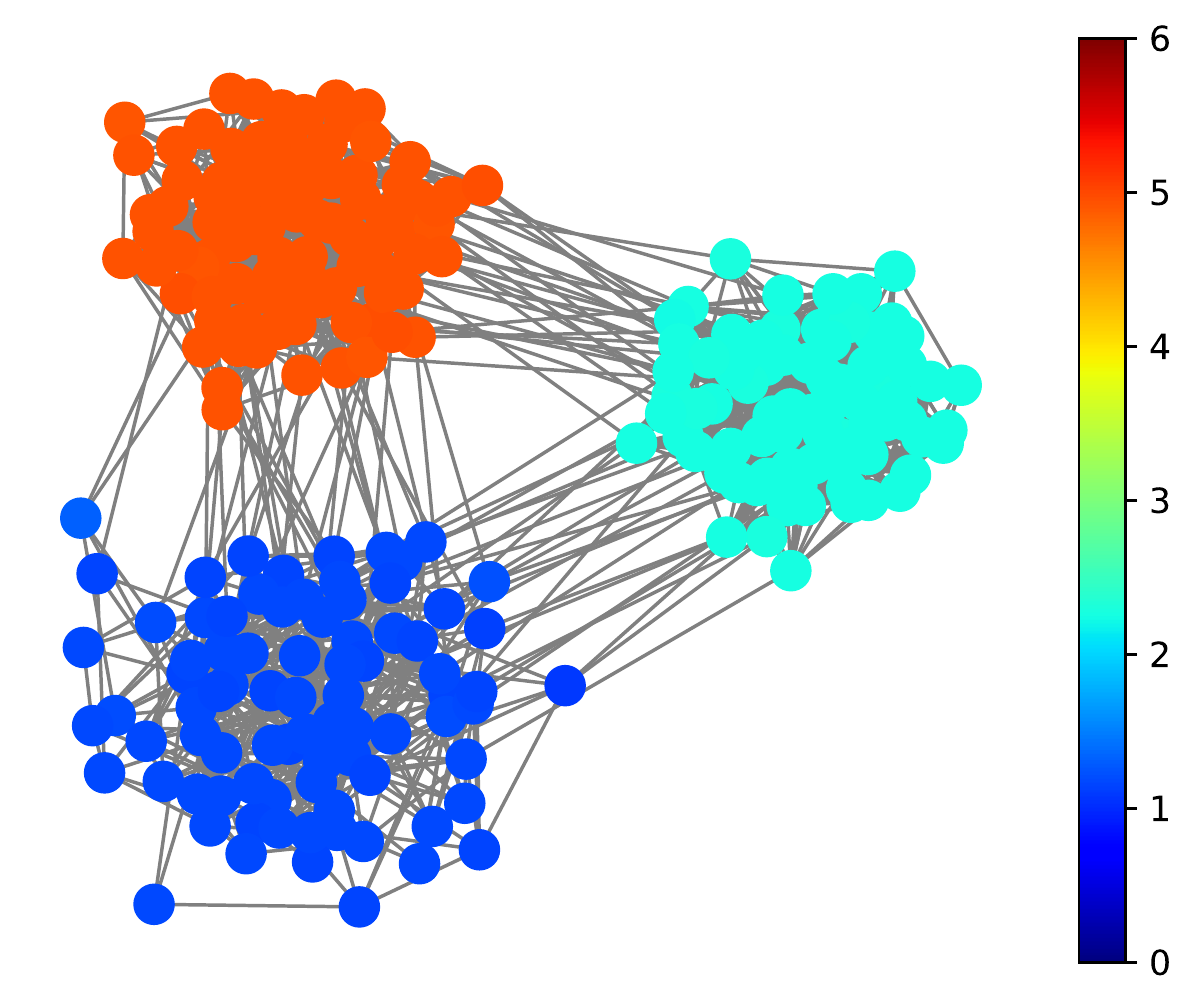}
    \subcaption{GraphDAU-TV-E}
  \end{minipage}
  \begin{minipage}[t]{.24\linewidth} \centering
    \includegraphics[width=\linewidth]{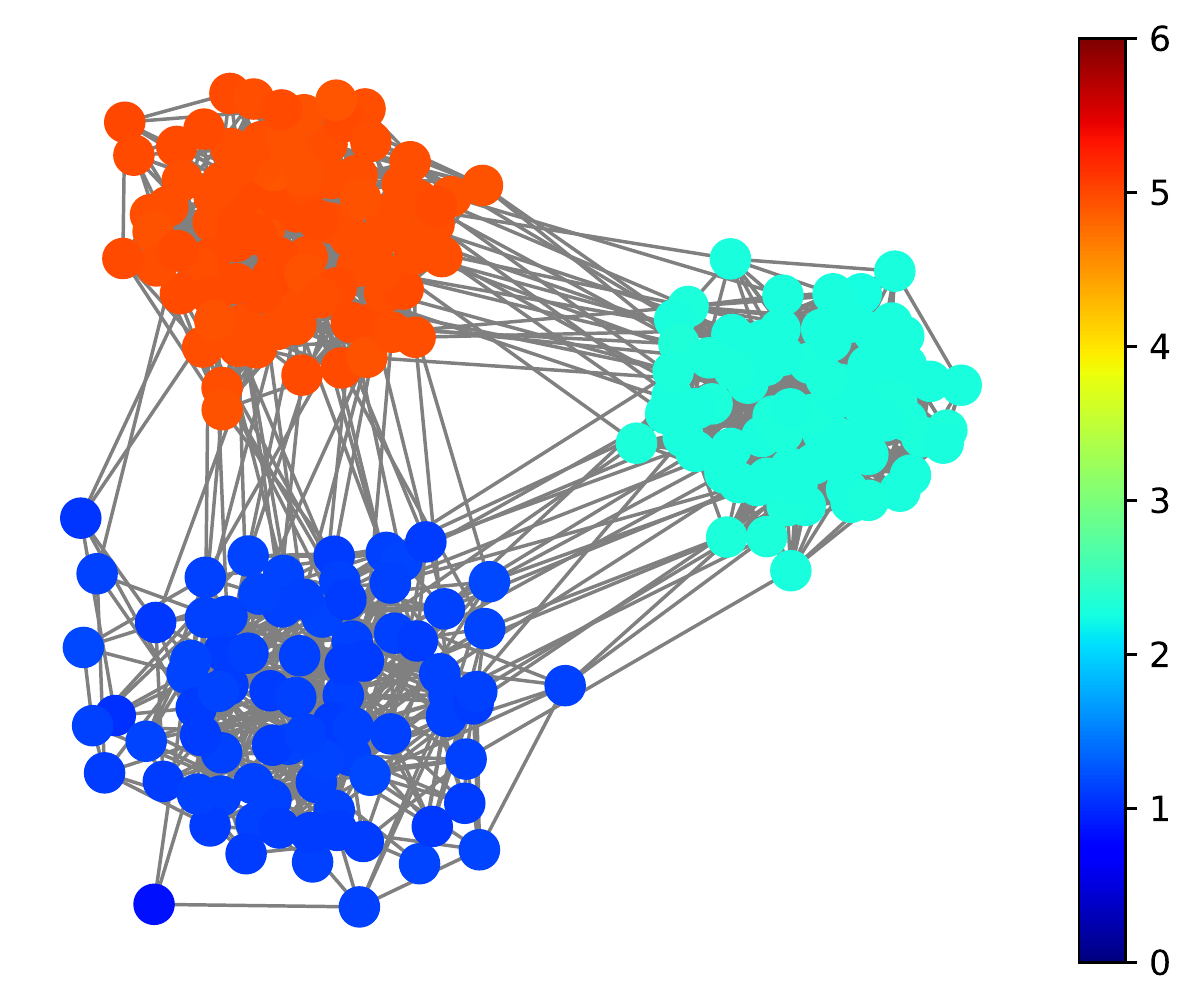}
    \subcaption{NestDAU-TV-C}
    \label{sfig:com_NEDAU}
  \end{minipage}
  \caption{Visualization: Denoising results of signals on community graph with $\sigma=1.0$.}
  \label{fig:community_img}
\end{figure*}
\begin{figure*}[!t]
  \centering
  \begin{minipage}[t]{.24\linewidth} \centering
    \includegraphics[width=\linewidth]{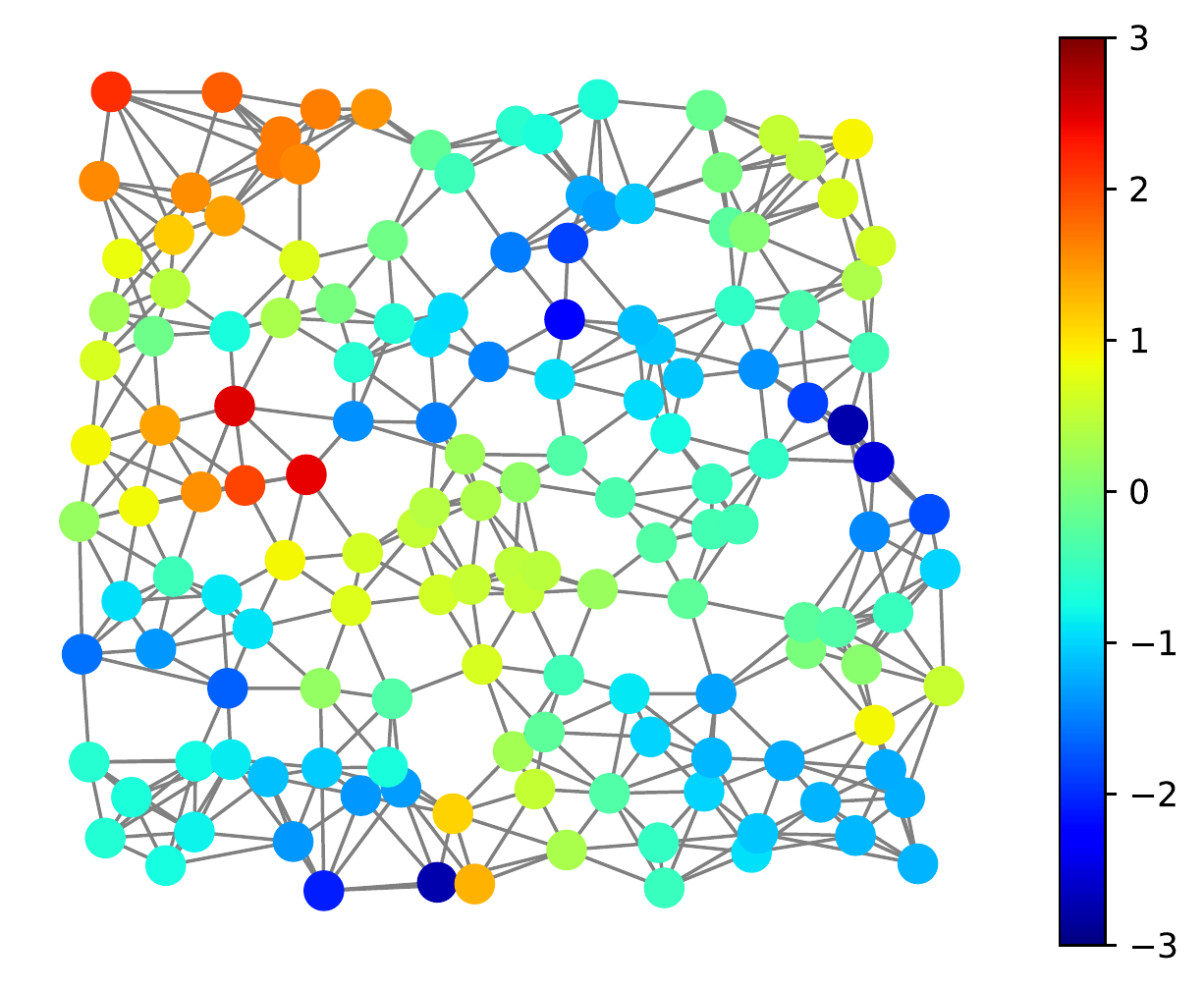}
    \subcaption{Ground truth} \label{sfig:sen_gt}
  \end{minipage}
  \begin{minipage}[t]{.24\linewidth} \centering
    \includegraphics[width=\linewidth]{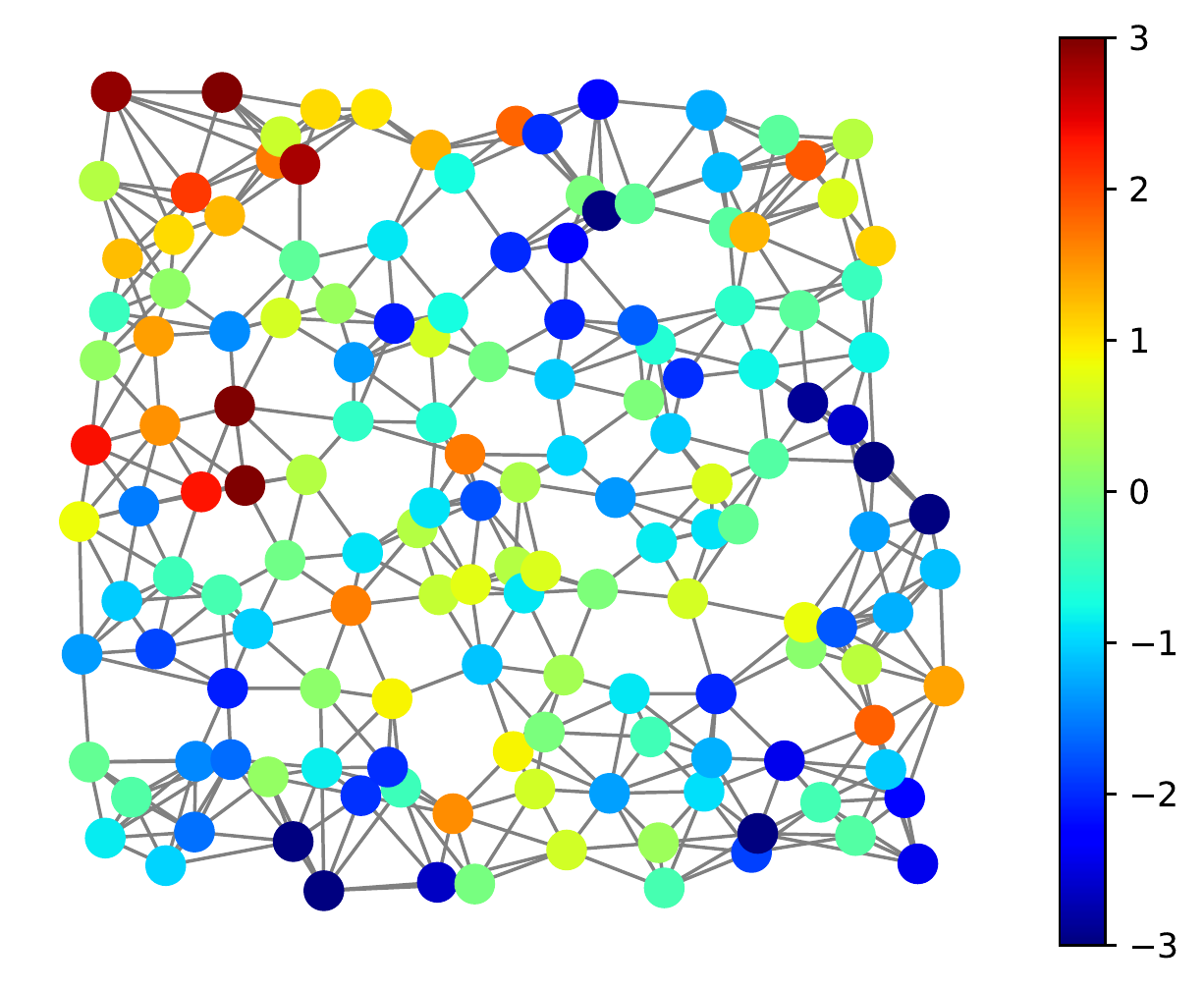}
    \subcaption{Noisy} \label{sfig:sen_noisy}
  \end{minipage}
  \begin{minipage}[t]{.24\linewidth} \centering
    \includegraphics[width=\linewidth]{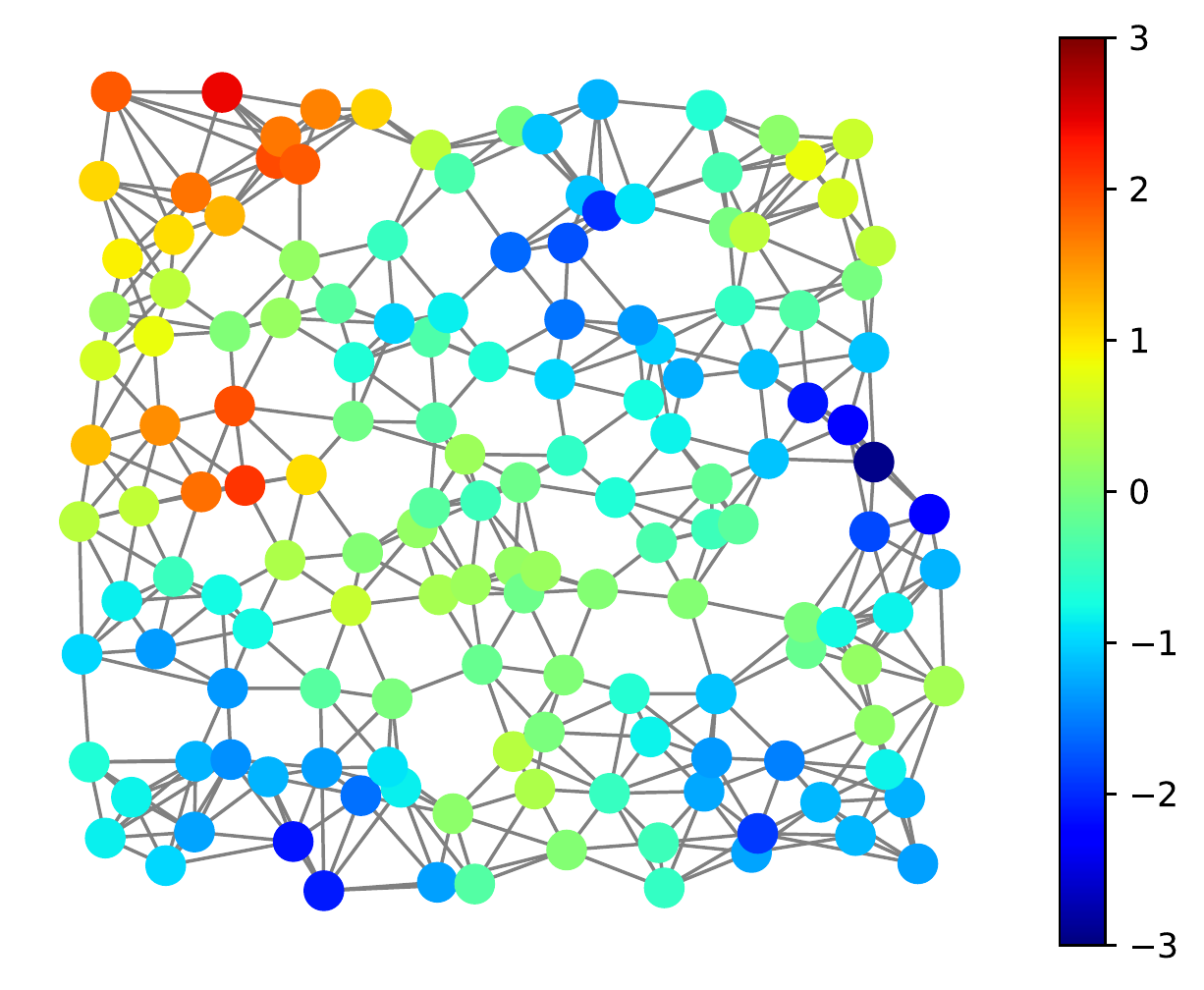}
    \subcaption{PnP-SGBF}
  \end{minipage}
  \begin{minipage}[t]{.24\linewidth} \centering
    \includegraphics[width=\linewidth]{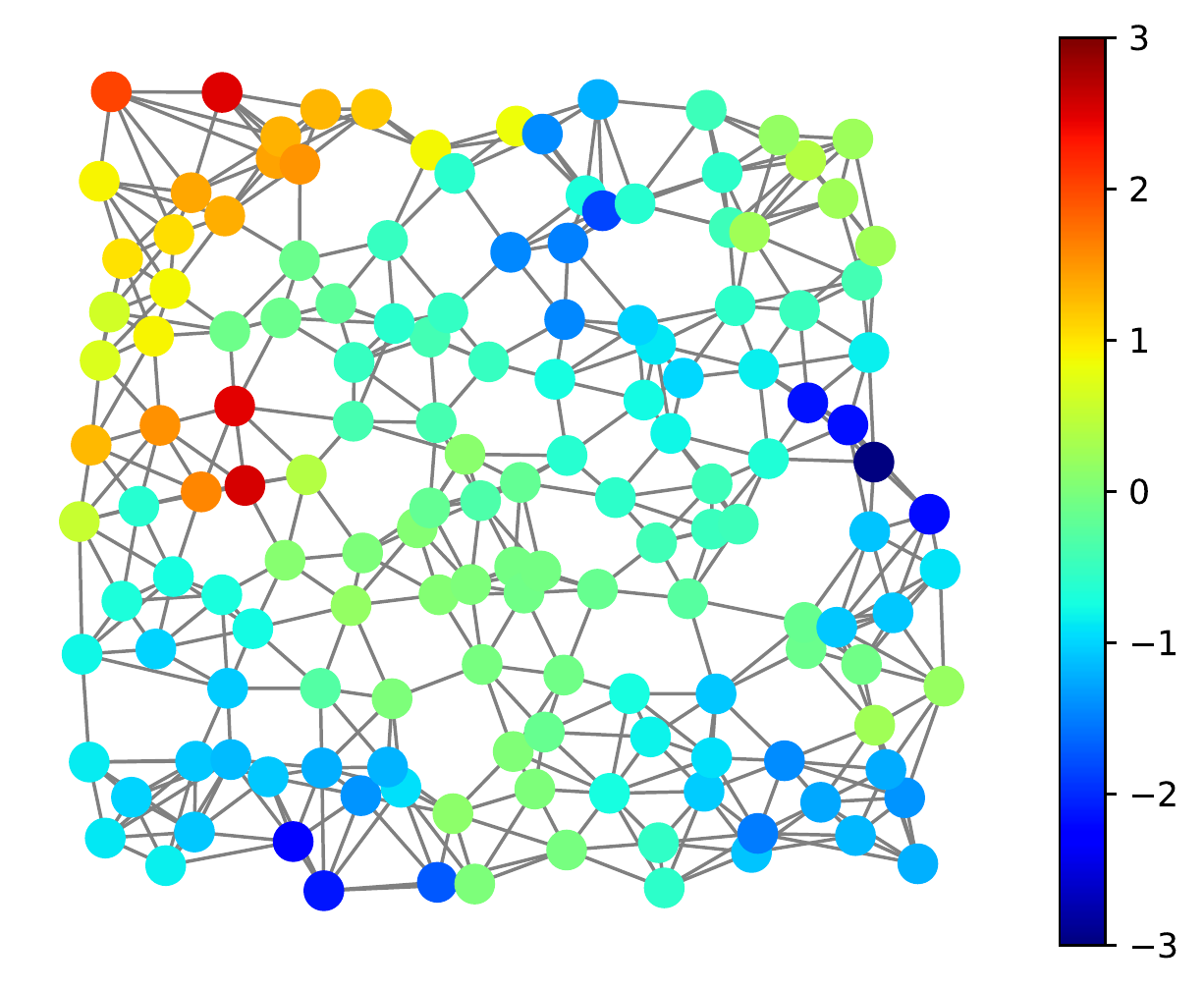}
    \subcaption{ADMM (GTV)}
  \end{minipage}
  \\
  \begin{minipage}[t]{.24\linewidth} \centering
    \includegraphics[width=\linewidth]{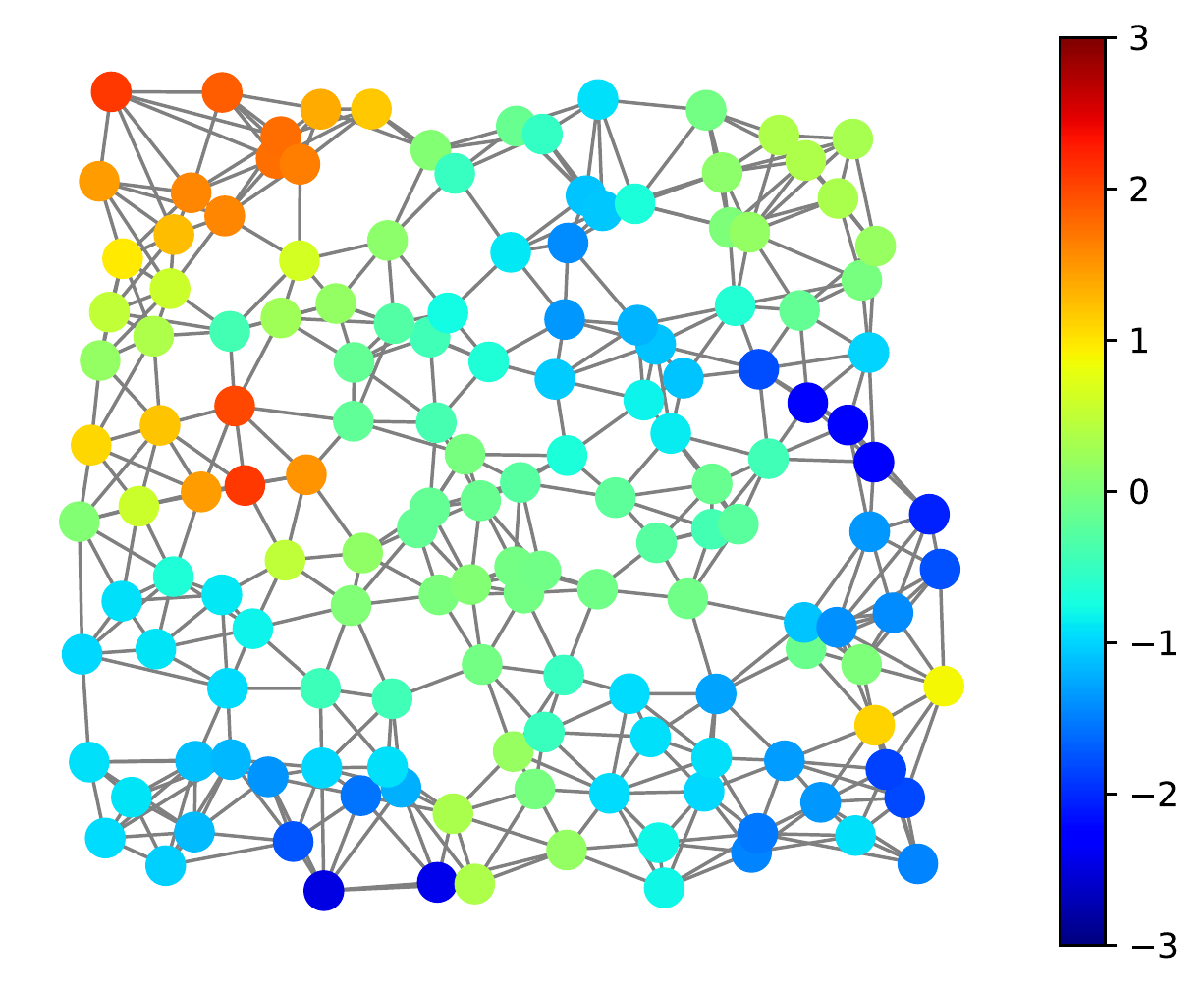}
    \subcaption{GUSC}
  \end{minipage}
  \begin{minipage}[t]{.24\linewidth} \centering
    \includegraphics[width=\linewidth]{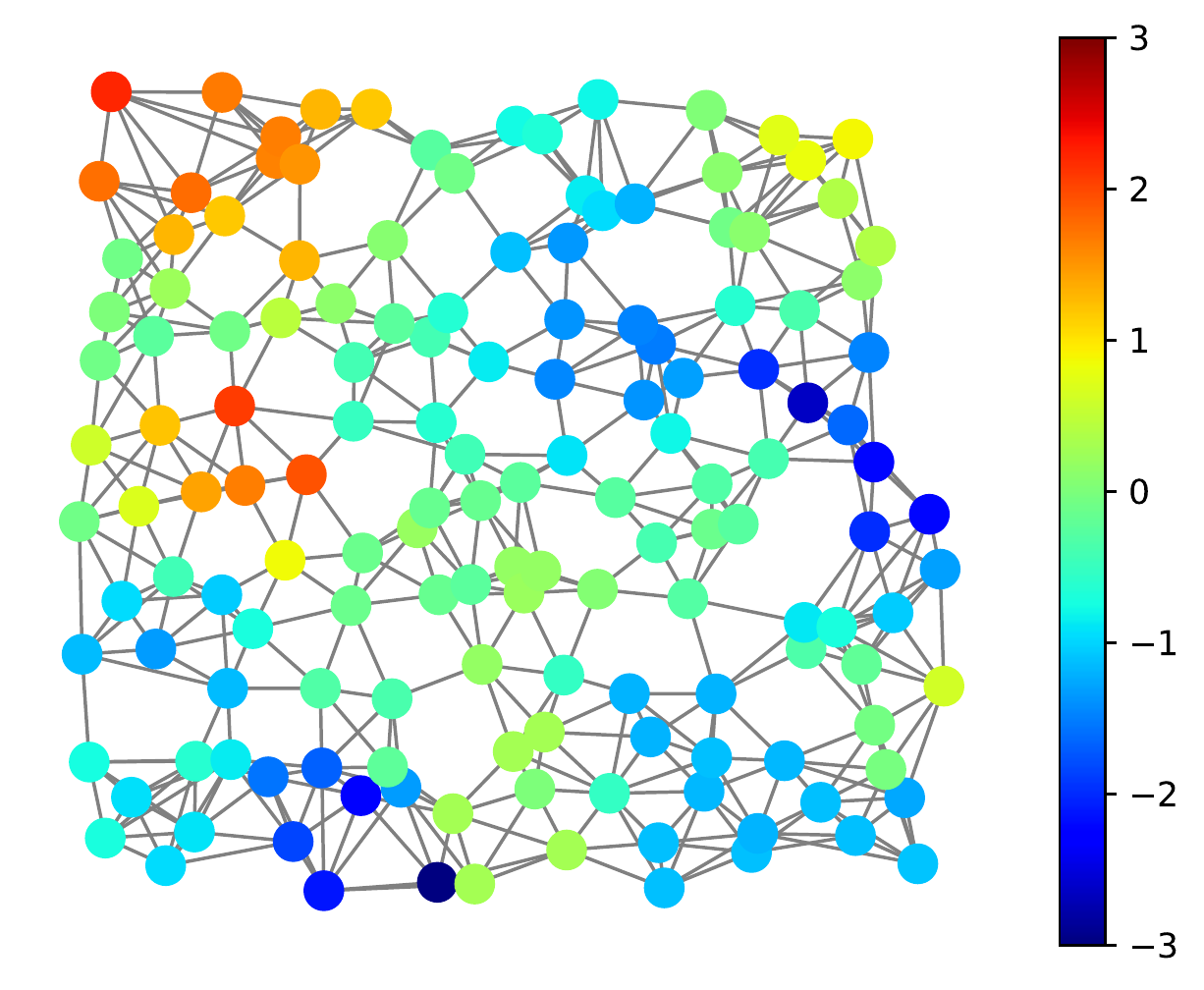}
    \subcaption{GUTF}
  \end{minipage}
  \begin{minipage}[t]{.24\linewidth} \centering
    \includegraphics[width=\linewidth]{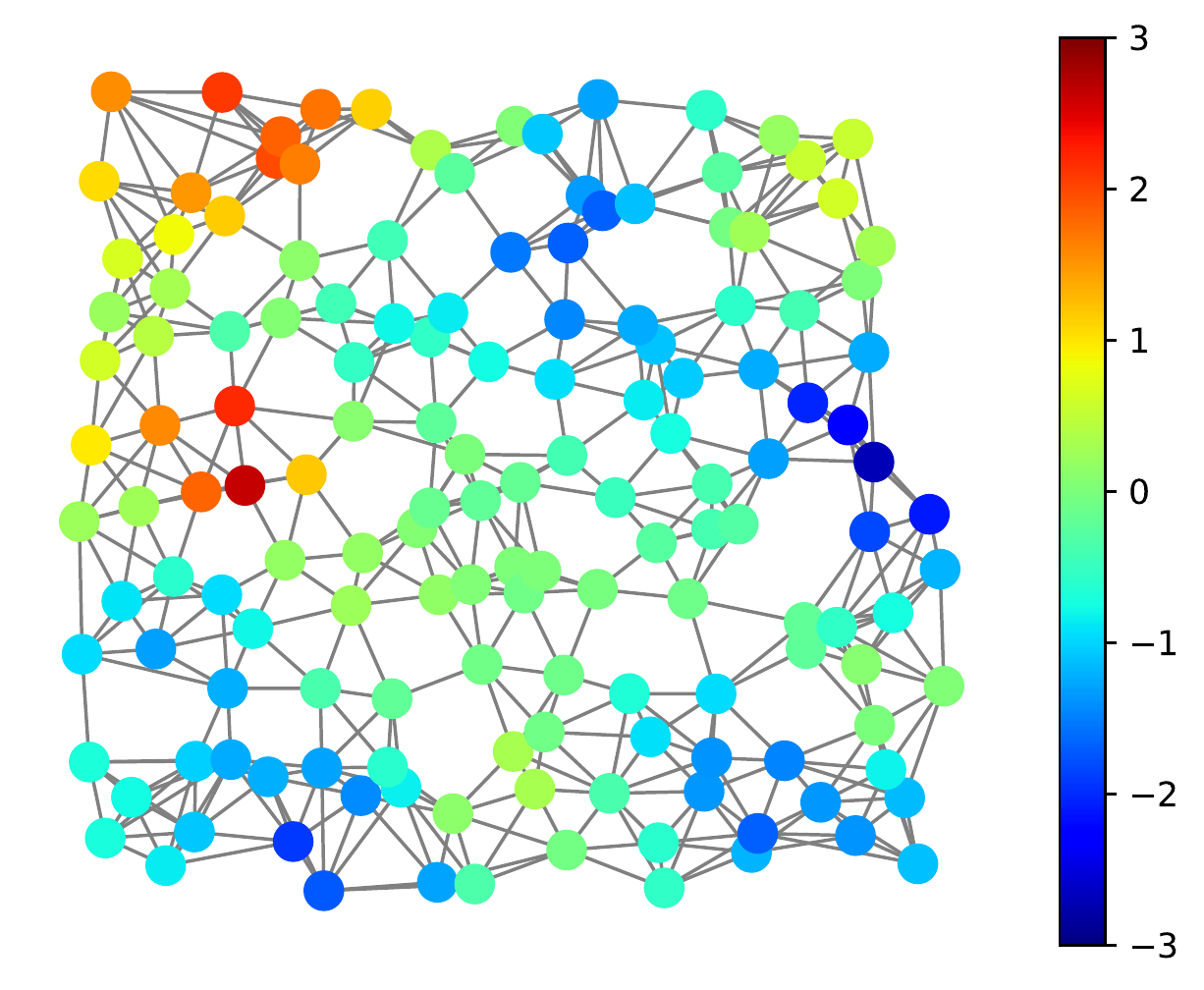}
    \subcaption{GraphDAU-EN-E}
  \end{minipage}
  \begin{minipage}[t]{.24\linewidth} \centering
    \includegraphics[width=\linewidth]{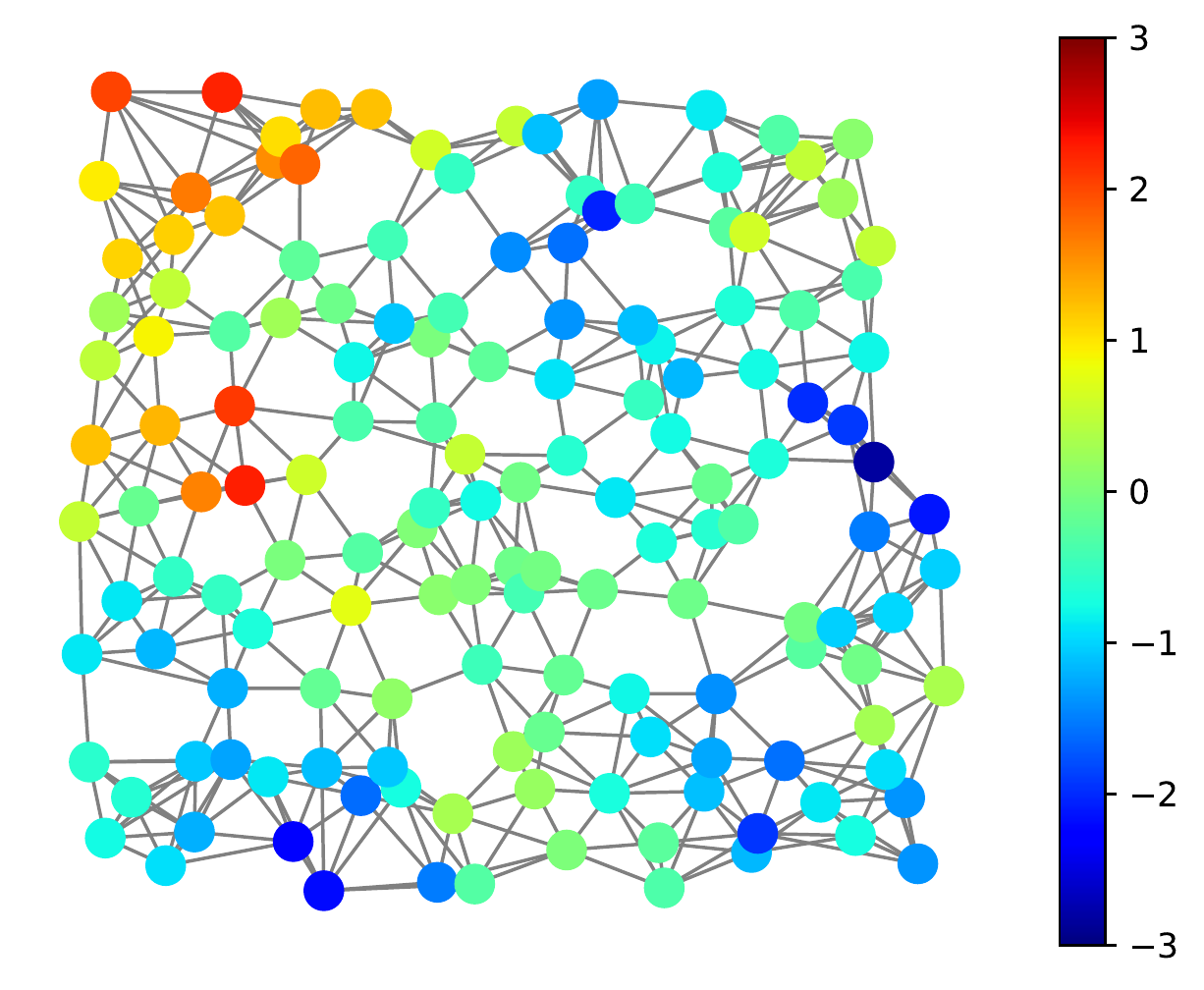}
    \subcaption{NestDAU-TV-C} \label{sfig:sensor_NDAU}
  \end{minipage}
  \caption{Denoising signals on random sensor graph ($\sigma=1.0$).}
  \label{fig:minn_img}
\end{figure*}

\subsection{Denoising Results: Fixed Graph}
\begin{figure*}[!t]  %
  \centering
  \begin{minipage}[t]{.24\linewidth} \centering
    \includegraphics[width=\linewidth]{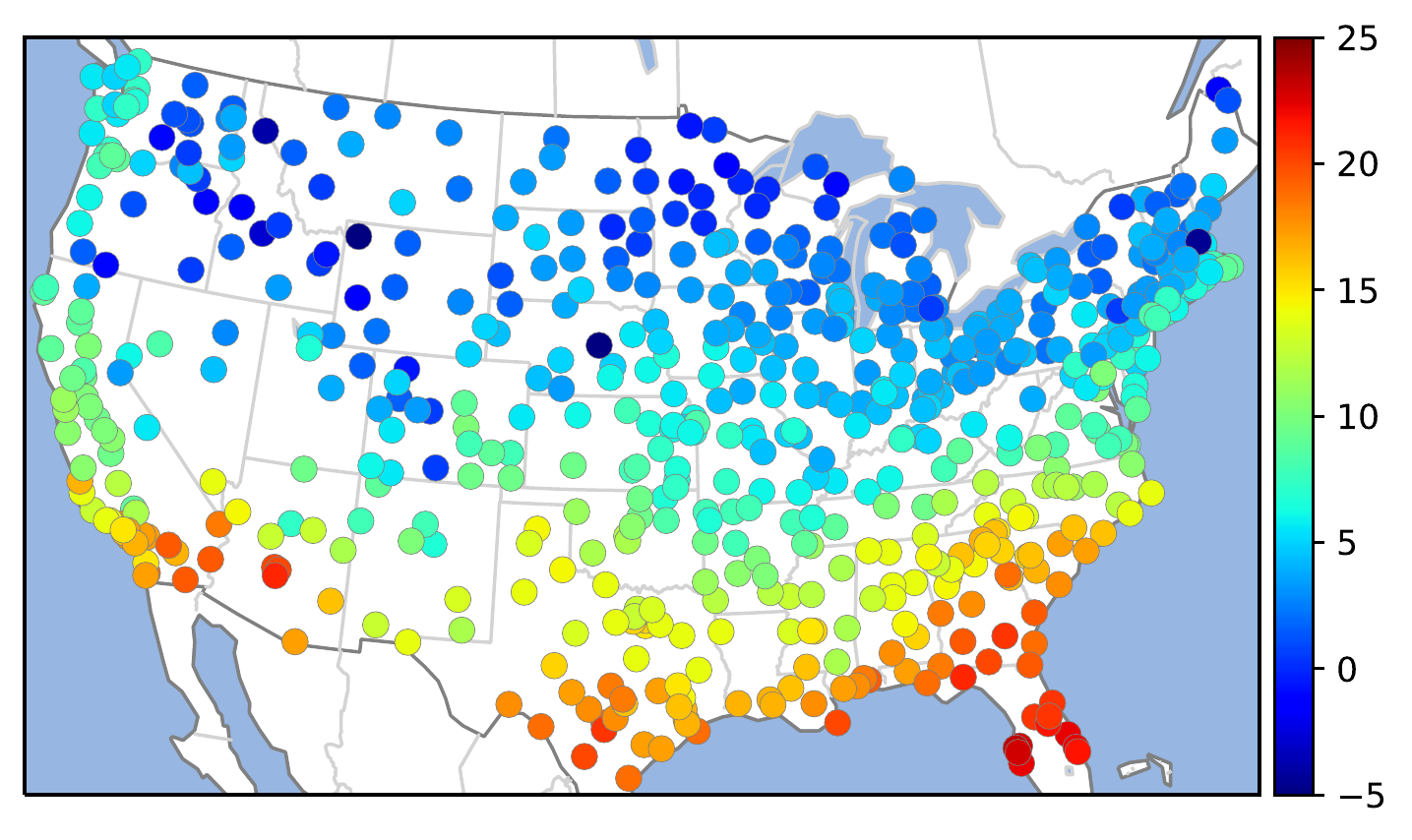}
    \subcaption{Ground truth} \label{sfig:temp_gt}
  \end{minipage}
  \begin{minipage}[t]{.24\linewidth} \centering
    \includegraphics[width=\linewidth]{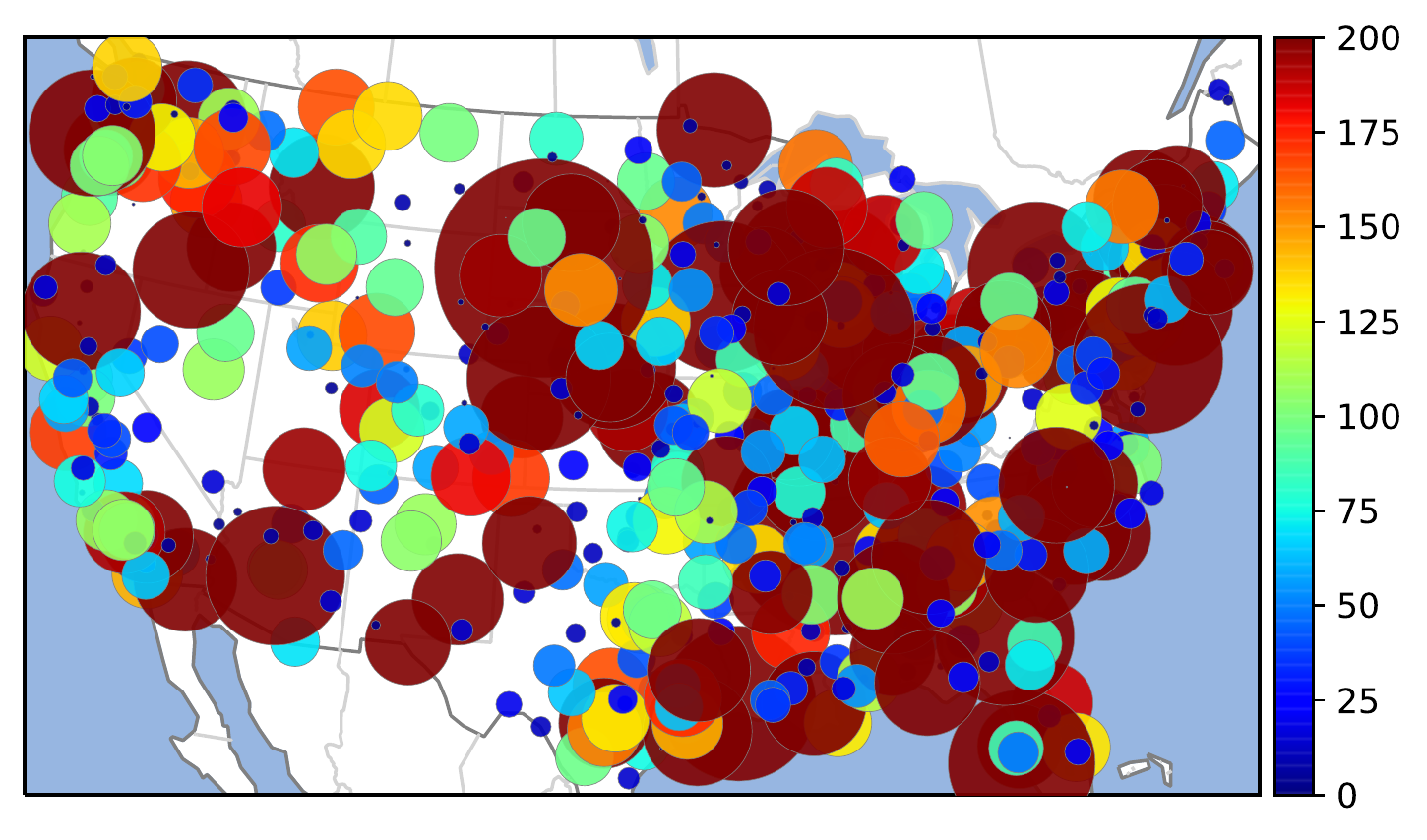}
    \subcaption{Noisy} \label{sfig:temp_noisy}
  \end{minipage}
  \begin{minipage}[t]{.24\linewidth} \centering
    \includegraphics[width=\linewidth]{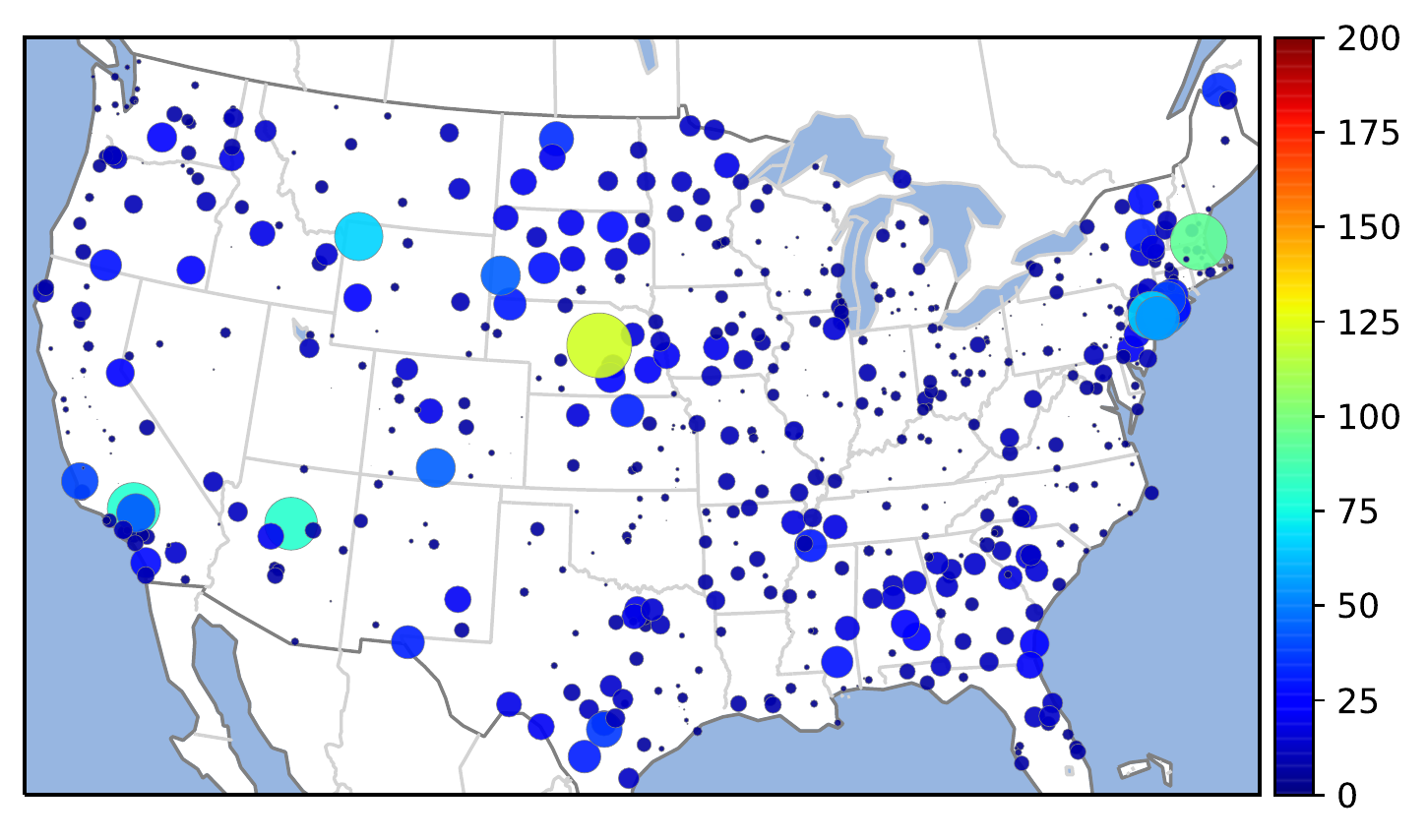}
    \subcaption{HD}
  \end{minipage}
  \begin{minipage}[t]{.24\linewidth} \centering
    \includegraphics[width=\linewidth]{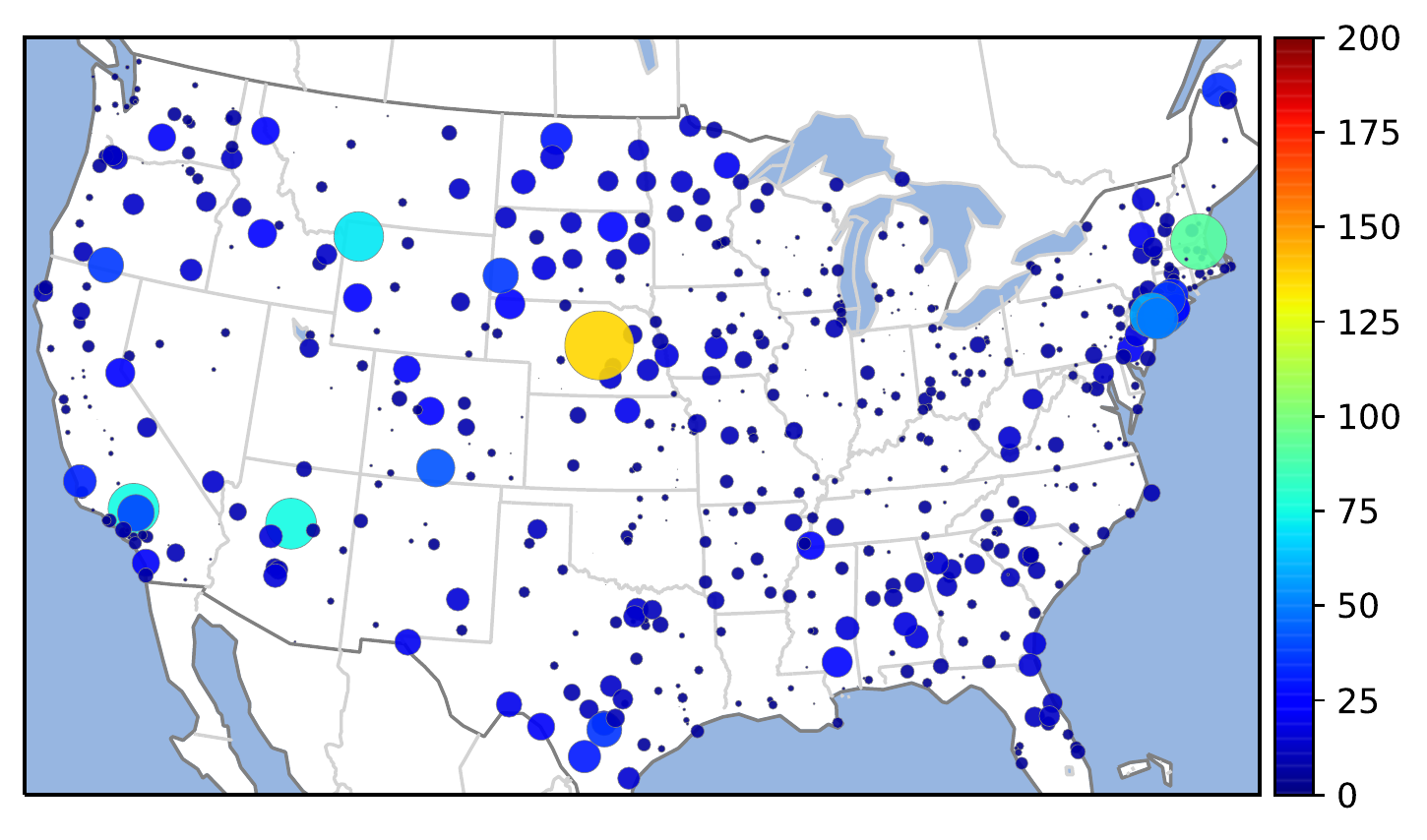}
    \subcaption{PnP-HD}
  \end{minipage}
  \\
  \begin{minipage}[t]{.24\linewidth} \centering
    \includegraphics[width=\linewidth]{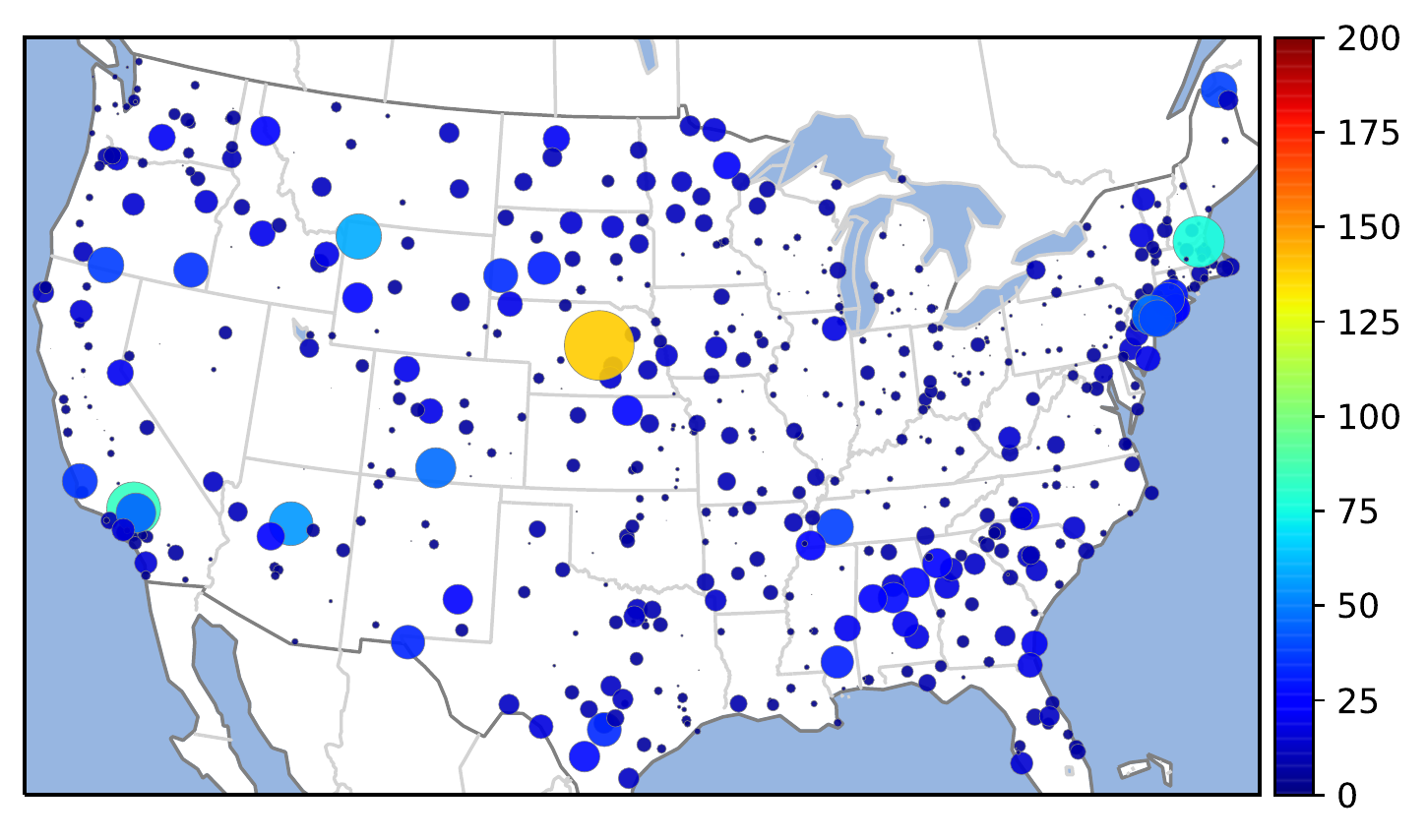}
    \subcaption{ADMM (GTV)}
  \end{minipage}
  \begin{minipage}[t]{.24\linewidth} \centering
    \includegraphics[width=\linewidth]{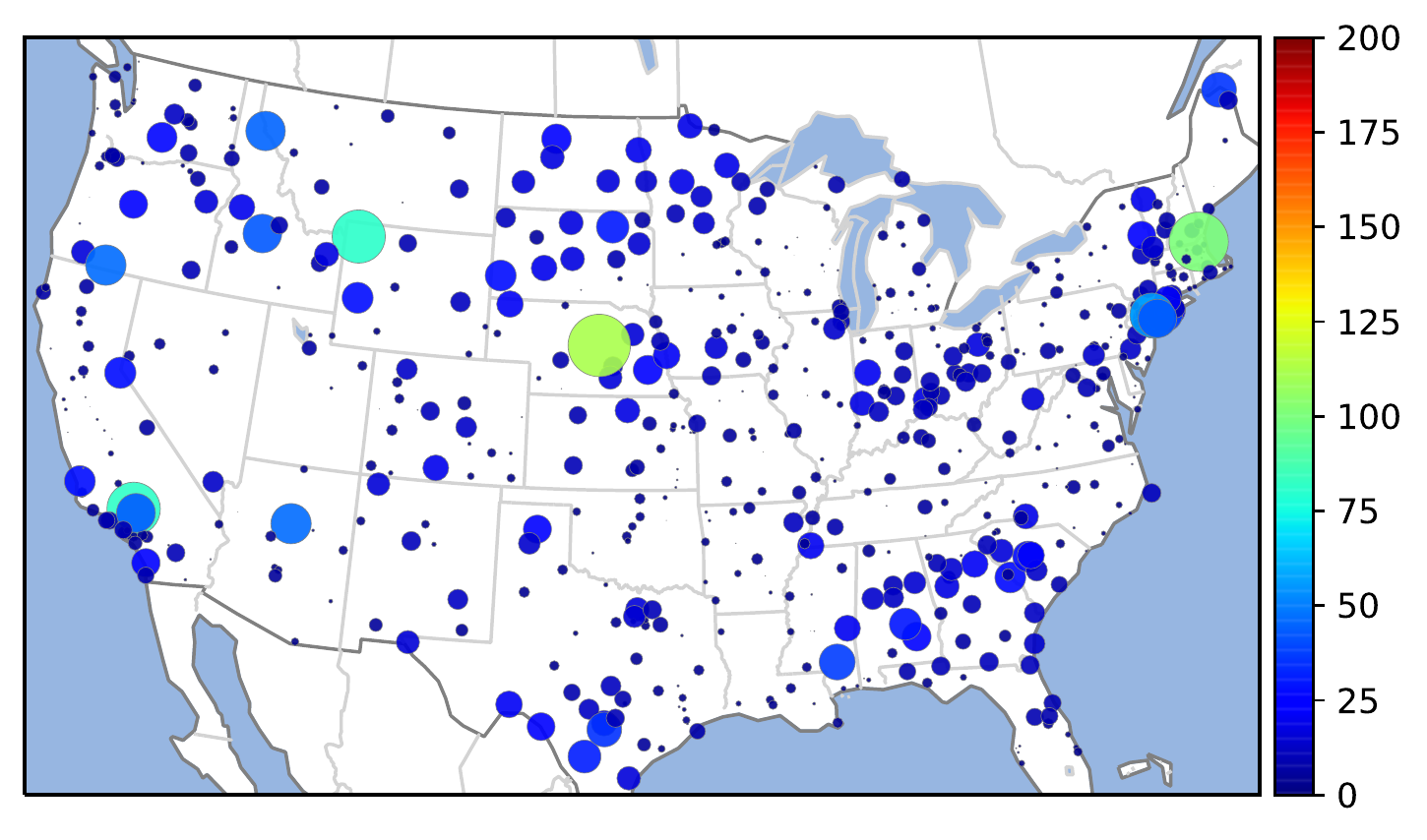}
    \subcaption{GUTF}
  \end{minipage}
  \begin{minipage}[t]{.24\linewidth} \centering
    \includegraphics[width=\linewidth]{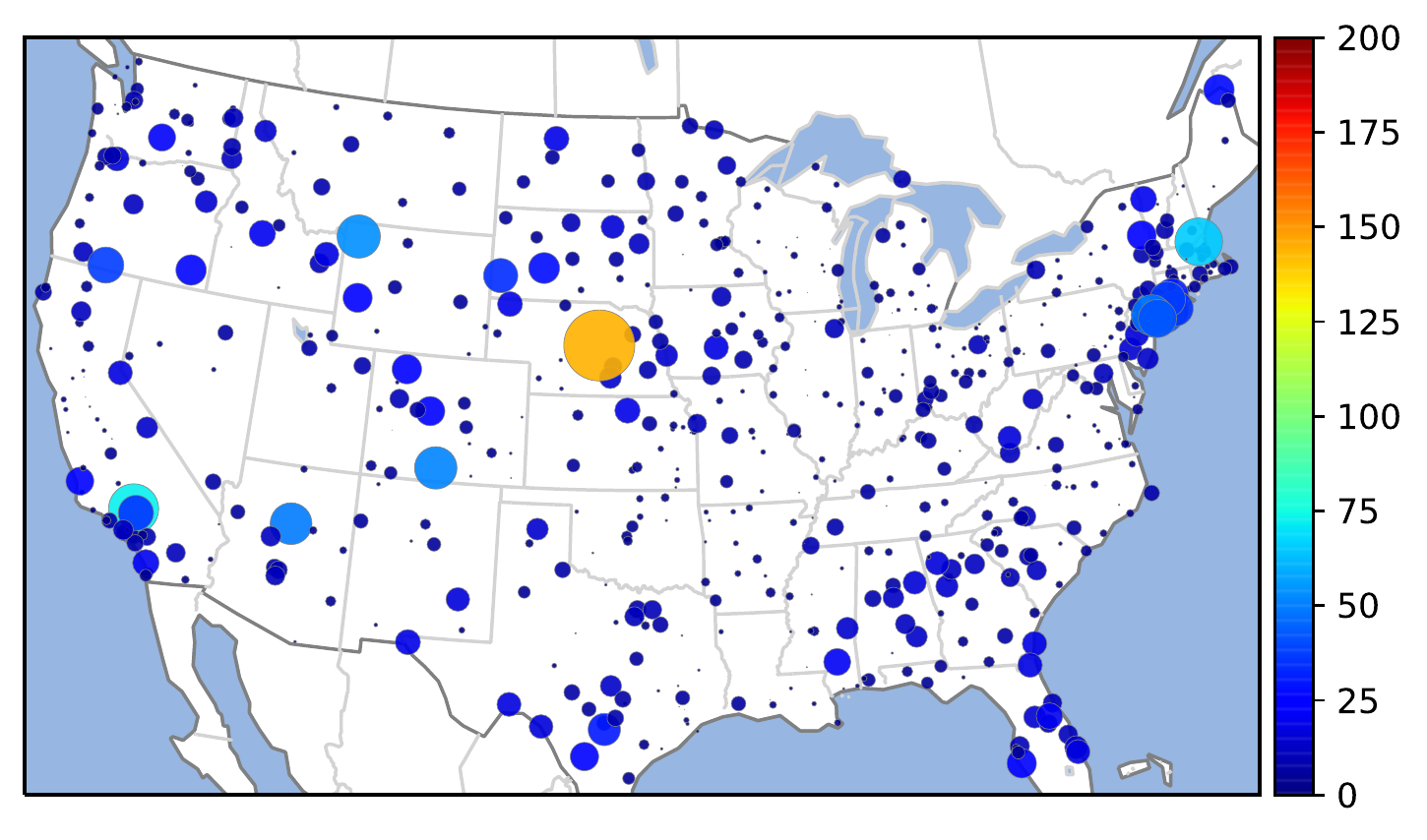}
    \subcaption{GraphDAU-EN-E}
  \end{minipage}
  \begin{minipage}[t]{.24\linewidth} \centering
    \includegraphics[width=\linewidth]{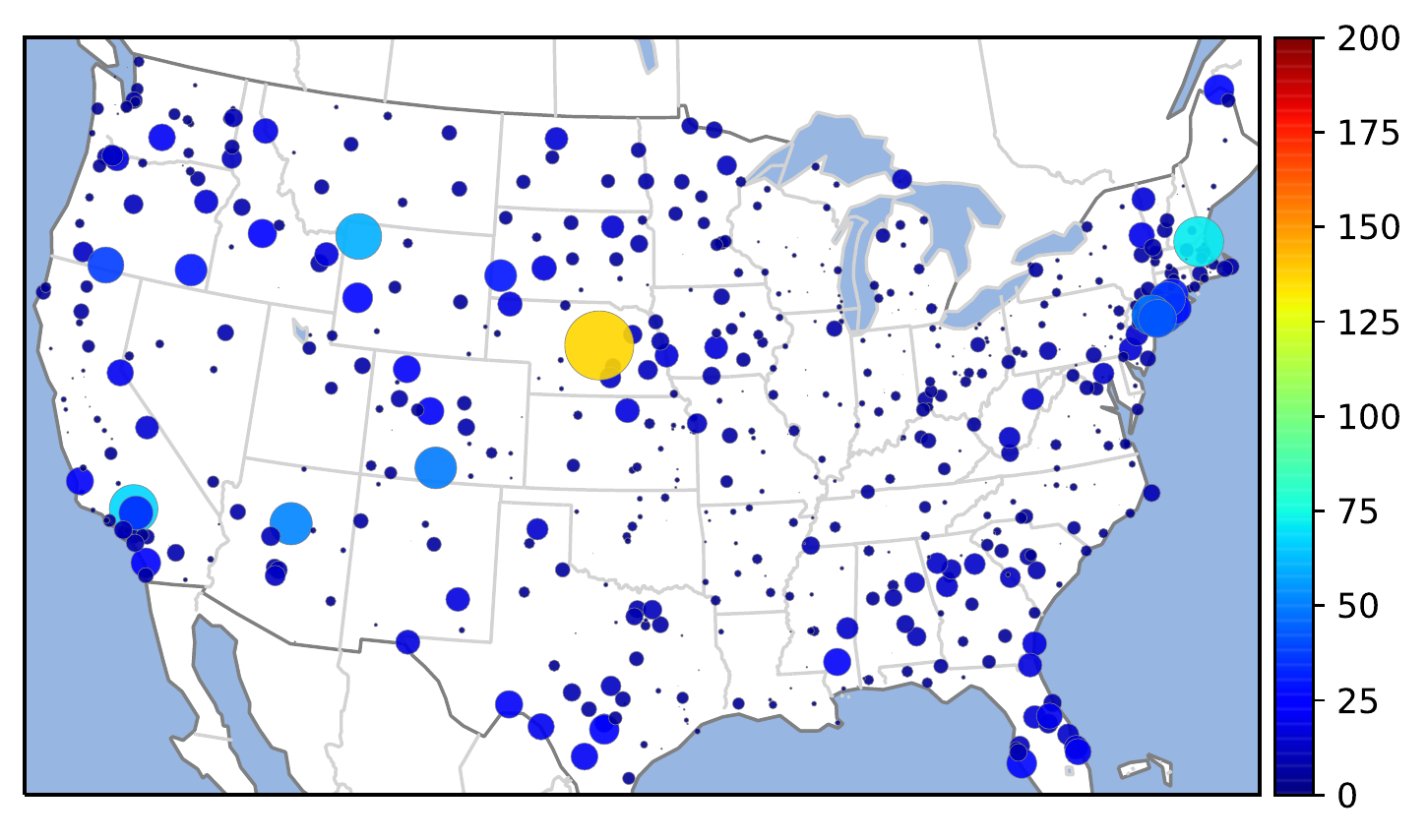}
    \subcaption{NestDAU-EN-C} \label{sfig:temp_NEDAU}
  \end{minipage}
  \caption{Denoising results of U.S.~temperature data ($\sigma=9.0$).
  (\subref{sfig:temp_gt}) is the original signal. The noisy and denoised results show the differences from the original signal: For visualization, the area of the nodes is proportional to the magnitude of the error, i.e., a large node has a large error.
  }
  \label{fig:temp_img}
\end{figure*}

The experimental results on the fixed graphs are summarized with the number of parameters in Table~\ref{tab:result_denoising}.
Visualizations of the denoising results are also shown in Figs.~\ref{fig:community_img}, \ref{fig:minn_img}, and \ref{fig:temp_img}.

In most cases, the proposed methods show RMSE improvements compared to all of the alternative methods.
It is observed that the proposed approach successfully restored graph signals having various characteristics.
Note that, in spite of the performance improvements, our methods have a significantly smaller number of parameters than the neural network-based approaches.

Although GraphDAU-TV and -EN outperform existing methods, NestDAUs provide even better performance by incorporating GraphDAUs as submodules of NestDAU.
These results imply that the nested structure is effective for graph signal restoration.
NestDAU using EVD often outperforms that using CPA in most datasets and conditions.

\subsection{Denoising Results: Graphs with Perturbation}
Table~\ref{tab:Graph_Perturbation} summarizes the results of the second experiment, denoising on graphs with small perturbation.
Overall, our algorithms outperform the alternatives as in the case for the fixed graph.
The proposed techniques show RMSE improvements for all of the signal types under consideration.
This implies that our methods effectively reflect the signal prior as the tuned parameters through training, leading to robustness against a slight change of graphs.
\begin{table*}[tpb] \centering
  \caption{Denoising results on graphs with perturbation (average RMSEs for test data)} \label{tab:Graph_Perturbation}
\begin{tabular}{c|ccc|cc|cc|cc|ccc}
\bhline{1.0pt}
 &  &  &  & \multicolumn{6}{c|}{Random sensor graph} & \multicolumn{3}{c}{3D Point Clouds} \\
 &  &  &  & \multicolumn{2}{c}{Piecewise constant} & \multicolumn{2}{c}{Piecewise smooth} & \multicolumn{2}{c|}{Globally smooth} & \multicolumn{3}{c}{RGB colors} \\ \hline
Methods & $L$ & $K$ & $P$ & $\sigma=0.5$ & $1.0$ & $0.5$ & $1.0$ & $0.5$ & $1.0$ & $20$ & $30$ & $40$ \\ \hline \hline
Noisy & - & - & - & 0.500 & 1.000 & 0.500 & 1.000 & 0.500 & 1.000 & 18.18 & 26.00 & 33.19 \\ \hline
HD & - & - & - & 0.412 & 0.627 & 0.417 & 0.646 & 0.357 & 0.537 & 11.84 & 14.74 & 17.42 \\
SGBF & - & - & - & 0.405 & 0.634 & 0.411 & 0.691 & 0.336 & 0.459 & 14.76 & 16.12 & 17.94 \\
ADMM (GTV) & 10 & - & - & 0.271 & 0.505 & 0.462 & 0.652 & 0.361 & 0.543 & 10.96 & 14.08 & 16.84 \\
PnP-HD & - & - & 8 & 0.415 & 0.645 & 0.413 & 0.645 & 0.343 & 0.538 & 11.89 & 14.97 & 17.78 \\
PnP-SGBF & - & - & 8 & 0.403 & 0.630 & 0.412 & 0.635 & 0.308 & 0.462 & 15.92 & 20.49 & 24.33 \\ \hline
GraphDAU-TV-E & 10 & - & - & 0.240 & 0.446 & 0.385 & 0.629 & 0.306 & 0.467 & 10.92 & 13.97 & 16.69 \\
GraphDAU-TV-C & 10 & 10 & - & 0.221 & 0.447 & 0.385 & 0.623 & 0.308 & 0.473 & 10.88 & 13.70 & 15.90 \\
GraphDAU-EN-E & 10 & - & - & \textbf{0.192} & 0.401 & 0.394 & 0.629 & 0.312 & 0.451 & \textbf{10.74} & 13.83 & 16.64 \\
GraphDAU-EN-C & 10 & 10 & - & 0.230 & 0.436 & 0.398 & 0.638 & 0.293 & \textbf{0.436} & 10.88 & \textbf{13.67} & 15.90 \\
NestDAU-TV-E & 10 & - & 8 & 0.206 & 0.407 & 0.368 & 0.613 & \textbf{0.290} & 0.439 & 10.83 & 13.75 & 16.06 \\
NestDAU-TV-C & 10 & 10 & 8 & 0.207 & 0.402 & 0.371 & \textbf{0.606} & \textbf{0.290} & 0.439 & 10.84 & 13.68 & \textbf{15.87} \\
NestDAU-EN-E & 10 & - & 8 & 0.206 & 0.410 & 0.371 & 0.612 & \textbf{0.290} & 0.441 & 10.89 & 13.92 & 16.45 \\
NestDAU-EN-C & 10 & 10 & 8 & 0.211 & \textbf{0.398} & \textbf{0.366} & 0.609 & 0.295 & \textbf{0.436} & 10.81 & 13.65 & 15.89 \\ \bhline{1.0pt}
\end{tabular}
\end{table*}

\subsection{Transferring Tuned Parameters for Different $N$} \label{ssec:transfer}
\begin{figure}[tb] %
  \centering
    \includegraphics[width=0.9\linewidth]{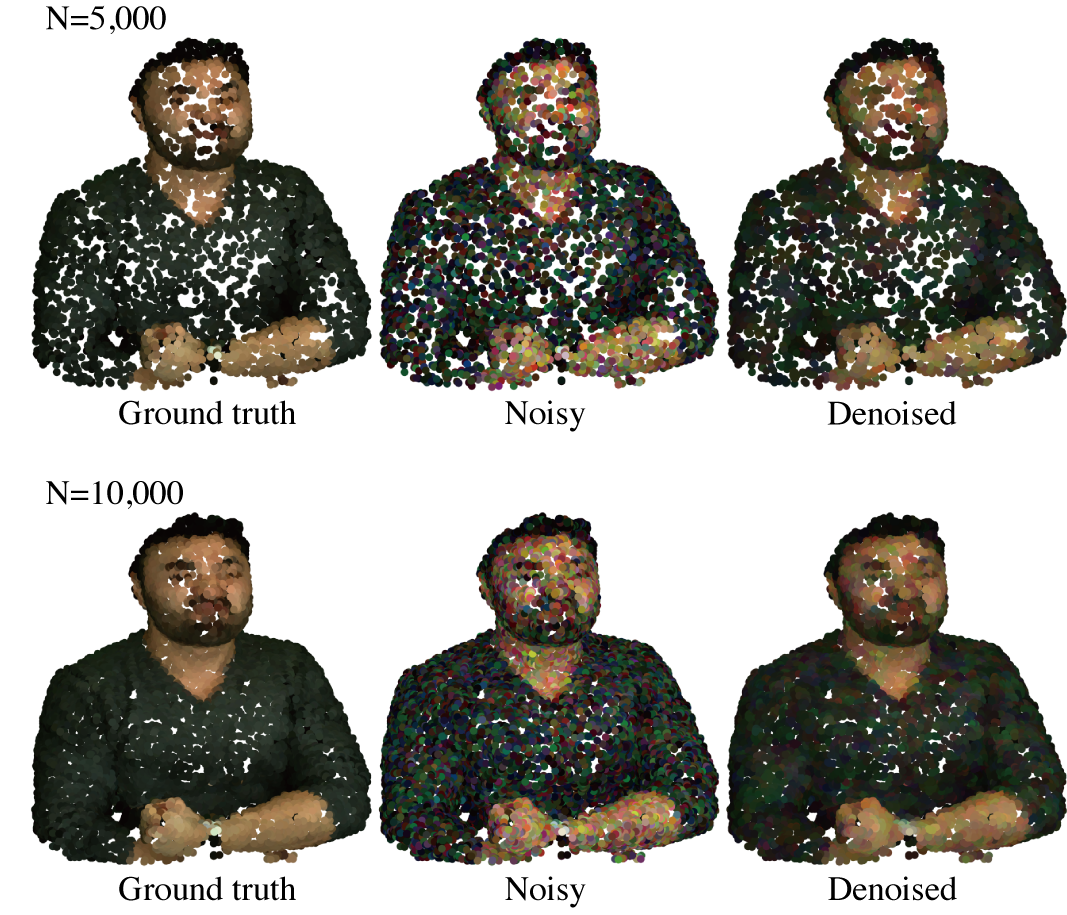} \label{a}
  \caption{\textbf{Parameter transfer:} The parameters of GraphDAU-TV-C with $L=10$ and $K=10$ are trained with $N=1,000$. Then, the model with trained parameters are applied to the larger number of points ($N=5,000$ (top row) and $N=10,000$ (bottom row)).
  }
  \label{fig:pointcloud_comparison}
\end{figure}

The results of the third experiment, transferring the tuned parameters to different $N$, are summarized in Table~\ref{tab:3DPC_transfer}.
This shows that even if the number of nodes increases, the proposed method works well as long as the signal and graph properties are similar.
Fig.~\ref{fig:pointcloud_comparison} shows the visualization of noisy and denoised results.
\begin{table}[tpb] \centering
  \caption{Parameter transfer of 3D point clouds (average RMSEs for test data)}
  \label{tab:3DPC_transfer}
    \begin{tabular}{l|c|ccc} \bhline{1.0pt}
                       & $N^{\prime}$ & \multicolumn{3}{c}{$N$} \\ \cline{2-5}
                       & 1,000 & 2,000 & 5,000 & 10,000 \\ \hline
  Noisy ($\sigma=30$) & 26.00 & 25.96 & 25.94 & 25.95 \\
  Denoised & 13.70 & 12.69 & 11.63 & 11.03 \\
  \bhline{1.0pt}
  \end{tabular}
\end{table}

\subsection{Performance Study: The Number of Layers} \label{ssec:analysis}
Along with the denoising results, the effect of the number of layers is studied here.
We use the dataset of the fixed community graph whose details are described in Section~\ref{sec:denoising_setup}.

\subsubsection{GraphDAU}%
Fig.~\ref{fig:influence_L} shows the performance analysis in terms of the number of layers $L$ of GraphDAU for $L \in \{1,\ldots,30\}$.
The average RMSE in the test data is reported.
As can be seen in the figure, the RMSE of GraphDAU rapidly decreases for $L \leq 10$, whereas there is a slight improvement for $L>10$.
We observed that GraphDAU-TV-E steadily decreases RMSEs while they are slightly oscillated for GraphDAU-EN-E.
Fig.~\ref{fig:influence_K} shows the influence of the polynomial order $K \in \{2, \ldots 30\}$ of GraphDAU-TV-C and -EN-C with $L=10$.
Both methods almost monotonically decrease RMSEs as $K$ becomes larger.

\subsubsection{NestDAU}
\begin{figure*}[t]
    \centering
  \begin{minipage}[t]{0.32\linewidth} \centering
    \includegraphics[width=\linewidth]{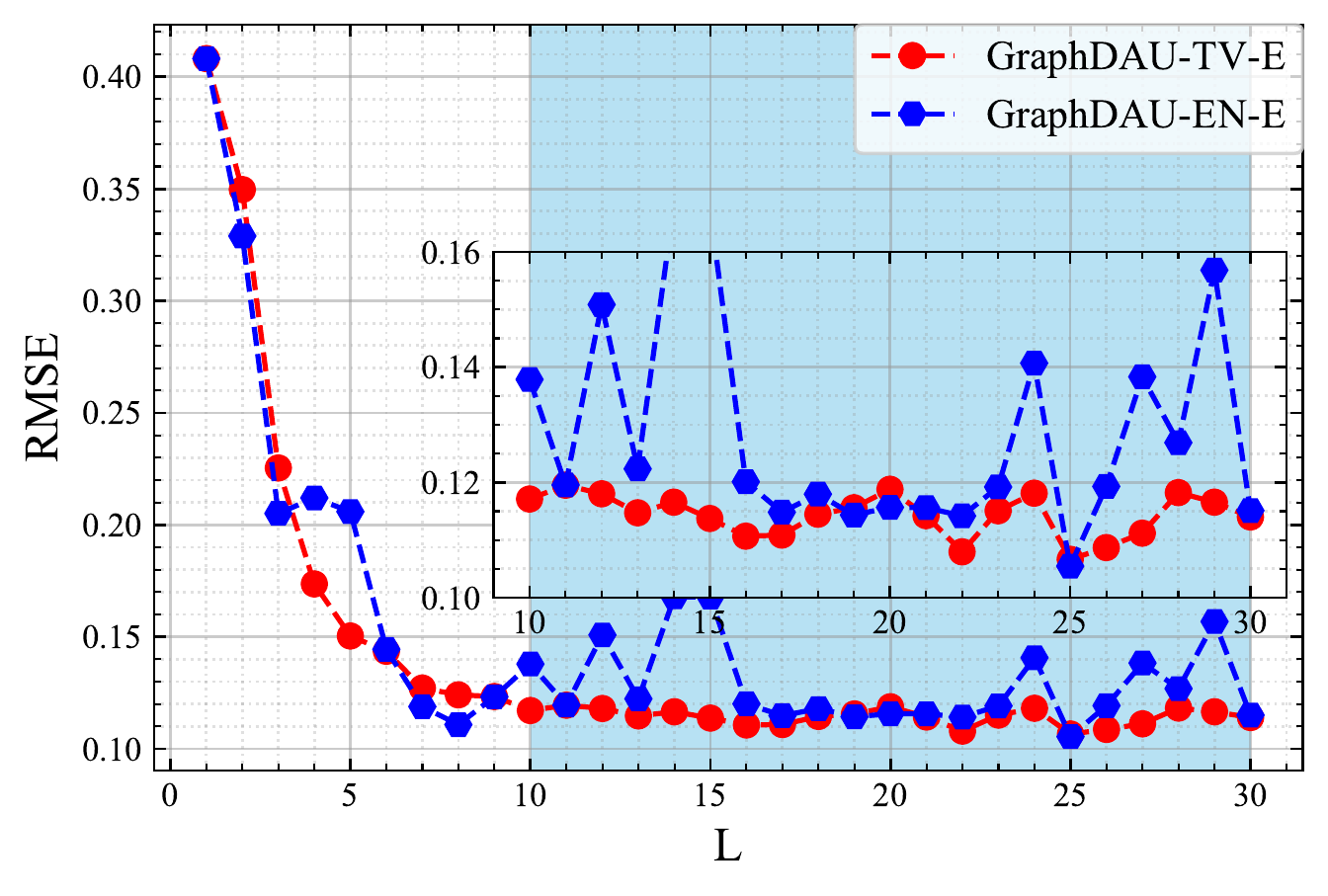}
    \subcaption{GraphDAU: $L$.} \label{fig:influence_L}
  \end{minipage}
  \begin{minipage}[t]{0.32\linewidth} \centering
    \includegraphics[width=\linewidth]{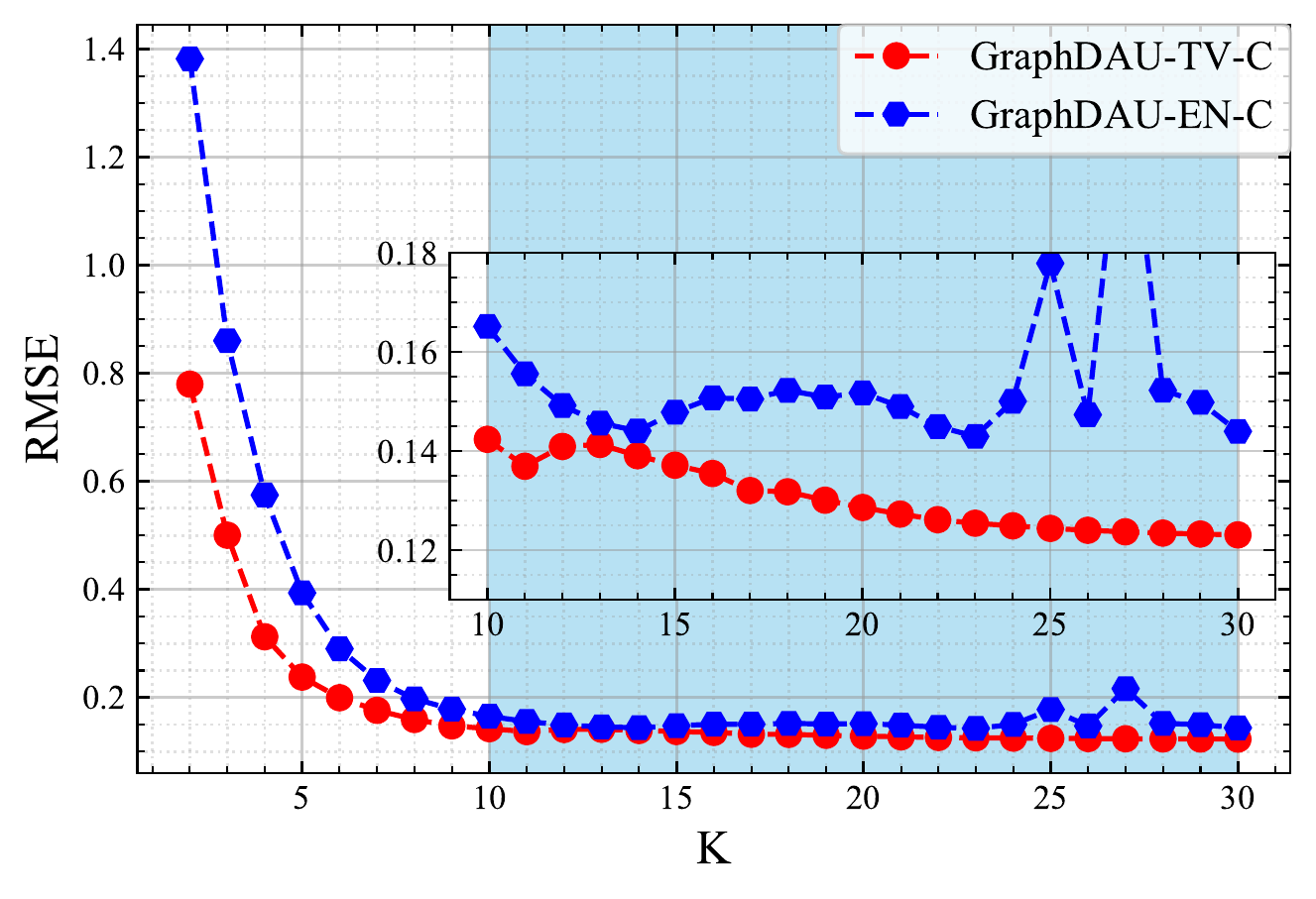}
    \subcaption{GraphDAU: $K$.} \label{fig:influence_K}
  \end{minipage}
  \begin{minipage}[t]{0.32\linewidth} \centering
    \includegraphics[width=\linewidth]{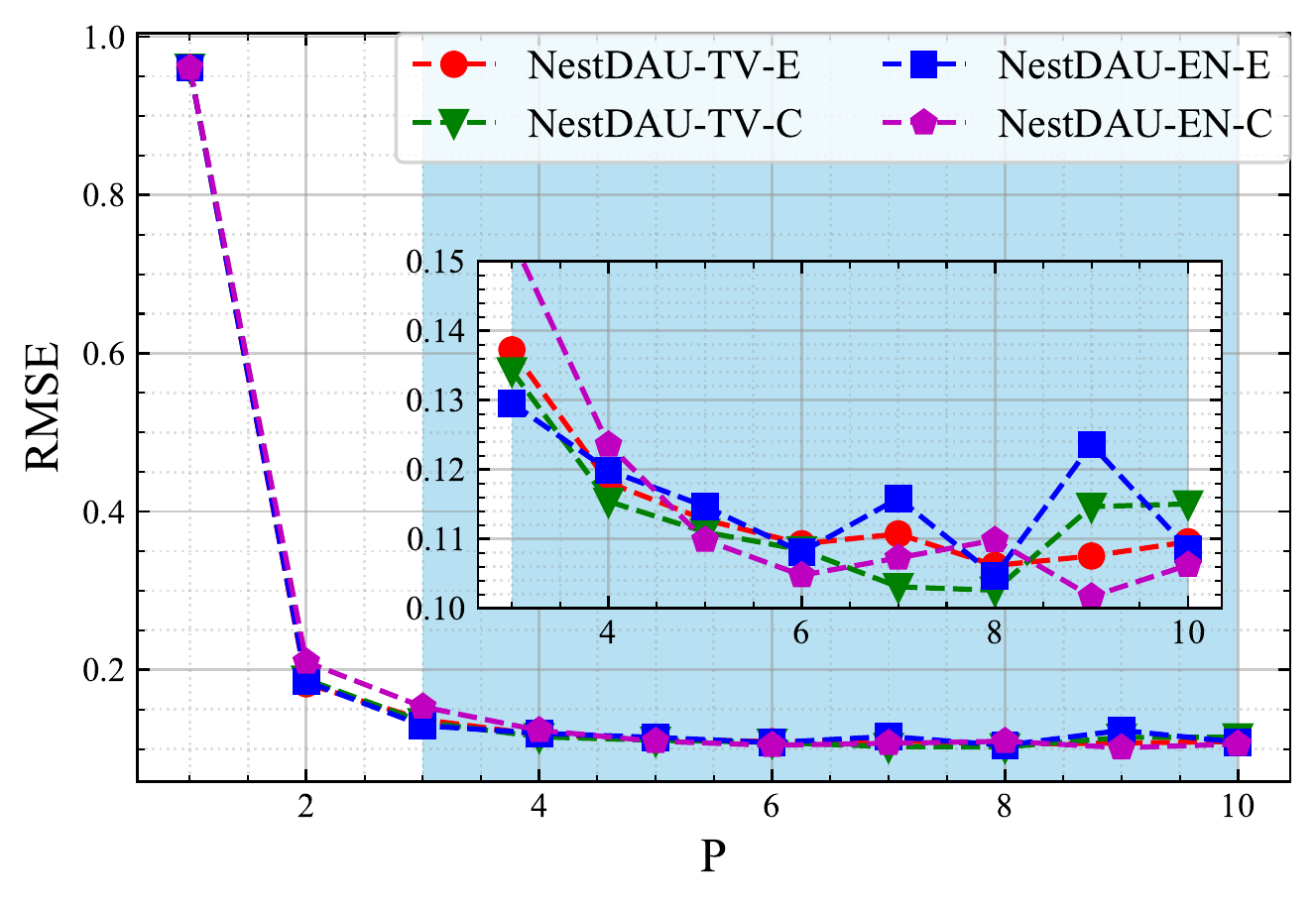}
    \subcaption{NestDAU: $P$.} \label{fig:influence_P}
  \end{minipage}
  \caption{Denoiser analysis with the results on community graph ($\sigma=1.0$).}
  \label{fig:effect_GraphDAU}
\end{figure*}

Fig.~\ref{fig:influence_P} shows the performance in terms of the number of layers $P$ of NestDAU.
The submodule GraphDAU contains $L=10$ for using EVD and $L=10$ and $K=10$ with that using CPA.
The number of layers is selected to $P \in \{1, \ldots,10\}$.
For NestDAU, all configurations are stable in terms of the layer size $P$.
Even if the in-loop denoisers are changed, the performances are almost equivalent.

\subsection{RMSE Analysis during Training}
\begin{figure}[!t]
    \centering
    \includegraphics[width=0.5\textwidth]{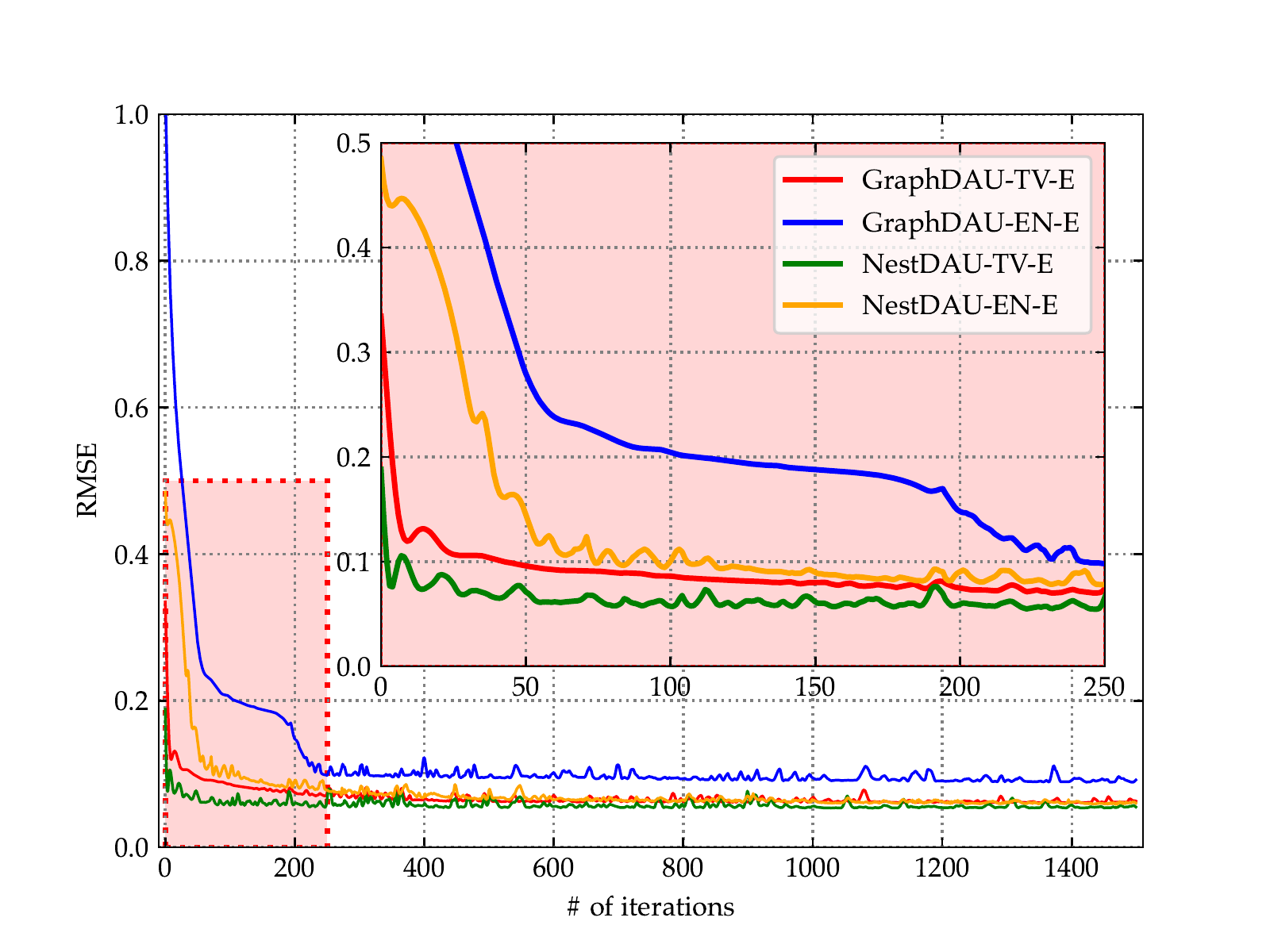}
    \caption{Average RMSEs of the validation data for graph signal denoising.
    The upper right box is the zoomed-in part of the red area at bottom left.}
    \label{fig:validation_plot}
\end{figure}

Fig.~\ref{fig:validation_plot} shows the average RMSEs of the validation data during training.
The data used are signals on community graphs ($\sigma=0.5$) described in Section~\ref{sec:denoising_setup}.
As shown in the figure, the RMSEs rapidly decrease with less than $250$ iterations (that is, the number of training data).
Furthermore, NestDAUs converge faster than their GraphDAU counterparts.

\section{Experimental Results: Interpolation} \label{sec:interpolation}
In this section, graph signal interpolation is performed and compared with the alternative methods.
We assume the nodes for missing signal values are known and they are set to zero.
This leads to a diagonal binary matrix $\mathbf{H} = \text{diag}\{0,1\}^N$ in \eqref{eq:problem} with various missing rates.

\subsection{Alternative Methods}
For interpolation, the following techniques are selected for comparison:
\begin{itemize}
  \item Bandlimited graph signal recovery based on graph sampling theory \cite{tanaka2020SamplingSignals}: Bandwidth is set to $N/10$;
  \item PnP-ADMM-based interpolation with fixed parameters with 8 iterations \cite{yazaki2019}: Its formulation is given in Section~\ref{ssec:DAU_pnp} and off-the-shelf denoisers are HD or SGBF;
  \item GUTF \cite{chen2020Graphunrolling};
  \item GUSC \cite{chen2020Graphunrolling}.
\end{itemize}
Although GUSC and GUTF are originally developed for a denoising task, we also include these methods to compare with neural network-based approaches.
The setup is the same as the previous section.

\subsection{Datasets and Setup}
\label{subsec:dataset_interpolation}
We used the following graph signals for interpolation:
\begin{itemize}
    \item Synthetic signals on a community graph having three clusters ($N=250$);
    \item Temperature data of the United States ($N=614$).
\end{itemize}
They are the same signals as those used in the denoising experiment in the previous section.

\noindent
\textbf{Characteristics of Graphs and Graph Signals:}
Synthetic graph signals on the community graph are generated in the same setup as that of the denoising experiment.
We then consider two interpolation conditions:
1) noiseless and 2) noisy (AWGN with $\sigma= 0.5$).
Three types of missing rate are considered: 30\%, 50\%, and 70\%.

The U.S.~temperature data are also used in this experiment as a real-world example.
In this case, AWGN ($\sigma=9.0$) are added onto the observed daily temperature data with the same setting as the denoising experiment.
Then, missing rates are set to 30\%, 50\%, and 70\% to validate the interpolation method.
Note that the missing nodes are randomly chosen, i.e., $\mathbf{H}$ are set to be different across all data.

\subsection{Interpolation Results}
\begin{figure*}[!tb]\centering
  \begin{minipage}[t]{.24\linewidth} \centering
    \includegraphics[width=\linewidth]{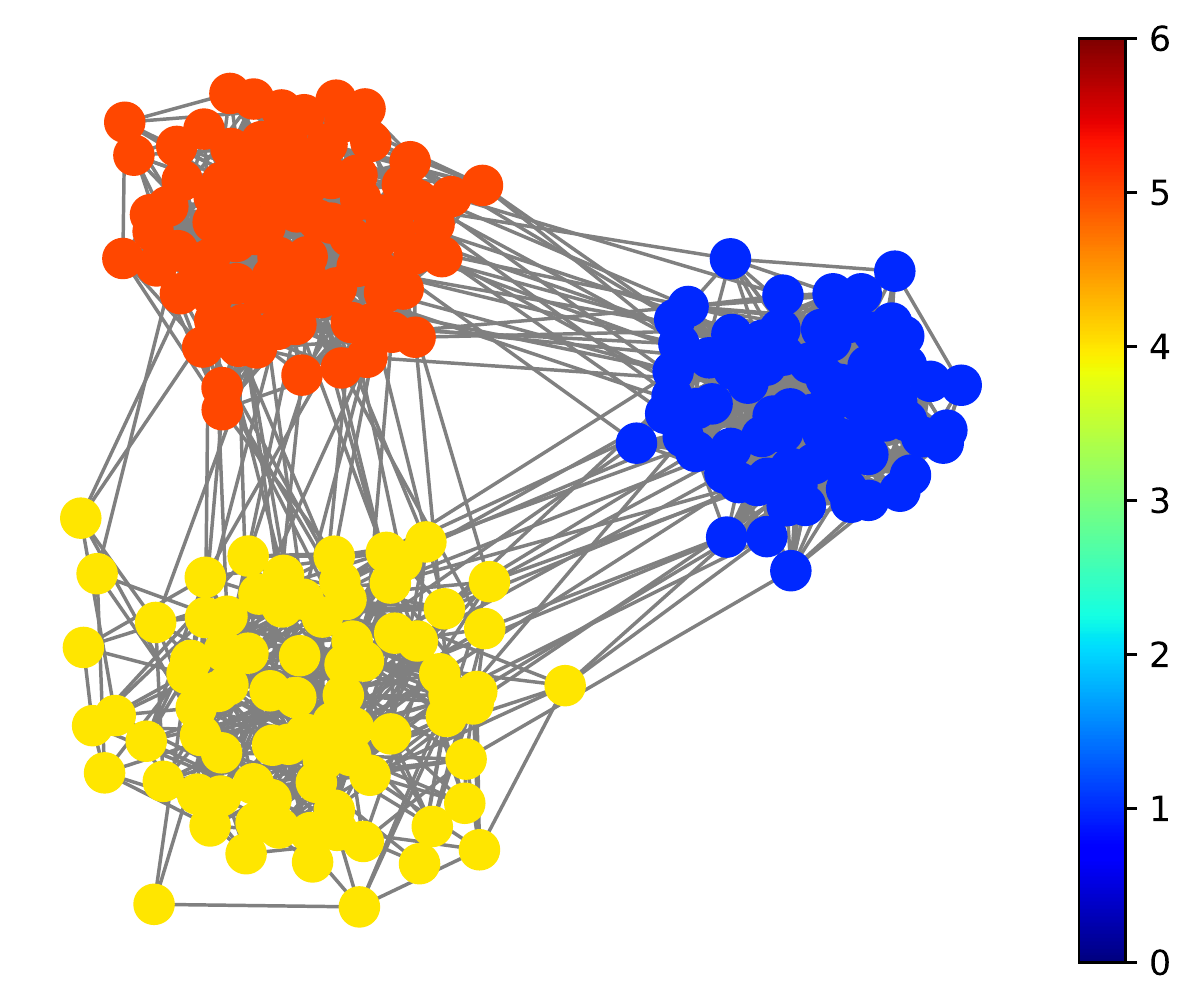}
    \subcaption{Ground truth} \label{sfig:comm_int}
  \end{minipage}
  \begin{minipage}[t]{.24\linewidth} \centering
    \includegraphics[width=\linewidth]{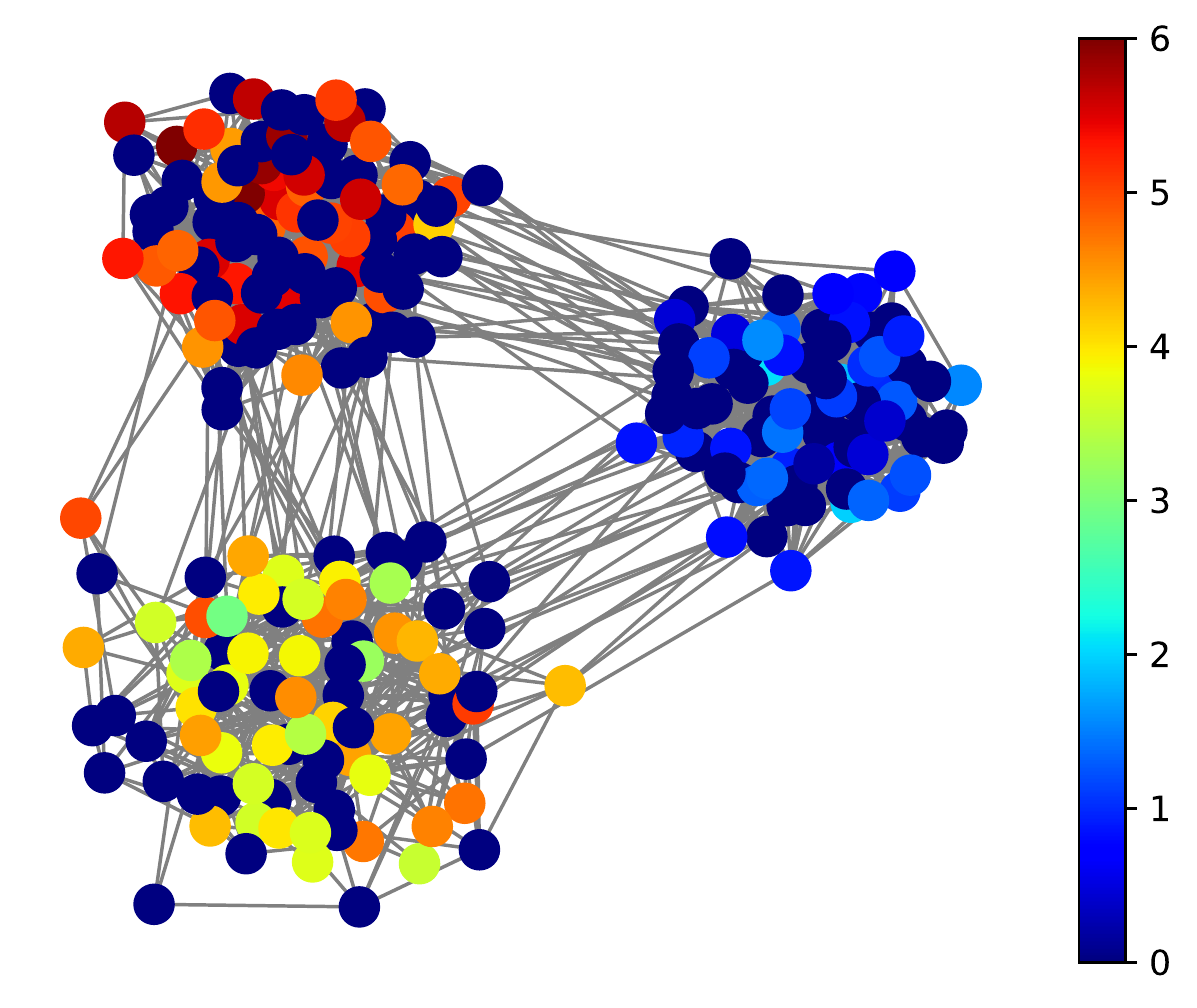}
    \subcaption{Noisy + missing ($50\%$)}
  \end{minipage}
  \begin{minipage}[t]{.24\linewidth} \centering
    \includegraphics[width=\linewidth]{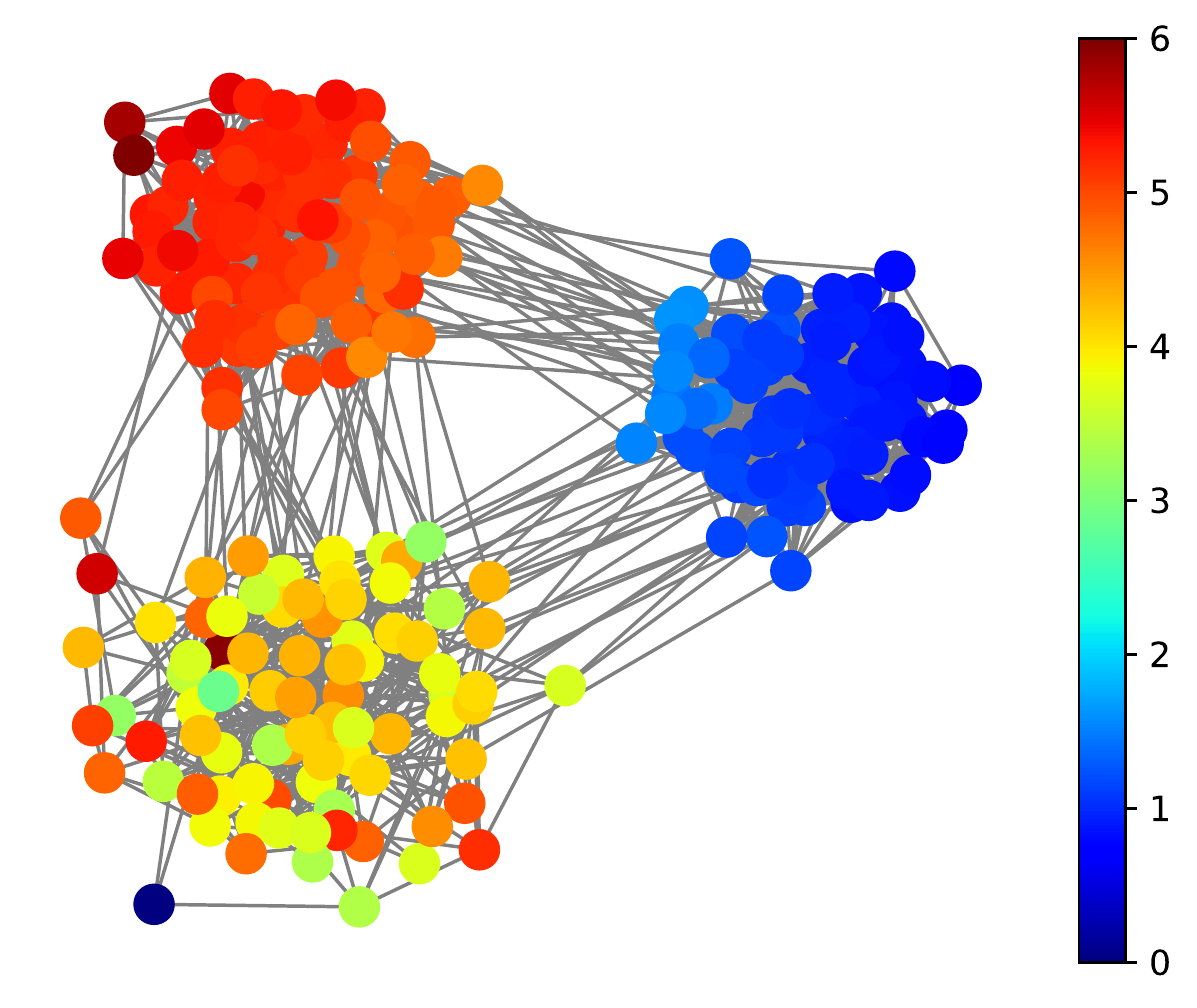}
    \subcaption{Graph sampling theory}
  \end{minipage}
  \begin{minipage}[t]{.24\linewidth} \centering
    \includegraphics[width=\linewidth]{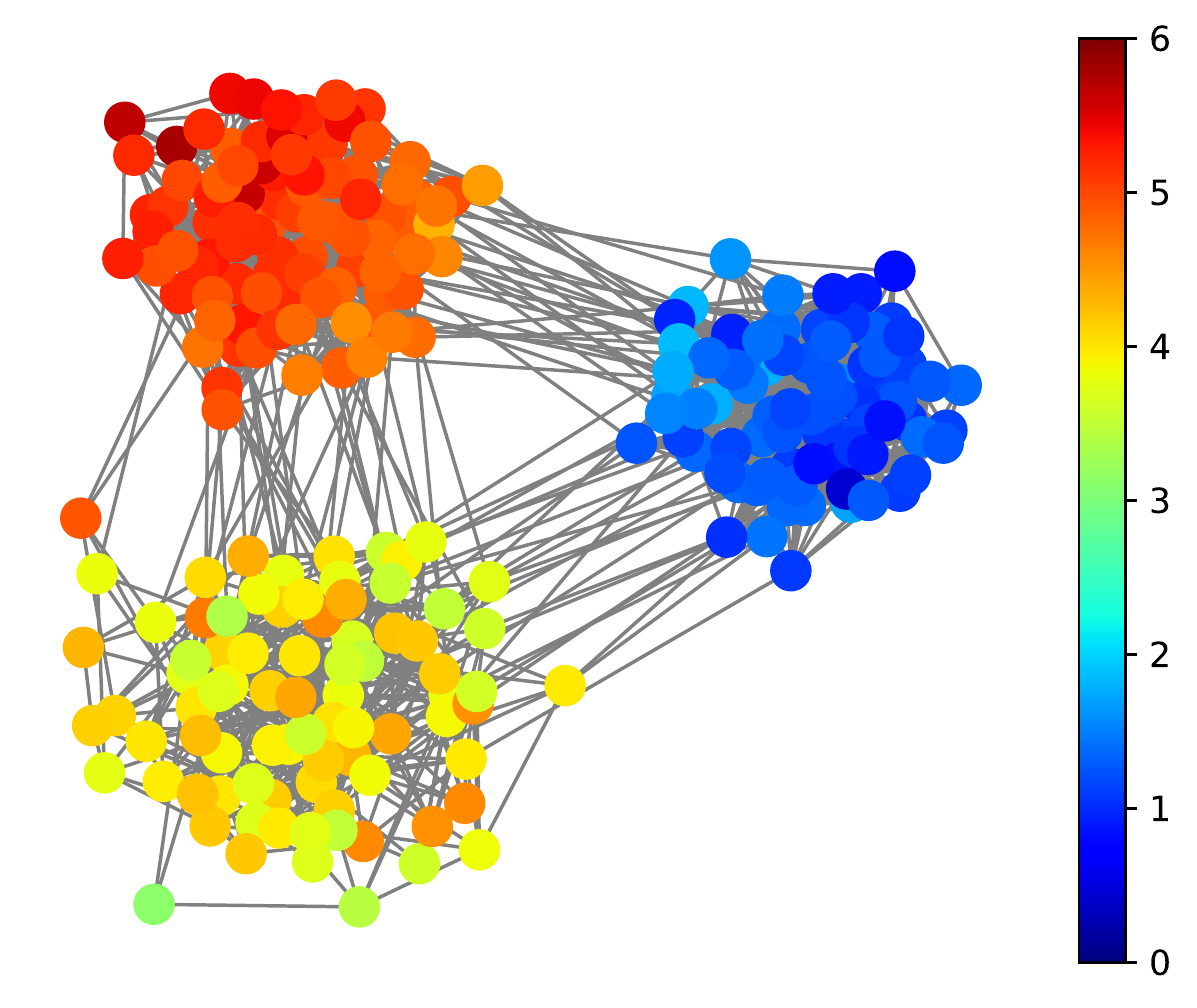}
    \subcaption{PnP-HD}
  \end{minipage}
  \\
  \begin{minipage}[t]{.24\linewidth} \centering
    \includegraphics[width=\linewidth]{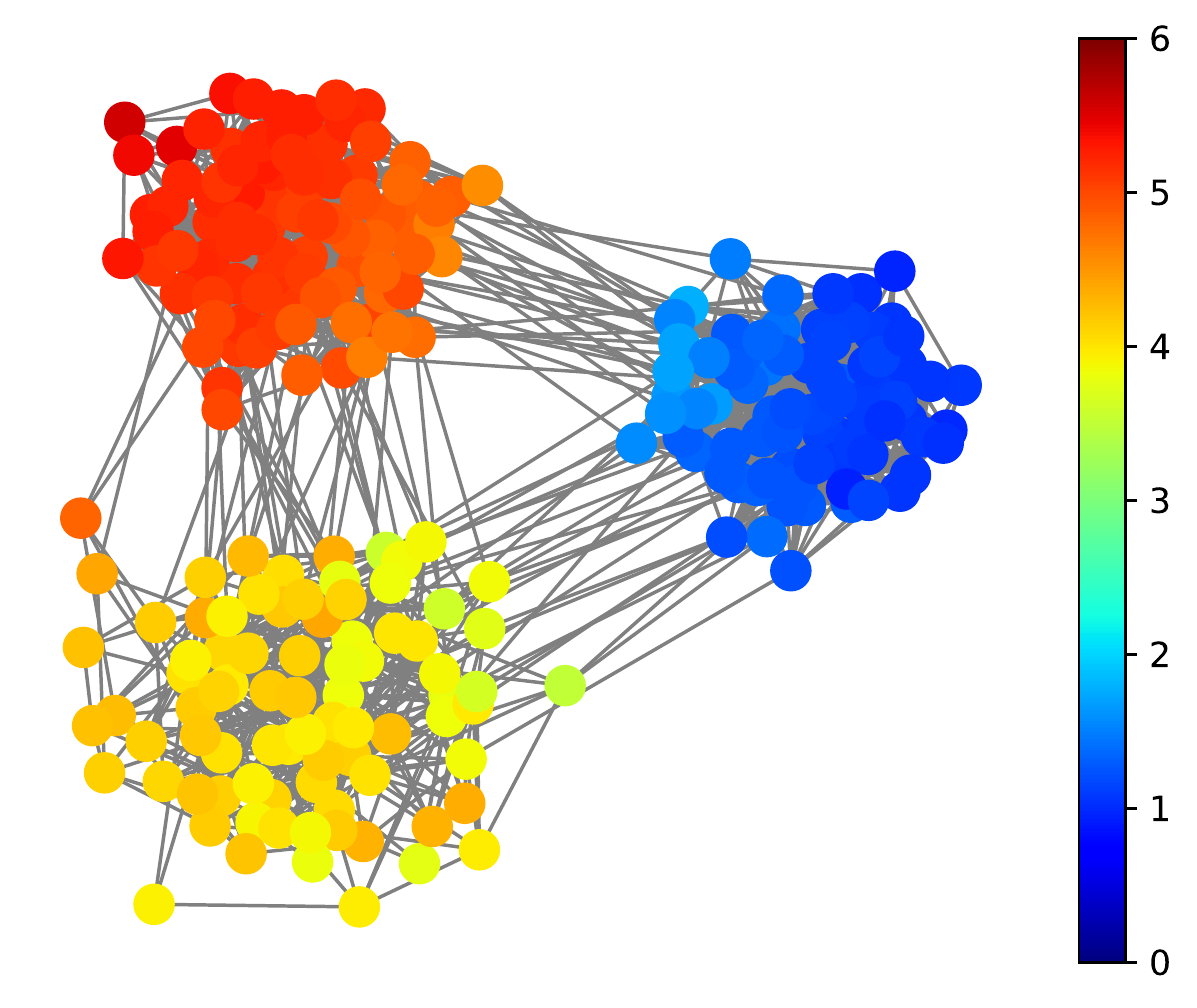}
    \subcaption{PnP-SGBF}
  \end{minipage}
  \begin{minipage}[t]{.24\linewidth} \centering
    \includegraphics[width=\linewidth]{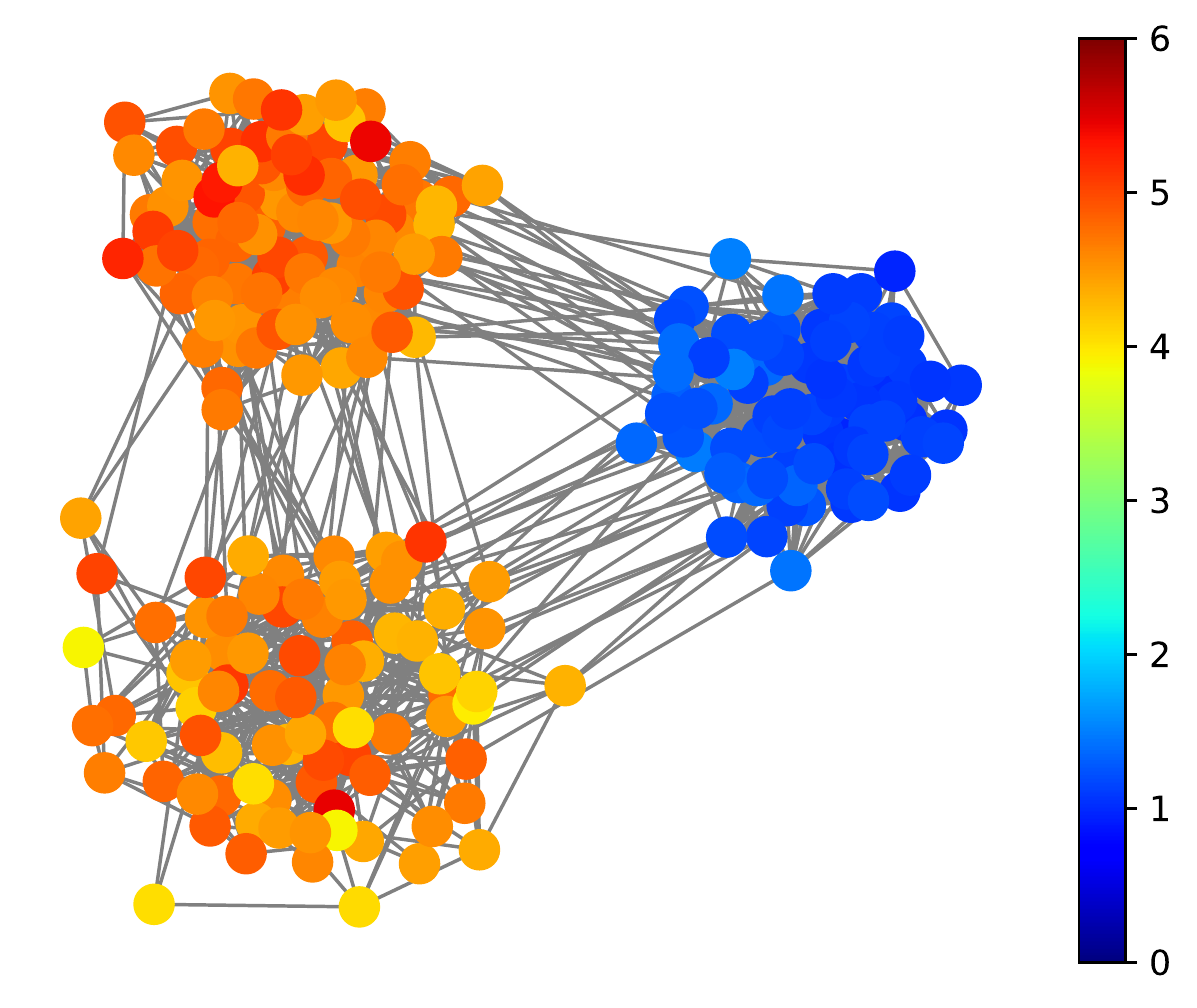}
    \subcaption{GUTF}
  \end{minipage}
  \begin{minipage}[t]{.24\linewidth} \centering
    \includegraphics[width=\linewidth]{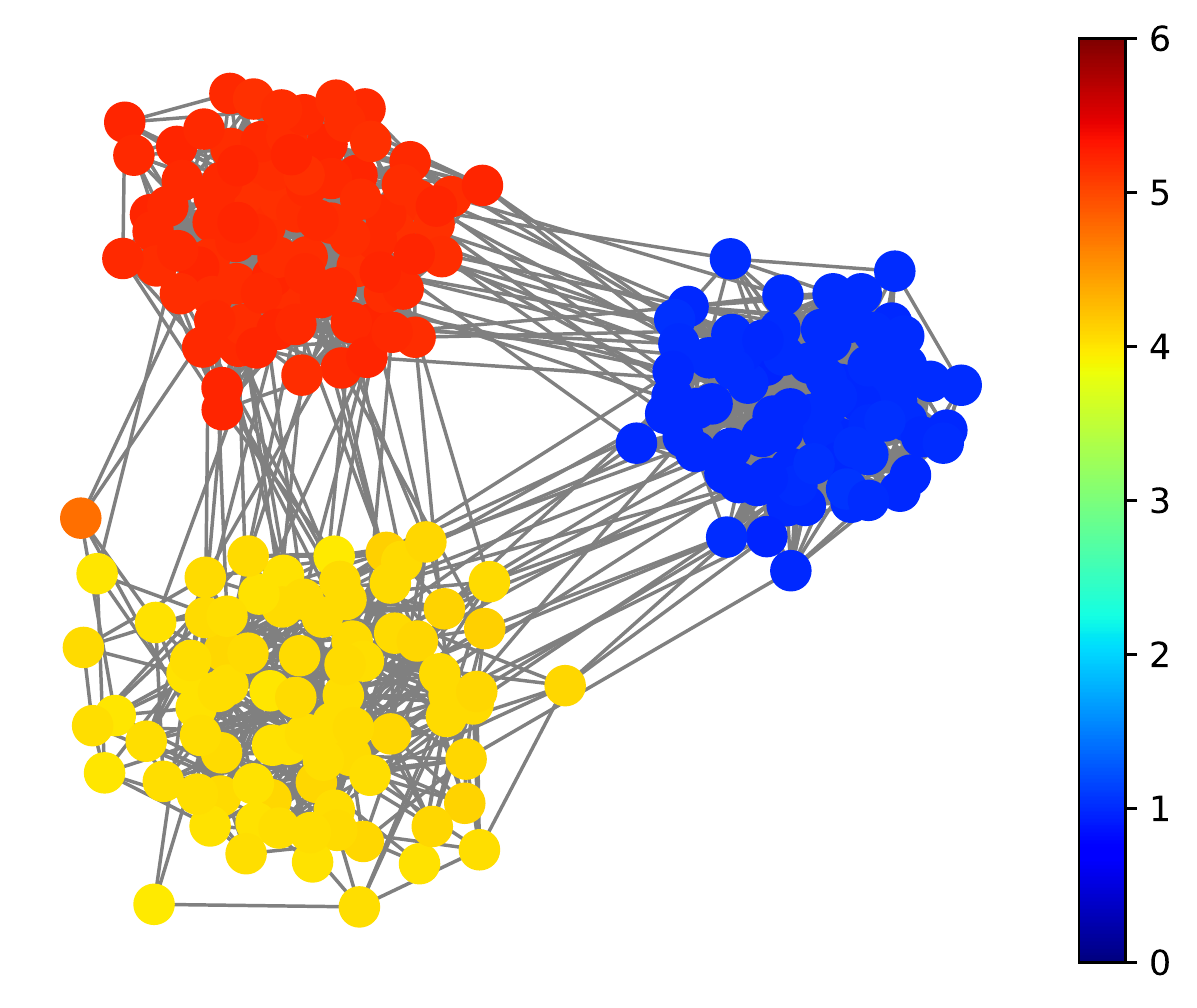}
    \subcaption{NestDAU-TV-E} \label{sfig:comm_NEDAU_T_int}
  \end{minipage}
  \begin{minipage}[t]{.24\linewidth} \centering
    \includegraphics[width=\linewidth]{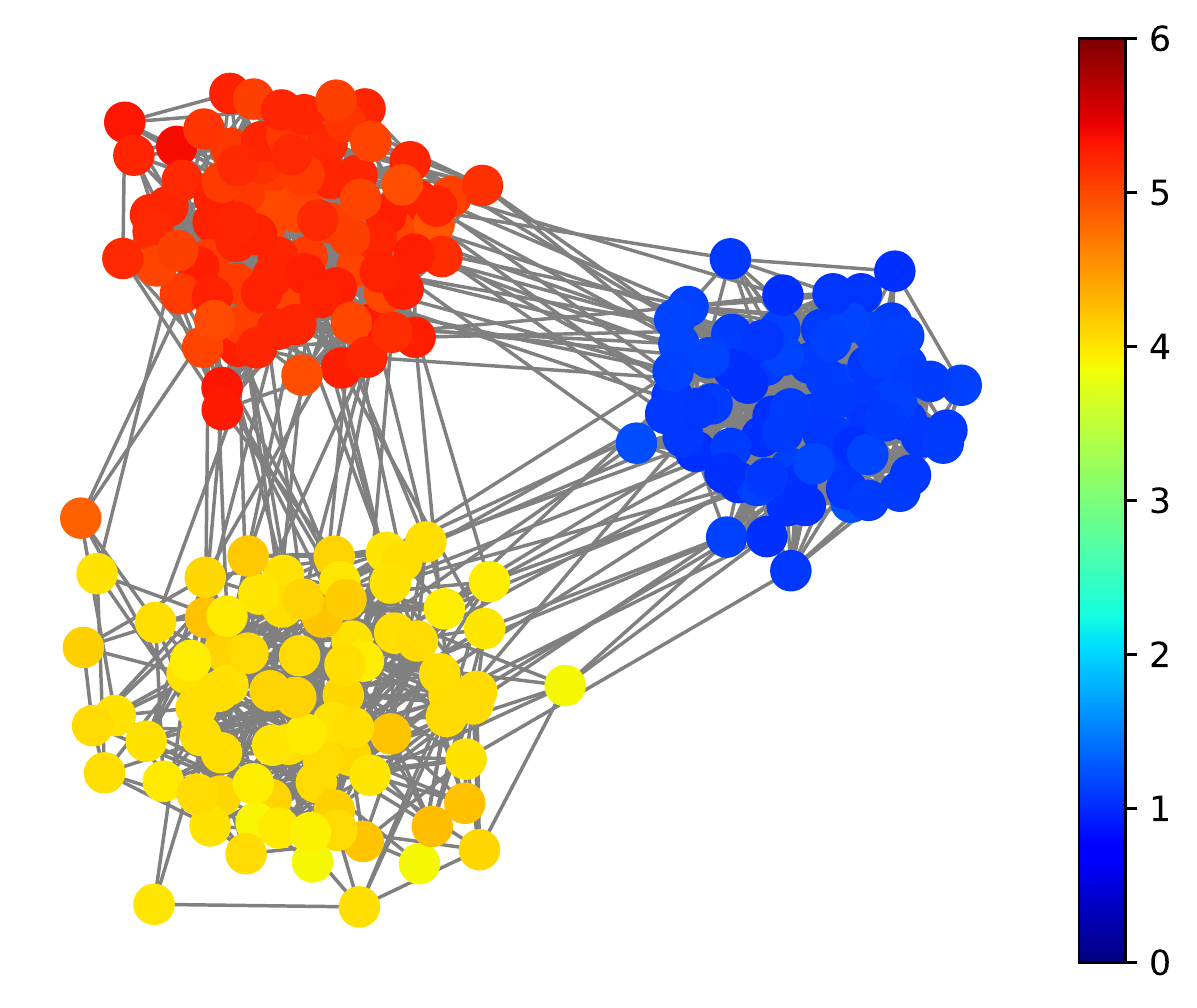}
    \subcaption{NestDAU-EN-E} \label{sfig:comm_NEDAU_E_int}
  \end{minipage}
  \caption{Interpolation results of a community graph (AWGN ($\sigma=0.5$) with 50\% missing).
  The proposed method ((\subref{sfig:comm_NEDAU_T_int}) and (\subref{sfig:comm_NEDAU_E_int})) captured the property of the ground truth signal.}
  \label{fig:comm_img_int}
\end{figure*}

\begin{figure*}[!tb] \centering
  \begin{minipage}[t]{.24\linewidth} \centering
    \includegraphics[width=\linewidth]{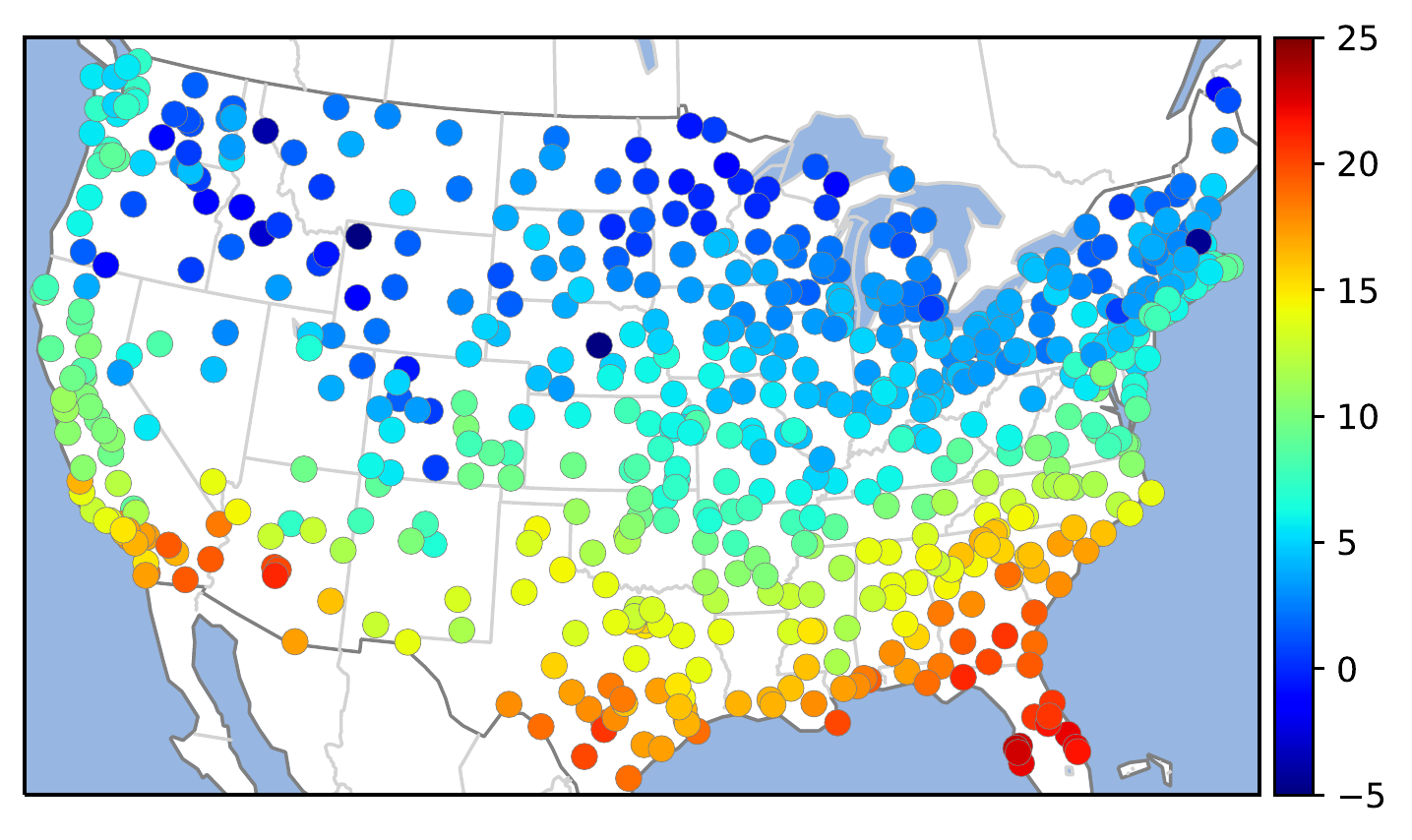}
    \subcaption{Ground truth} \label{sfig:temp_int}
  \end{minipage}
  \begin{minipage}[t]{.24\linewidth} \centering
    \includegraphics[width=\linewidth]{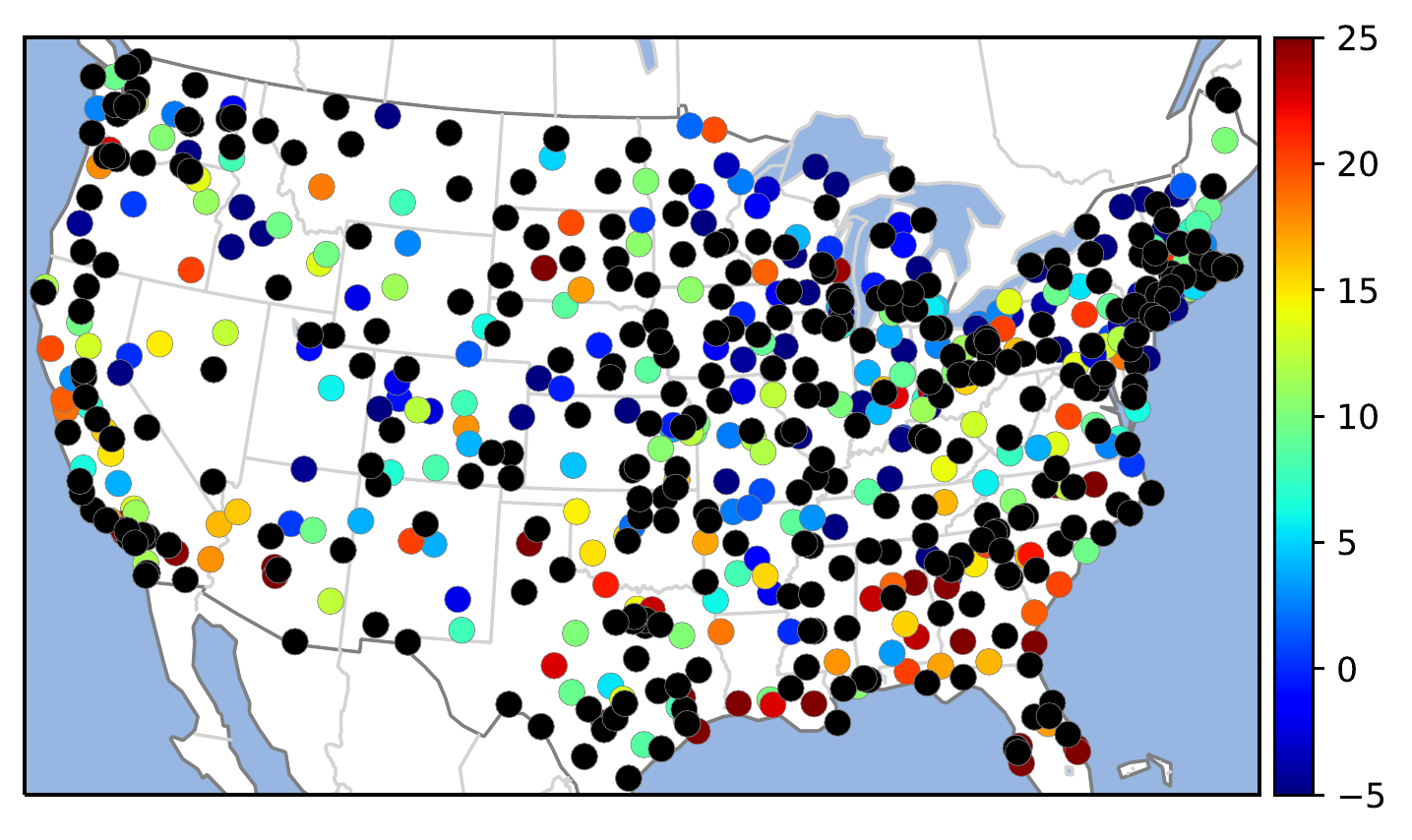}
    \subcaption{Noisy + missing ($50\%$)}  \label{sfig:noise}
  \end{minipage}
  \begin{minipage}[t]{.24\linewidth} \centering
    \includegraphics[width=\linewidth]{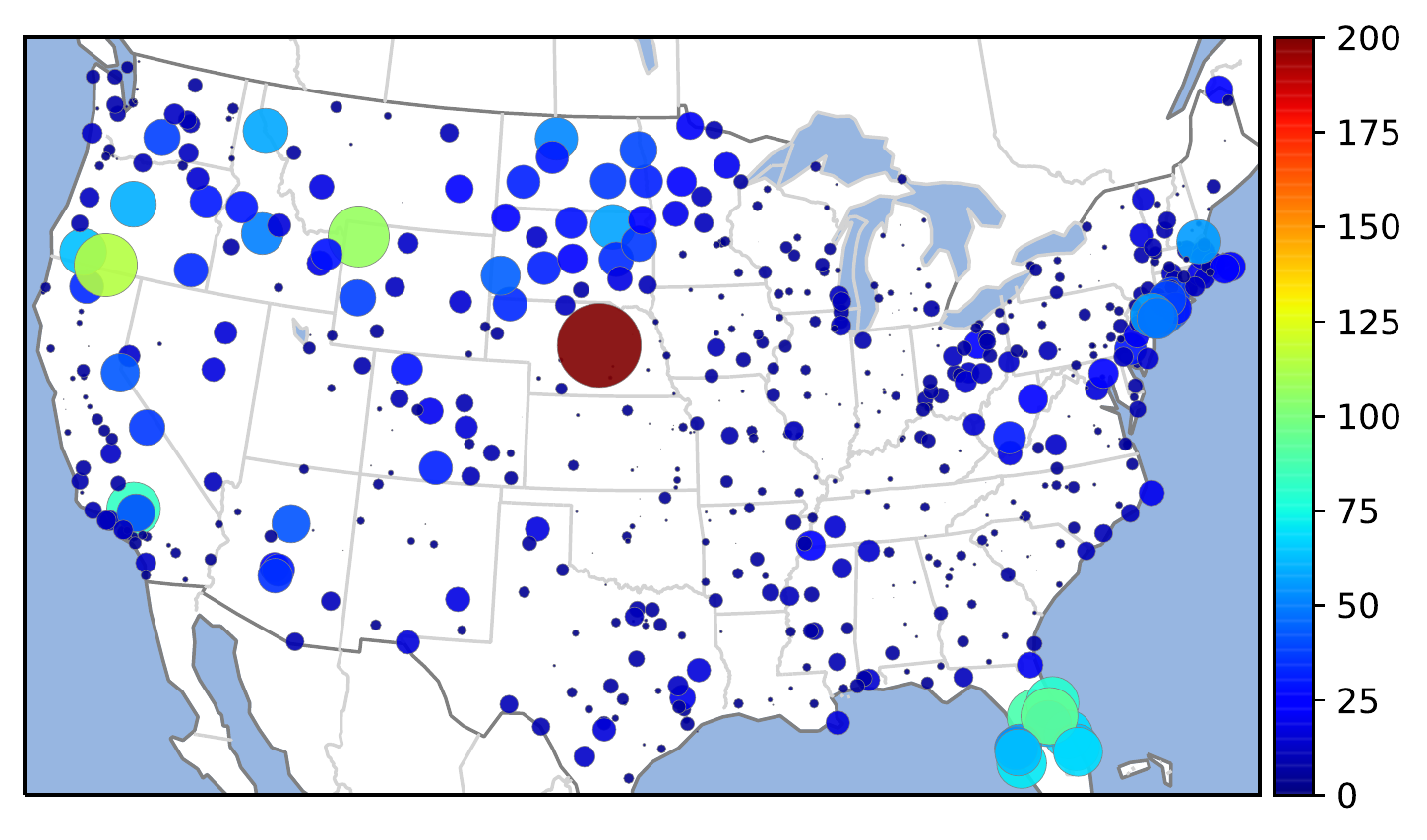}
    \subcaption{PnP-HD}
  \end{minipage}
  \begin{minipage}[t]{.24\linewidth} \centering
    \includegraphics[width=\linewidth]{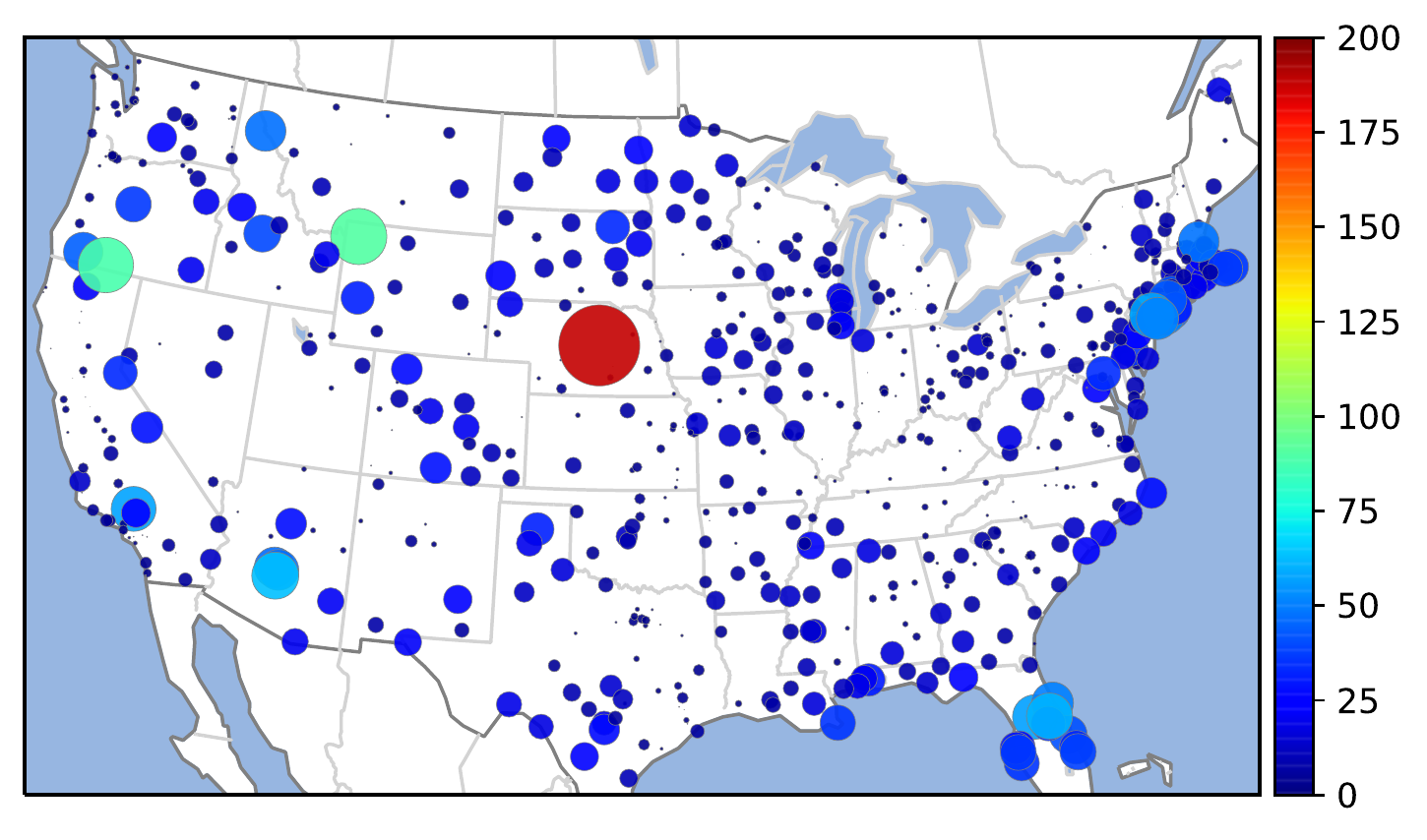}
    \subcaption{PnP-SGBF}
  \end{minipage}
  \\
  \begin{minipage}[t]{.24\linewidth} \centering
    \includegraphics[width=\linewidth]{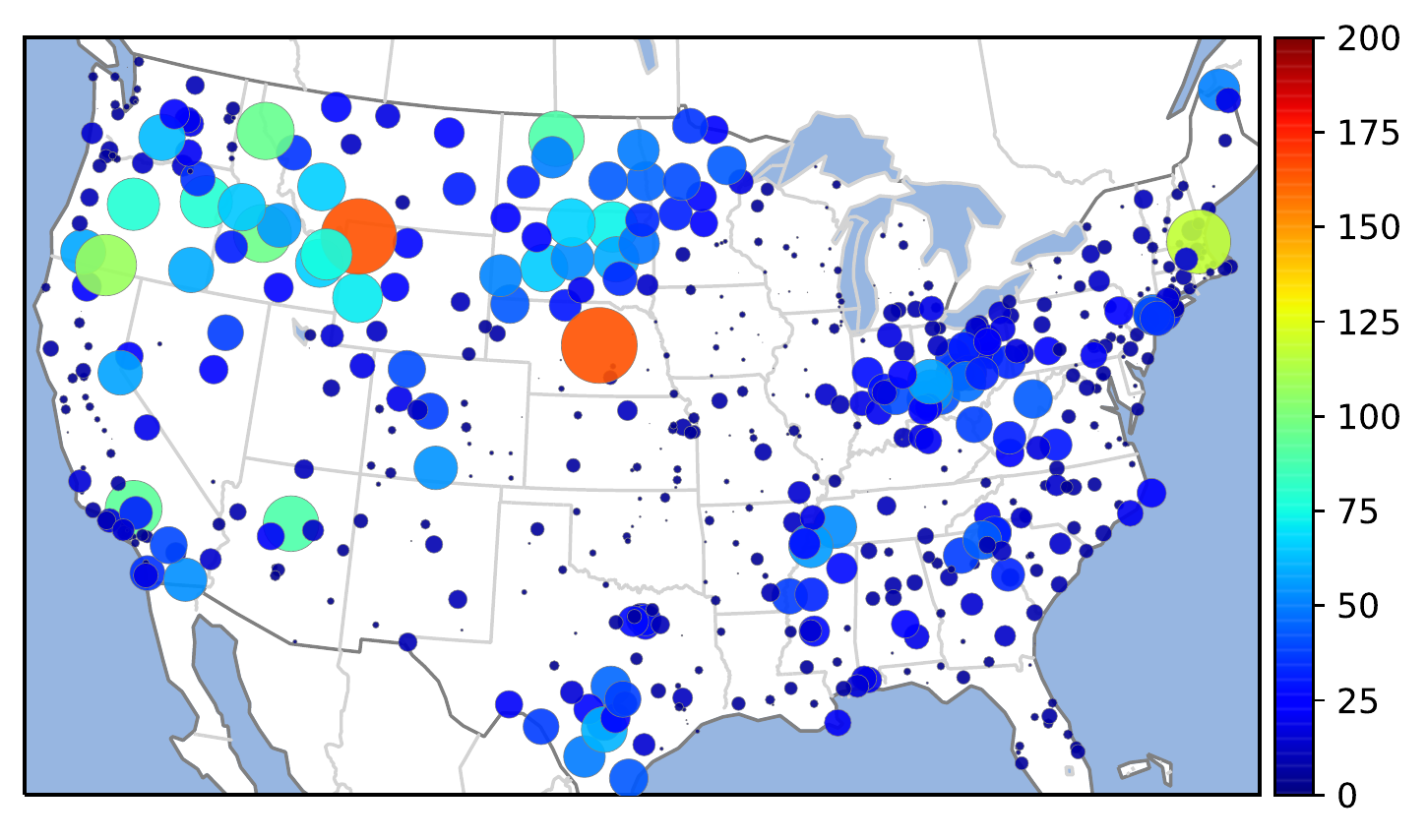}
    \subcaption{GUSC}
  \end{minipage}
  \begin{minipage}[t]{.24\linewidth} \centering
    \includegraphics[width=\linewidth]{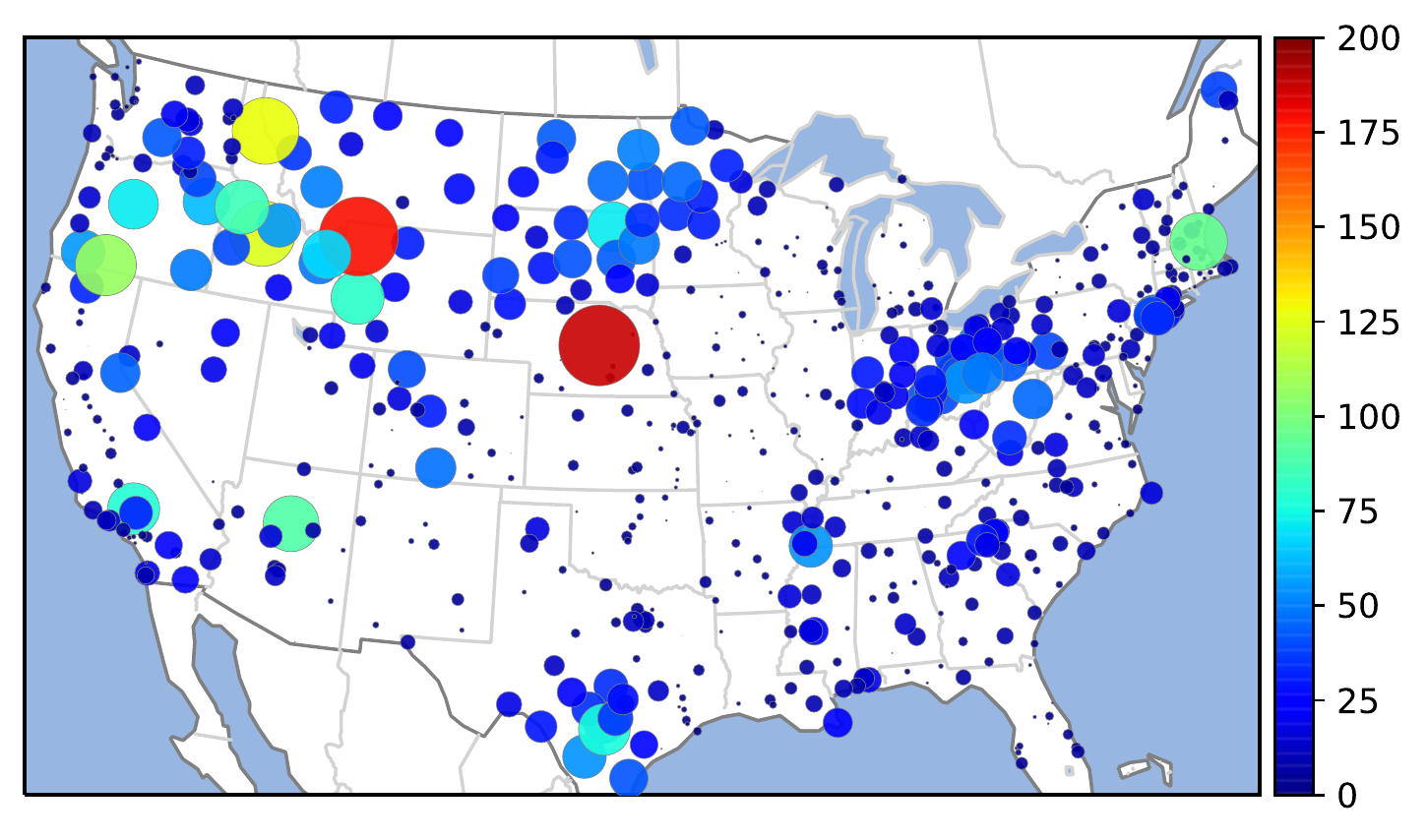}
    \subcaption{GUTF}
  \end{minipage}
  \begin{minipage}[t]{.24\linewidth} \centering
    \includegraphics[width=\linewidth]{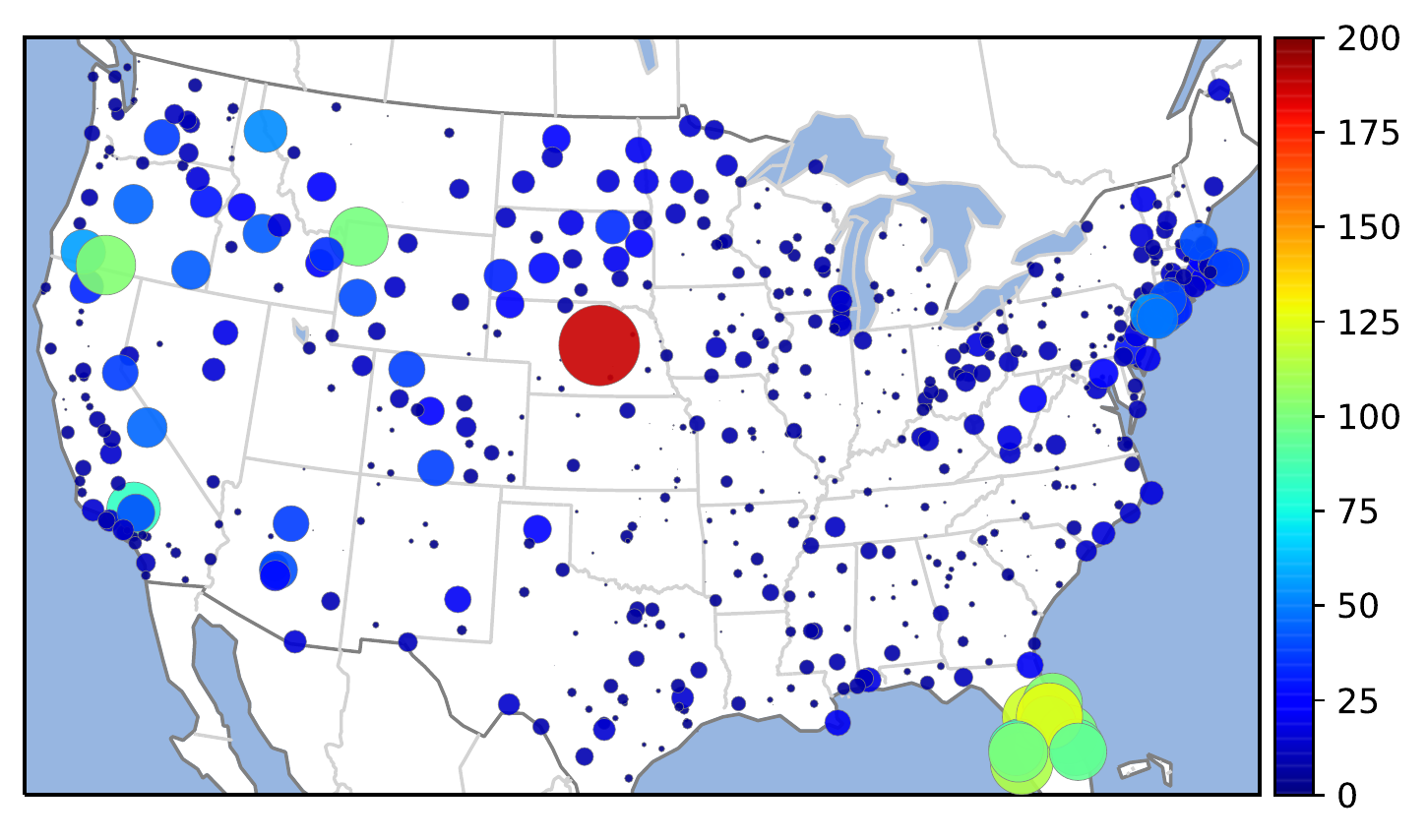}
    \subcaption{NestDAU-TV-E}
  \end{minipage}
  \begin{minipage}[t]{.24\linewidth} \centering
    \includegraphics[width=\linewidth]{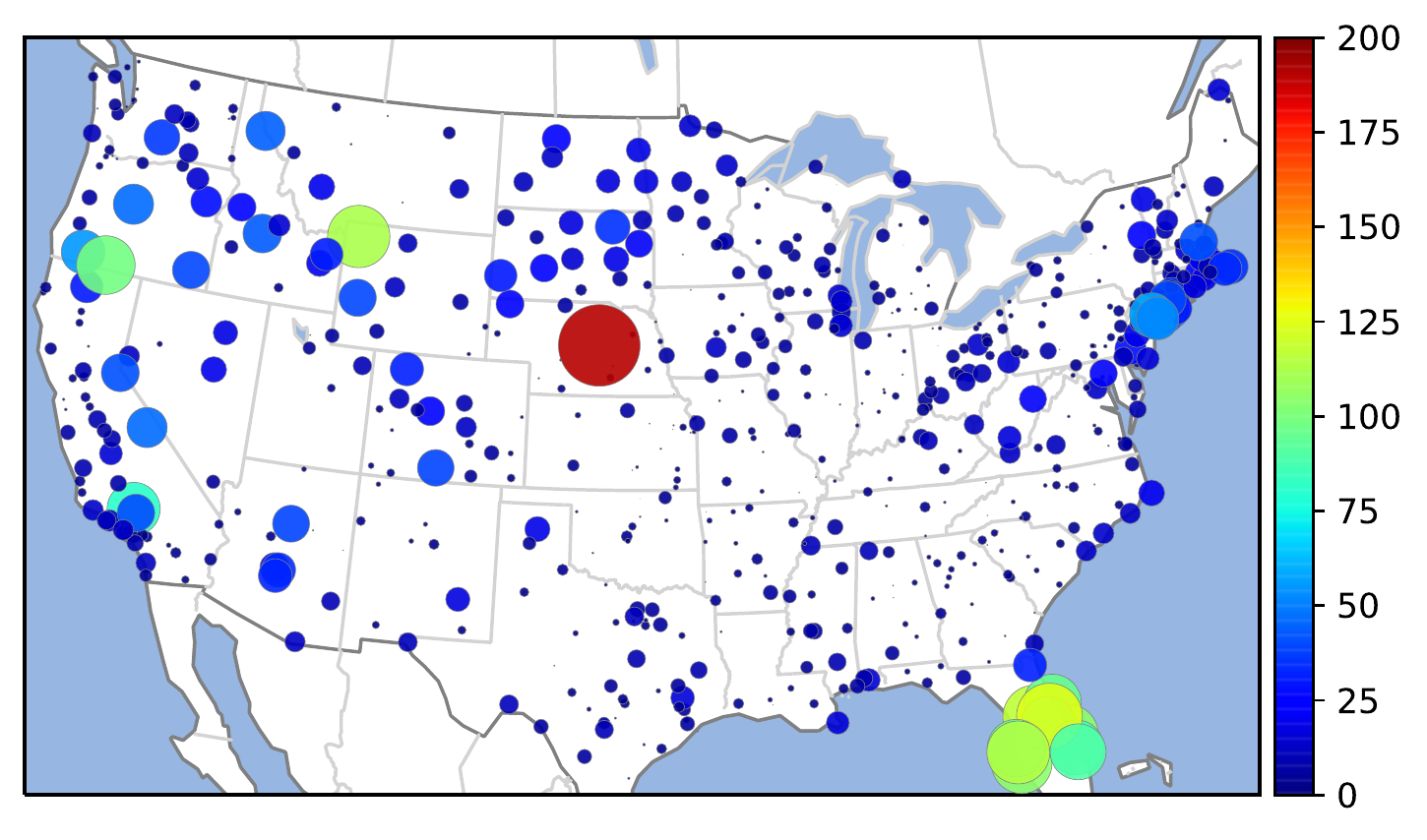}
    \subcaption{NestDAU-EN-C} \label{sfig:temp_NEDAU_int}
  \end{minipage}
  \caption{Interpolation results of U.S~temperature data. (\subref{sfig:temp_int}) and (\subref{sfig:noise}) are shown in the original scale. For clear visualization, the node size of the interpolation results are set to be proportional to the magnitude of the error.
  }
  \label{fig:temp_img_int}
\end{figure*}

The RMSE results obtained by the proposed and existing methods are summarized in Table~\ref{tab:summary_interpolate}.
The visualizations of the interpolation results are also shown in Figs.~\ref{fig:comm_img_int} and \ref{fig:temp_img_int}.

As can be seen, the proposed approaches show better RMSE than the alternatives.
For the community graph, NestDAU-TV shows better results than NestDAU-EN.
This is because NestDAU-TV reflects the prior of the graph signals, i.e., piecewise constant.
For the U.S. temperature data, NestDAU-EN is better than NestDAU-TV because the temperature data tend to be very smooth on the graph.
In particular, NestDAU-EN-C outperforms the others in all missing rates.
This implies that the proposed NestDAU presents its effectiveness beyond denoising.

\begin{table*}[tb] \centering
  \caption{Interpolation results. (average RMSEs for test data)}
  \label{tab:summary_interpolate}
  \begin{tabular}{c|c|c|ccc|ccc|ccc} \bhline{1.0pt}
  &  &  & \multicolumn{6}{c|}{Community graph} & \multicolumn{3}{c}{U.S.~temperature} \\ \cline{4-12}
  &  &  & \multicolumn{3}{c|}{\textit{noiseless}} & \multicolumn{3}{c|}{$\sigma=0.5$} & \multicolumn{3}{c}{$\sigma=9.0$} \\ \hline
   & $P$ & params/miss(\%) & 30 & 50 & 70 & 30 & 50 & 70 & 30 & 50 & 70 \\\hline\hline
   Noisy + missing & - & - & 2.034 & 2.633 & 3.092 & 2.099 & 2.691 & 3.155 & 9.143 & 9.169 & 9.170 \\ \hline
   Graph sampling theory &	- &	-	 &0.214 &	0.279 &	0.571 &	0.341 &	0.465	& 1.411	& 3.961	& 4.851	& 10.06 \\
   PnP-HD & 8 & - & 0.131 & 0.179 & 0.386 & 0.263 & 0.316 & 0.462 & 2.938 & 3.155 & 3.558 \\
   PnP-SGBF & 8 & - & 0.136 & 0.174 & 0.268 & 0.218 & 0.236 & 0.331 & 2.982 & 3.293 & 3.512 \\ \hline
   GUSC & - & 11,270 & 0.384 & 0.543 & 0.706 & 0.414 & 0.551 & 0.722 & 3.853 & 4.921 & 6.657 \\
   GUTF & - & 19,397 & 0.309 & 0.470 & 0.641 & 0.338 & 0.472 & 0.636 & 3.478 & 4.395 & 6.162 \\ \hline
NestDAU-TV-E & 8 & 168 & 0.013 & 0.025 & \textbf{0.059} & 0.077 & 0.094 & \textbf{0.140} & 2.940 & 3.177 & 3.491 \\
NestDAU-TV-C & 8 & 168 & \textbf{0.012} & \textbf{0.022} & 0.082 & \textbf{0.072} & \textbf{0.093} & 0.148 & 2.968 & 3.179 & 3.501 \\
NestDAU-EN-E & 8 & 248 & 0.077 & 0.140 & 0.160 & 0.107 & 0.185 & 0.311 & 2.907 & 3.116 & 3.466 \\
NestDAU-EN-C & 8 & 248 & 0.084 & 0.120 & 0.175 & 0.123 & 0.128 & 0.321 & \textbf{2.903} & \textbf{3.105} & \textbf{3.461} \\
\bhline{1.0pt}
  \end{tabular}
\end{table*}

\section{Concluding Remarks} \label{sec:conclusion}
In this paper, we proposed graph signal denoising and restoration methods based on ADMM and Plug-and-Play ADMM with deep algorithm unrolling, respectively.
The ADMM-based unrolled denoiser automatically controls its regularization strengths by tuning its parameters from training data.
The PnP-ADMM-based unrolled restoration is applicable to any linear degradation matrix and contains the proposed ADMM-based denoiser in its sub-module, leading to a nested DAU structure.
The unrolled restoration methods provide fully interpretable structures and have a small number of parameters with respect to fully parameterized neural networks.
The techniques only tune layer-wise trainable parameters in the iterative algorithm and do not include fully-connected neural networks.
This implies that we only need a small set of training data: It is beneficial especially for graph signals because their underlying structures often change.
In extensive experiments, the proposed methods experimentally outperform various alternative techniques for graph signal restoration.
Furthermore, we can reuse the learned parameters for graphs with different sizes.

\appendix
Here, we present some non-trivial gradient computations of the trainable parameters with respect to the learnable parameters in GraphDAU.

First, let $\mathcal{L}$ be the loss function.
Its partial derivatives with respect to parameters are given as follows using the chain rule:
\begin{align}
\frac{\partial \mathcal{L}}{\partial \gamma_{\ell}} &=
\frac{\partial \mathcal{L}}{\partial \mathbf{x}^{(\ell+1)}}
\cdot
\frac{\partial \mathbf{x}^{(\ell+1)}}{\partial \gamma_{\ell}},
\label{eq:gamma_derivative}\\
\frac{\partial \mathcal{L}}{\partial \beta_{\ell}} &=
\frac{\partial \mathcal{L}}{\partial \mathbf{v}^{(\ell+1)}}
\cdot
\frac{\partial \mathbf{v}^{(\ell+1)}}{\partial \beta_{\ell}},
\label{eq:beta_derivative}\\
\frac{\partial \mathcal{L}}{\partial \alpha_{\ell}} &=
\frac{\partial \mathcal{L}}{\partial \mathbf{v}^{(\ell+1)}}
\cdot
\frac{\partial \mathbf{v}^{(\ell+1)}}{\partial \alpha_{\ell}}.
\label{eq:alpha_derivative}
\end{align}
From \eqref{eq:admm_graph_re1}, $\mathbf{x}^{(\ell+1)}$ is given as
\begin{equation}
  \begin{split}
\mathbf{x}^{(\ell+1)} =&
\mathbf{U}
    \text{diag}\left(
        \frac{\gamma_{\ell}}{\gamma_{\ell}+ \lambda_1},
        \ldots,
        \frac{\gamma_{\ell}}{\gamma_{\ell}+ \lambda_{N}}
    \right)
\mathbf{U}^{\top} \\
&\times
\left(
    \mathbf{y} + \frac{1}{\gamma_{\ell}}
    \mathbf{M}^{\top} (
        \mathbf{v}^{(\ell)} - \mathbf{u}^{(\ell)}
    )
\right) \\
=&
\mathbf{U}
\left[
    \text{diag}
    \left(
        \frac{\gamma_{\ell}}{\gamma_{\ell}+\lambda_1},
        \ldots,
        \frac{\gamma_{\ell}}{\gamma_{\ell}+\lambda_{N}}
    \right)
    \mathbf{U}^{\top} \mathbf{y} \right. \\
    &+
    \left.
    \text{diag}
    \left(
        \frac{1}{\gamma_{\ell}+\lambda_1},
        \ldots,
        \frac{1}{\gamma_{\ell}+\lambda_{N}}
    \right)
    \widetilde{\mathbf{x}}^{(\ell)}
\right],
\nonumber
\end{split}
\end{equation}
where $\widetilde{\mathbf{x}}^{(\ell)} = \mathbf{U}^{\top} \mathbf{M}^{\top} (\mathbf{v}^{(\ell)} - \mathbf{u}^{(\ell)})$.
Then, $\frac{\partial \mathbf{x}^{(\ell+1)}}{\partial \gamma_{\ell}}$ in \eqref{eq:gamma_derivative} is calculated as follows:
\begin{equation}
  \begin{split}
\frac{\partial \mathbf{x}^{(\ell+1)}}{\partial \gamma_{\ell}}
=&
\mathbf{U}
\left[
    \frac{\partial}{\partial \gamma_{\ell}}
    \text{diag}
    \left(
        \frac{\gamma_{\ell}}{\gamma_{\ell}+\lambda_1},
        \ldots,
        \frac{\gamma_{\ell}}{\gamma_{\ell}+\lambda_{N}}
    \right)
    \mathbf{U}^{\top} \mathbf{y} \right. \\
    &+
    \left.
    \frac{\partial}{\partial \gamma_{\ell}}
    \text{diag}
    \left(
        \frac{1}{\gamma_{\ell}+\lambda_1},
        \ldots,
        \frac{1}{\gamma_{\ell}+\lambda_{N}}
    \right)
    \widetilde{\mathbf{x}}^{(\ell)}
\right] \\
=&
\mathbf{U}
\left[
    \text{diag}
    \left(
        \frac{\lambda_1}{(\gamma_{\ell}+\lambda_1)^2},
        \ldots,
        \frac{\lambda_{N}}{(\gamma_{\ell}+\lambda_{N})^2}
    \right)
    \mathbf{U}^{\top} \mathbf{y} \right. \\
    &-
    \left.
    \text{diag}
    \left(
        \frac{1}{(\gamma_{\ell}+\lambda_1)^2},
        \ldots,
        \frac{1}{(\gamma_{\ell}+\lambda_{N})^2}
    \right)
    \widetilde{\mathbf{x}}^{(\ell)}
\right].
\end{split}
\end{equation}

We also derive the partial derivatives with respect to $\beta_{\ell}$ for both GraphDAU-TV and EN.
Let $\widetilde{\mathbf{v}}^{(\ell)}=\mathbf{M}\mathbf{x}^{(\ell+1)}+\mathbf{u}^{(\ell)}$ and $\sigma(\cdot)$ be the ReLU activation function.
The soft-thresholding operator can be represented with two ReLU functions as
\begin{equation}
\begin{split}
S_{\beta_{\ell}}(\widetilde{\mathbf{v}}^{(\ell)}_i)
=
\sigma(\widetilde{\mathbf{v}}^{(\ell)}_i - \beta_{\ell})
- \sigma(- \widetilde{\mathbf{v}}^{(\ell)}_i - \beta_{\ell}).
\nonumber
\end{split}
\end{equation}
Therefore, the auxiliary variable $\mathbf{v}^{(\ell+1)}$ is explicitly given by $\mathbf{v}^{(\ell+1)}
= S_{\beta_{\ell}}(\widetilde{\mathbf{v}}^{(\ell)})$,
and its gradient for $\beta_{\ell}$ is shown as follows:
\begin{equation}
\begin{split}
\frac{\partial \mathbf{v}_i^{(\ell+1)}}{\partial \beta_{\ell}}
=
- \sigma^{\prime}(\widetilde{\mathbf{v}}_i^{(\ell)} - \beta_{\ell})
+
\sigma^{\prime}(- \widetilde{\mathbf{v}}_i^{(\ell)} - \beta_{\ell}),
\end{split}
\end{equation}
where $\sigma^{\prime}(\cdot)$ is the derivative of $\sigma(\cdot)$.
GraphDAU-EN also contains the parameter $\alpha_{\ell}$.
By taking the partial derivative of $\mathbf{v}^{(\ell+1)}$ with respect to $\alpha_{\ell}$ is given as follows:
\begin{equation}
\begin{split}
\mathbf{v}_i^{(\ell+1)}
=
\alpha_{\ell} \left(
\sigma(\widetilde{\mathbf{v}}_i^{(\ell)} - \beta_{\ell})
- \sigma(- \widetilde{\mathbf{v}}_i^{(\ell)} - \beta_{\ell})
\right),
\nonumber
\end{split}
\end{equation}

\begin{equation}
\begin{split}
\frac{\partial \mathbf{v}_i^{(\ell+1)}}{\partial \alpha_{\ell}}
=
\sigma(\widetilde{\mathbf{v}}_i^{(\ell)} - \beta_{\ell})
- \sigma(- \widetilde{\mathbf{v}}_i^{(\ell)} - \beta_{\ell}).
\end{split}
\end{equation}

\newcommand{\noopsort}[1]{}

\end{document}